\documentclass{dmathesis}
\usepackage{fancyhdr}
\usepackage{graphicx}
\usepackage{epsfig}
\usepackage{cite}
\usepackage{slashed}
\usepackage{amsmath}
\usepackage{theorem}
\usepackage{amssymb}
\usepackage{latexsym}
\usepackage{hyperref}
\usepackage{pdfpages}
\usepackage{stackrel}
\graphicspath{{fig/}}
\usepackage{showlabel}
\usepackage[slovak,english]{babel} 
\usepackage[nottoc]{tocbibind}

\newcounter{ind}

{\theorembodyfont{\rmfamily}}
{\theorembodyfont{\rmfamily}}
{\theorembodyfont{\rmfamily}}
{\theorembodyfont{\rmfamily}}
{\theorembodyfont{\rmfamily}}
{\theorembodyfont{\rmfamily}}

\newcommand{\be}{\begin{equation}}
\newcommand{\ee}{\end{equation}}
\newcommand{\beqn}{\begin{eqnarray}}
\newcommand{\eeqn}{\end{eqnarray}}
\newcommand{\ket}[1]{{\left|{#1}\right\rangle }}

\usepackage{amsbsy} 
\newcommand{\bkav}[3]{{\left\langle{#1}\middle|{#2}\middle|{#3} \right\rangle }}
\newcommand{\av}[1]{{\left\langle{#1} \right\rangle }}
\def \Lag {{\cal L}} 
\def \FD {{\cal D}} 
\def \Ope {{\cal O }} 
\def \Tr {\text{Tr}}

\newcommand{\im}{\mathop{\mbox{Im}}}

\newcommand{\tinymsbar}{{\overline{\mbox{\tiny\rm{MS}}}}}

\begin{document}
\pagenumbering{roman}
\setcounter{page}{1}
\newpage
\thispagestyle{empty}

\begin{center}
{\large UNIVERSIT\"AT BIELEFELD}
\vglue0.2cm

{\large Fakult\"at f\"ur Physik}
\end{center}

\begin{figure}[!h]
  \begin{center}
    \includegraphics[width=4cm]{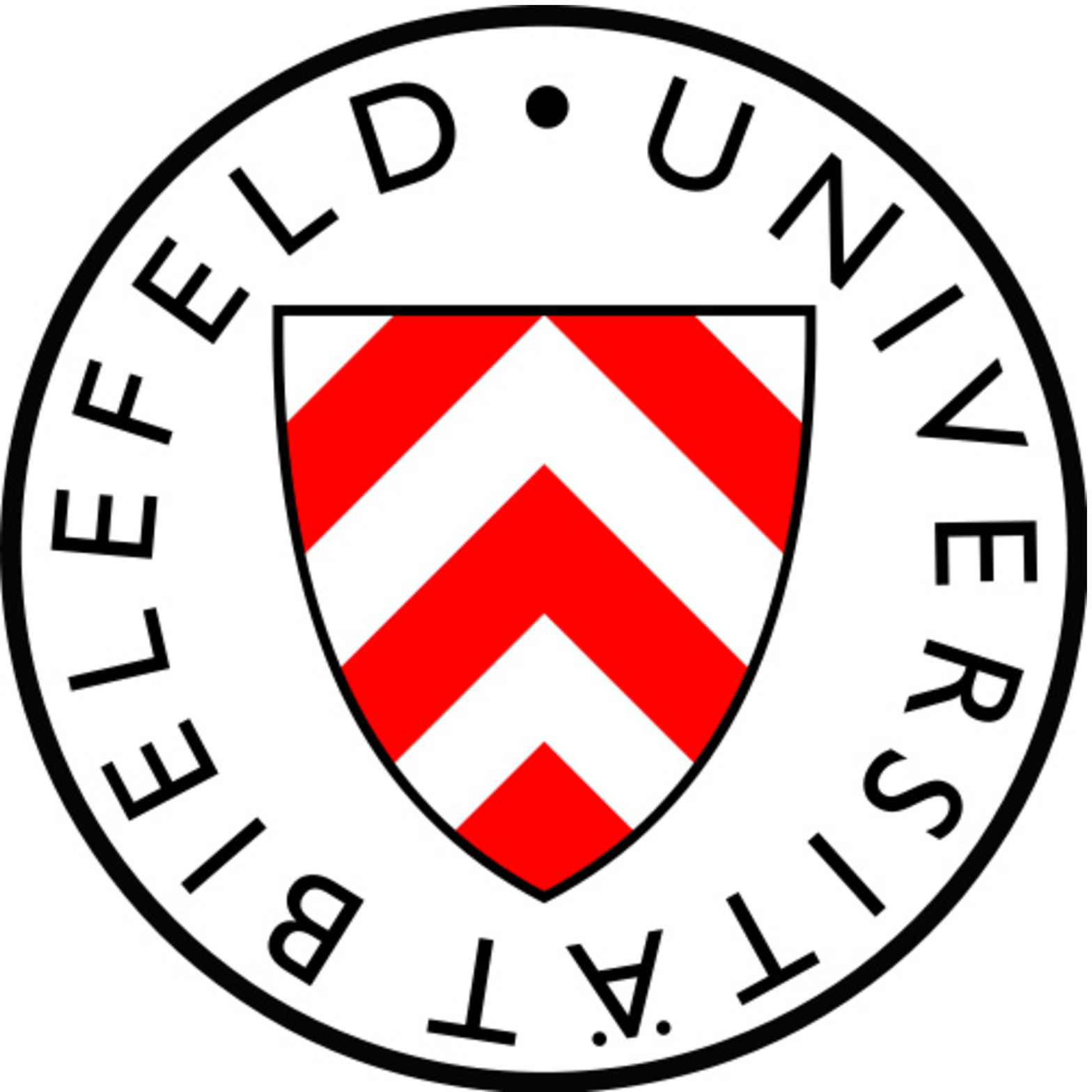}
  \end{center}
\end{figure}

\begin{center}
\vglue1.5cm
{ \LARGE{Energy Momentum Tensor Correlators}\\
\vglue0.15cm
       \LARGE{ in Improved Holographic QCD}}
\end{center}

\vglue2cm
{
\vglue1cm
\begin{center}
In Attainment of the Academic Degree\\
Doctor rerum naturalium

\end{center}
\vglue4.5cm

\begin{center}
{\large Martin Kr\v s\v s\'ak}\\

{\large July 2013} 

\end{center}

\newpage
\thispagestyle{empty}
\addcontentsline{toc}{chapter}{\numberline{}Abstract}
\begin{center}
\textbf{ {\Large Energy Momentum Tensor Correlators in Improved Holographic QCD}}\\\vspace*{0.4cm}
\large Martin Kr\v s\v s\'ak\\
\vspace*{0.4cm}
Supervisor: PD Dr. Aleksi Vuorinen\\
\vspace*{0.4cm}
{\Large \textbf{Abstract} }
\end{center}

In this thesis, we study the physics of the quark gluon plasma (QGP) using holographic methods borrowed from string theory. We start our discussion by motivating the use of such machinery, explaining how recent experimental results from the LHC and RHIC colliders suggests that the created QGP should be described as a strongly coupled liquid with small but nonvanishing bulk and shear viscosities. We argue that holographic dualities are a very efficient framework for studying transport properties in such a medium.

Next, we introduce the underlying physics behind all holographic dualities, the AdS/CFT correspondence, and then motivate the necessity of implementing conformal invariance breaking in them. After this, we  present the phenomenologically most successful holographic model of the strong interactions --- Improved Holographic QCD (IHQCD). 

Working within IHQCD, we next move on to calculate energy momentum tensor correlators in the bulk and shear channels of large-$N_c$ Yang-Mills theory. In the shear channel, we confront our results with those derived in strongly coupled ${\mathcal N}=4$ Super Yang-Mills theory as well as weakly interacting ordinary Yang-Mills theory. Close to the critical temperature of the deconfinement transition, we observe significant effects of conformal invariance breaking. In the bulk channel, where the conformal result is trivial, we make comparisons with both perturbative and lattice QCD. We observe that lattice data seem to favor our holographic prediction over the perturbative one over a wide range of temperatures.

\newpage
\vspace*{2cm}
\noindent The work presented here is based upon the journal papers
\begin{itemize}
\item [1]  K.~Kajantie, M.~Kr\v s\v s\'ak, A.~Vuorinen,\textit{ Energy momentum tensor correlators in hot Yang-Mills theory: holography confronts lattice and perturbation theory},  JHEP \textbf{1305 }(2013) 140, arXiv:1302.1432 [hep-ph],
\item [2]K.~Kajantie, M.~Kr\v s\v s\'ak, M.~Veps\"al\"ainen and A.~Vuorinen, \textit{Frequency and wave number dependence of the shear correlator in strongly coupled hot Yang-Mills theory}, Phys. Rev. D {\bf 84}, 086004 (2011), arXiv:1104.5352 [hep-ph],
\end{itemize} 
as well as the works published in conference proceedings
\begin{itemize}
\item [3] M.~Kr\v s\v s\'ak,
  \textit{Viscosity Correlators in Improved Holographic QCD},  Proceedings of the \textit{Barcelona Postgrad Encounters on Fundamental Physics},  arXiv:1302.3181 [hep-ph],
\item [4] A.~Vuorinen, M.~Kr\v s\v s\'ak, Y.Zhu,
\textit{Bulk and shear spectral functions in weakly and strongly coupled Yang-Mills theory}, Proceedings of the workshop \textit{Xth Quark Confinement and the Hadron Spectrum}, PoS ConfinementX (2012) 191, arXiv:1301.3449 [hep-ph]. 
\end{itemize}

\tableofcontents
\newpage

\pagenumbering{arabic}
\setcounter{page}{1}
\setcounter{equation}{0}
\chapter{Motivation\label{ch1}}

\begin{figure}
\centering
\includegraphics[width=1\textwidth]{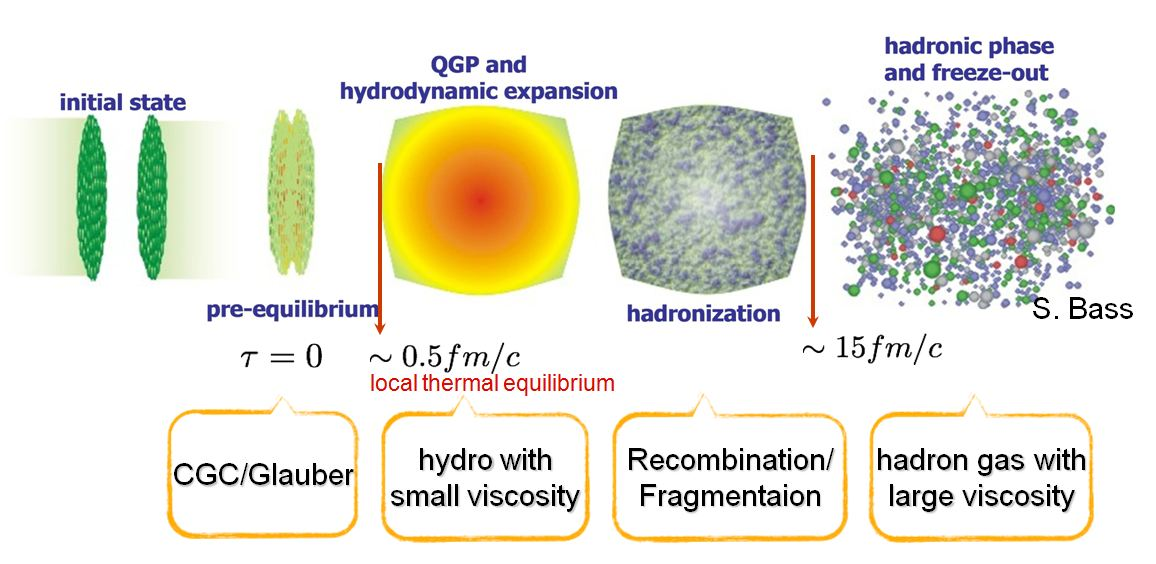}
\label{fig1}
\caption{A schematic cartoon of a heavy ion collision. The figure has been taken from a presentation of S.~Bass. }
\end{figure}
We live in an exciting time for particle physics --- and particle physicists. Only recently did the largest particle accelerator ever built, the Large Hadron Collider (LHC) of CERN, start its TeV scale proton-proton and heavy ion runs, with its numerous experiments, such as ATLAS, CMS and ALICE, measuring and analyzing the collision products. Of these collaborations, ATLAS and CMS are primarily dedicated to the study of p-p collisions, aiming to provide answers to two fundamental questions in elementary particle physics. The first of these is to verify the existence of the Higgs boson, which by now seems to indeed have been accomplished \cite{Chatrchyan:2012ufa,Aad:2012tfa}. The second question on the other hand has to do with the extensions of the Standard Model, in particular the possible existence of supersymmetry at the TeV scale. If discovered, supersymmetry would not only have the potential to solve many phenomenologically important problems in modern high energy physics, but is also a direct prediction of the most promising candidate for a unified `theory of everything' --- string theory. Unfortunately, so far there have been no traces of any beyond the Standard Model physics in the ATLAS and CMS results \cite{Strege:2012bt}, indicating that if present in the Nature at all, supersymmetry can most likely only be restored at energies above 2 TeV.

In contrast to ATLAS and CMS, the ALICE experiment is mainly dedicated to study the collisions of heavy ions, i.e.~the process schematically depicted in fig.~\ref{fig1}. Here, one starts with two colliding (lead) nuclei that are highly Lorentz contracted as illustrated by their pancake shapes in the figure. Immediately after the collision, the system is  thought to be in a complicated initial state, the \textit{color glass condensate}, that can be described using the so-called McLerran-Venugopalan model (see e.g.~\cite{Iancu:2003xm}). Through their mutual  interactions, the  constituents of the nuclei then exchange energy and momentum, eventually leading to the expanding system achieving local thermal equilibrium. This state of matter, where the quarks and gluons are liberated and can move freely, is called the \textit{quark gluon plasma} (QGP). The fireball consisting of the QGP then undergoes a rapid expansion, described most efficiently through relativistic hydrodynamic simulations, and finally leads to recombination and hadronization once the temperature of the system falls below a certain critical value.

Of our interest in this thesis is the description of a near-thermal QGP, which is believed to have existed also in the early universe, shortly after the Big Bang. Its creation in a laboratory environment was first achieved by the experiments at the Relativistic Heavy Ion Collider (RHIC) of the Brookhaven National Laboratory in 2004 \cite{Adams:2005dq,Back:2004je,Arsene:2004fa,Adcox:2004mh}, and later of course by the ALICE experiment at the LHC \cite{Muller:2012zq}. 

A common conclusion from the analysis of all experiments conducted so far \cite{Tannenbaum:2012ma,Muller:2012zq} has been that a successful hydrodynamical description of the expanding fireball requires the incorporation of a small but nonvanishing shear viscosity. This poses a direct challenge to the theory community, as one would clearly like to have a first principles prediction for this and other transport coefficients. The smallness of the viscosity --- the experiments suggesting its ratio to the entropy being of the order $\eta/s\lesssim 0.2$ --- has in addition been somewhat of a surprise, considering that
expectations from perturbative QCD calculations were suggesting parametrically larger values $\eta/s\sim 1/(g^4\ln\,g)\gtrsim 1$ \cite{Aarts:2002cc,Arnold:2003zc}.

By now it has been rather widely accepted that observations such as the smallness of the shear viscosity and the rapid apparent thermalization of the heavy ion collision product indicate that the created medium may in fact not be amenable to a description via weakly interacting quasiparticles. Rather, it may be more natural to think of the QGP as a strongly coupled liquid, indicating that fundamentally nonperturbative methods should be used to determine its properties.  Due to the restriction of lattice QCD to the Euclidean formulation of the theory, this has unfortunately meant that many interesting dynamical quantities, and even most transport coefficients, are out of the reach of traditional field theory tools (though some progress has lately been achieved in the lattice determination of e.g.~the shear viscosity \cite{Meyer:2007ic,Meyer:2007dy,Meyer:2011gj}.

Whereas lattice simulations are demanding numerical calculations requiring considerable computer power, a complementary and rather elegant approach to strongly coupled field theory was proposed some 15 years ago. This is of course the by now famous \textit{AdS/CFT correspondence} \cite{Maldacena:1998im}, which relates the physics of strongly interacting field theories to a weakly coupled gravitational theory in one dimension higher. Using this method, many physically interesting quantities have been computed in the infinitely strongly coupled limit of the conformal ${\mathcal N}=4$ Super Yang-Mills theory, with some predictions even conjectured to be rather universal. One famous example of this universality is the ratio of the shear viscosity to entropy, which obtains the value $\eta/s=1/(4\pi)$ in all theories with holographic duals (with two-derivative gravitational actions).

While the $1/(4\pi)$ prediction is consistent with experimental results, one needs to exercise caution in translating such insights to real world QCD (and QGP). This is primarily due to the conformality of the ${\mathcal N}=4$ theory, which is a property of QCD only at very high energies, and obviously calls for the development of non-conformal holographic models. This is exactly what we aim to do in this thesis, namely study the \textit{Improved Holographic QCD} (IHQCD) model of \cite{Gursoy:2007cb,Gursoy:2007er,Gursoy:2010fj}, which not only incorporates a dynamical breaking of conformality, but whose entire structure is designed to systematically mimic the most crucial properties of QCD. In the later chapters of this thesis, we will use this model to study the transport properties of the QGP, and in particular illustrate the effects of the broken conformal invariance on the predictions of holography.

This thesis is organized as follows.  
\begin{itemize}
\item In chapter~\ref{sec2}, we start with an introduction to a selected class of issues in quantum field theory that are required to understand the physics of strong interactions at finite temperature. 
\item In chapter~\ref{sec3}, we present an introduction to holography. We start with a discussion of classical general relativity, and by studying the thermodynamics of black holes motivate the holographic principle. Next, we move on to the best understood realization of the holographic principle obtained from string theory, the AdS/CFT correspondence.
\item In chapter~\ref{sec4}, we illustrate how the IHQCD model is developed to mimic the crucial properties of QCD. We discuss briefly the thermodynamic properties of IHQCD and our method of numerical integration of the equations of motion within the model.
\item In chapter~\ref{ch6}, we present our original research, which amounts to the calculation of various energy momentum tensor correlators in IHQCD. We confront our results with those of perturbative and lattice QCD where available. 
\item In chapter~\ref{ch7}, we finally draw conclusions.
\end{itemize}

\setcounter{equation}{0}
\chapter{Elements of Quantum Field Theory\label{sec2}}
To begin, we present a  brief review some of the most important elements of Quantum Field Theory (QFT), with emphasis on concepts that we will encounter in the following chapters from a holographic viewpoint. We begin our treatment from scalar field theories, and then move on to gauge fields and their interactions, covering briefly also the large-$N_c$ limit of QCD as well as supersymmetry, before ending up with the basics of thermal field theory. The main references for this chapter are the excellent textbooks \cite{Peskin:1995ev,Kiselev:2000ts} on a zero-temperature QFT, as well as the classic introduction to finite-temperature field theory, \cite{Kapusta:2006pm}.
\section{Quantum Fields at Zero Temperature}
Consider a real scalar field $\phi(x)$, with $x=(t,\vec{x})$ denoting the coordinate of a four-dimensional Minkowskian spacetime with signature $(-,+,+,+)$. We write the corresponding action in the form\begin{equation}
S=\int \Lag (\phi,\partial_\mu \phi) d^4x, \label{scalaraction}
\end{equation}
where $\Lag (\phi,\partial_\mu \phi)$ stands for the (as of yet unspecified) \textit{Lagrangian density} for the scalar field. 
\subsection{Green's functions}
The most elementary called \textit{Green's function} or \textit{correlator}, standing for the amplitude for a particle to propagate from a spacetime point $x$ to $y$, is given by
\begin{equation}
G(x,y)=\bkav{0}{\phi(x)\phi(y)}{0} \equiv \av{\phi(x)\phi(y)}, \label{green1}
\end{equation}
where $\ket{0}$ denotes the ground (vacuum) state of the field theory.  This function can, however, be seen to obtain nonzero values also outside the lightcone, marking a violation of causality. Thus, we often rather work with the vacuum expectation value of the commutator $\av{[\phi(x),\phi(y)]}$, for which the noncausal parts can be seen to cancel. 

Further, we can define two \textit{causual correlators}, the \textit{retarded} and \textit{advanced} Green's functions, corresponding to particles moving forward (backward) in time,
\begin{equation}
G^R(x,y)=\theta(x^0-y^0)\av{[\phi(x),\phi(y)]}, \label{greenret}
\end{equation}
\begin{equation}
G^A(x,y)=\theta(y^0-x^0)\av{[\phi(y),\phi(x)]}, \label{greenadv}
\end{equation}
where $\theta$ is the Heaviside step function. Another important correlator, playing an important role in QFT, is the \textit{Feynman Green's function} or \textit{time-ordered correlator}
\begin{equation}
G^F(x,y)= \av{T \phi(x)\phi(y)}=\theta(x^0-y^0)\av{\phi(x)\phi(y)} +\theta(y^0-x^0)\av{\phi(y)\phi(x)},
\end{equation}
where $T$ is the time-ordering symbol.

The most convenient way to evaluate the time-ordered correlator in the \textit{path integral} formulation of QFT is using the \textit{generating functional} of Green's functions, $Z[J]$. This quantity is defined by
\begin{equation}
Z[J]= \int \FD \phi \exp \left\lbrace i \int d^4 x [\Lag + J(x)\phi(x)]\right\rbrace, \label{genfunc}
\end{equation}
where $\FD \phi$ denotes functional integration over all field configurations. Using this function, we can calculate e.g.~the vacuum expectation value of the field operator via
\begin{equation}
\av{ \phi(x)}=\left. \frac{1}{Z_0}\left(-i\frac{\delta}{\delta J(x)} \right)Z[J]\right|_{J=0}, \label{expvalue}
\end{equation}
where $Z_0=Z[J=0]$ and $\delta / \delta J$ denotes a functional derivative. The time-ordered correlation function is further obtained as a second functional derivative
\begin{equation}
\av{T \phi(x)\phi(y)}=\left. \frac{1}{Z_0}\left(-i\frac{\delta}{\delta J(x)} \right)\left(-i\frac{\delta}{\delta J(y)} \right)Z[J]\right|_{J=0}. \label{propagator}
\end{equation}
We can see that these results can be easily generalized to higher-point correlator functions by simply taking more functional derivatives, and even for arbitrary (usually local) operators, writing
\begin{equation}
\av{T \Ope(x_1),\dots,\Ope(x_n)}=\left. \frac{1}{Z_0}\left(-i\frac{\delta}{\delta J(x_1)} \right)\dots\left(-i\frac{\delta}{\delta J(x_n)} \right)Z[J]\right|_{J=0}. \label{npoint}
\end{equation}
For us, particlularly important operators $\Ope$, will be components of the  energy momentum tensor $T_{\mu\nu}$, as well as the field strength tensor squared, $F^2$, in the case of gauge theory.  
\subsection{Renormalization and Running Coupling}
The simplest QFT one can write down is clearly a theory for one non-interacting scalar field, described by the Lagrangian
\begin{equation}
\Lag_0=\frac{1}{2}(\partial_u \phi)^2+ \frac{1}{2} m \phi^2 .
\end{equation}
This is an exactly solvable model, and, as it turns out, actually the only example of a theory that can be fully analytically solved using standard methods. For example, if we add a quartic interaction to the Lagrangian
\begin{equation}
\Lag=\Lag_0+\Lag_I=\frac{1}{2}(\partial_u \phi)^2 +\frac{1}{2} m \phi^2 +\frac{g_B}{4!}\phi^4, \label{lagphi4}
\end{equation}  
the functional integral in (\ref{genfunc}) are no longer straightforwardly solvable. What we can do in this case is, however, to use the machinery of \textit{perturbation theory}, where we expand integrand of the generating functional $Z[J]$ in a power series in the coupling constant. This perturbative approach, however, works only if the coupling constant can be considered small.

When performing explicit calculations in a given QFT, one very often runs into ultraviolet (UV) divergences in loop diagrams, encountered when perturbative expansions are carried out beyond their leading orders.  In order to subvert the UV divergences, the parameters (and sometimes even the fields) of the theory have to be redefined in a process called \textit{renormalization}. This leads to the emergence of a corrected, or renormalized, action, which is free of divergent quantities. The coupling constant of this new action, called the \textit{renormalized coupling} $g$, is in general different from the original \textit{bare coupling} $g_B$, appearing in (\ref{lagphi4}). Upon subtracting the divergences of the latter, the renormalized coupling typically becomes a function of an energy-like parameter, the \textit{renormalization scale} $\mu$, giving rise to the term \textit{running coupling}. This dependence is encoded in the \textit{beta function} of the theory,
\begin{equation}
\beta(g)=\mu\frac{\partial g}{\partial \mu},
\end{equation}
which e.g.~ for a scalar theory with a quartic interaction reads
\begin{equation}
\beta(g)=\frac{3}{16\pi^2}g^2 + O(g^3),
\end{equation}
We observe that in this case, the coupling $g$ increases with energy, implying that at high-enough energies perturbation theory will inevitably fail. As we will see, this behavior is exactly opposite to that encountered in the theory of the strong nuclear interaction, where the coupling constant in fact decreases with energy. 
\subsection{Conformal Field Theory}
As we have seen, quantum field theories are usually defined over Minkowski spacetime, and thus they have to respect the symmetry group of this space. The symmetry group of the Minkowski space is found as the symmetry group under which the metric
\begin{equation}
ds^2=g_{\mu\nu}dx^\mu dx^{\nu},
\end{equation}
is invariant. This is the famous Poincare symmetry, which we know consists of translations as well as special orthogonal $SO(1,3)$ transformations, i.e.~the Lorentz transformations.

A conformal field theory is a theory that is invariant under more general transformations that leave the metric invariant up to a scale change
\begin{equation}
ds^2= \Omega(x)g_{\mu\nu}dx^\mu dx^{\nu}.\label{confransdef}
\end{equation}
It turns out that the conformal symmetry group is the product of translations and the group $SO(2,4)$, representing the Lorentz and scaling symmetries. The presence of the scaling symmetry means that the theory must be free of dimensionfull parameters, such as mass scales. As a consequence, the beta function of conformal field theories has to vanish.

Conformal symmetry also constraints the form of correlation functions much more than Poincare symmetry alone. For example, we find that a general two-point correlation function in conformal field theory is given by 
\begin{equation}
\av{T \Ope(x_1),\Ope(x_2)} \approx
\frac{1}{(x_1-x_2)^{2\Delta}}, \label{confcorr}
\end{equation}
where $\Delta$ is the scaling dimension of the operator $\Ope(x)$.

\subsection{Energy Momentum Tensor}
Consider next in somewhat more detail the spacetime translations 
\begin{equation}
x^{\mu} \to x^\mu + a^\mu(x),
\end{equation}
under which the scalar field transforms as 
\begin{equation}
\phi \to \phi + a^\mu(x)\partial_\mu \phi.
\end{equation}
Using \textit{Noether's  theorem}, which states that to each symmetry of a Lagrangian corresponds a conserved current, we find that varying the action (\ref{scalaraction}) with respect to $a^\mu$ provides us with a conservation law 
\begin{equation}
\partial_\nu \theta^{\mu\nu}=0, \label{consveq}
\end{equation} 
where 
\begin{equation}
\theta^{\mu\nu}=\frac{\partial \Lag}{\partial \left( \partial_\nu \phi \right)} \partial^\mu\phi-g^{\mu\nu}\Lag \label{canten}
\end{equation}
is the \textit{canonical energy momentum tensor}. We note that the canonical energy momentum tensor is not symmetric by definition, which can be seen to result from the freedom to add any divergenceless quantity to $\theta_{\mu\nu}$. Adding a properly defined such quantity, we can however define a \textit{symmetric} energy momentum tensor $T_{\mu\nu}$. It can be shown, that this symmetric energy momentum tensor can be alternatively derived through a variation of the action with respect to the spacetime metric,
\begin{equation}
T^{\mu\nu}= -\frac{2}{\sqrt{-g}}\frac{\delta \sqrt{-g}\Lag}{\delta g_{\mu\nu}} \label{metem}
\end{equation}
where $g$ is the determinant of $g_{\mu\nu}$. We call the energy momentum tensor derived in this way the \textit{metric energy momentum tensor}. From now on, unless specified otherwise, when we talk about the energy momentum tensor, we will have the symmetric, or metric, energy momentum tensor in mind.

\section{Gauge Theory}
Nearly all fundamental forces of the nature can be understood through theories with a gauge symmetry. The simplest example of this is  electromagnetism, where we have only one gauge field 
\begin{equation}
A_\mu=(\phi,\vec{A}),
\end{equation} 
with the field strength 
\begin{equation}
F_{\mu\nu}=\partial_\mu A_\nu -\partial_\nu A_\mu.
\end{equation}
The electric and magnetic fields are then given by
\begin{equation}
E_i=-F_{i0}, \ \ \ \ B_i=\frac{1}{2}\epsilon_{ijk}F_{jk},
\end{equation}
while the Lagrangian density reads 
\begin{equation}
\Lag=- \frac{1}{4} (F_{\mu\nu})^2.
\end{equation}
This theory has a gauge symmetry under the $U(1)$ group, meaning that the field strength $F_{\mu\nu}$ is invariant under the local gauge transformation 
\begin{equation}
A_{\mu}(x)\to A_{\mu}(x)-\frac{1}{e}\partial_\mu \alpha(x). \label{gaugetransf}
\end{equation}

The above relations define a pure Maxwell theory of photons, to which we have to add fermions (electrons) and specify the coupling between these two. Using the \textit{minimal coupling} prescription, we obtain form here Quantum Electrodynamics, described by the Lagrangian density
\begin{equation}
\Lag_{QED}=\bar{\psi}(i  \slashed{D}-m)\psi -\frac{1}{4} (F_{\mu\nu})^2,
\end{equation}
where $\slashed{D}=\gamma^\mu D_\mu$, with $\gamma^\mu$ the so-called Dirac matrices  and $D_\mu$ the \textit{gauge covariant derivative}
\begin{equation}
D_\mu = \partial_\mu + ieA_\mu .
\end{equation}
The gauge symmetry in this case is the requirement that $\Lag_{QED}$ be invariant under the simultaneous transformations (\ref{gaugetransf}) of the gauge field and of the following transformation of the fermions 
\begin{equation}
\psi(x)\to e^{i \alpha(x)}\psi(x). \label{rot4}
\end{equation}
The symmetry group of the transformation (\ref{rot4}) is the unitary group $U(1)$, and hence we will call electrodynamics a \textit{gauge theory with a $U(1)$ symmetry group}.
\subsection{Yang-Mills Theory}
If we want to have more gauge fields in our theory, we have to consider their respective commutation relations. If all fields commute with each other, we call such a theory \textit{abelian}. Electrodynamics is an example of an \textit{abelian gauge theory}, since we have only one gauge field $A_{\mu}$ that trivially commutes with itself. The generalization of this construction to the non-abelian case is in general dubbed \textit{Yang-Mills Theory}, where we have set of Yang-Mills fields $A_\mu^a$, with $a$ an index belonging to some nonabelian gauge group. The gauge group is spanned by the generators $T^{a}$ belonging some representation and satisfying the Lie algebra of the group,
\begin{equation}
[T_a,T_b]=f_{ab}^{\ \ c}T_c, \label{liealg}
\end{equation}
where $f_{ab}^{\ \ c}$ are the corresponding structure constants.

The most important examples of non-abelian symmetry groups are the special orthogonal group $SO(N_c)$ and the special unitary group $SU(N_c)$. We will be interested primarily in the last one, since $SU(3)$ is the gauge symmetry group of Quantum Chromodynamics, i.e.~the theory of strong interactions. In this case, the index $a$ is called a \textit{color index} and runs from $1$ to $N_c^2-1$. The generators $T^a$ of the fundamental representation can be represented by $N_c\times N_c$ traceless Hermitian matrices, i.e.~the Pauli matrices for $SU(2)$ and the Gell-Mann matrices for $SU(3)$.

The adjoint representation of a simple Lie group such as $SU(N_c)$ is defined through $(T^a)_{bc}=f^a_{\ bc}$. This is important because the gluon fields $A_\mu^a$ live in the adjoint representation, giving rise to the field strength 
\begin{equation}
 F^a_{\mu\nu} \,\equiv\,  \partial_\mu A^a_\nu - \partial_\nu A^a_\mu + g_{YM} f^{abc} A^b_\mu A^c_\nu\, .
\end{equation}
Under a gauge transformation $\alpha_a(x)$, the gluon fields transform as 
\begin{equation}
A^{\mu}_a(x)\to A^\mu_{a}(x) + f_{abc}A^{\mu}_b\alpha_c(x) +\frac{1}{g_{YM}}\partial^\mu \alpha_a(x), \label{gaugetransfYM}
\end{equation}
implying that the field strength $ F^a_{\mu\nu}$ is \textit{not} invariant under gauge transformation, but instead 
\begin{equation}
 F^a_{\mu\nu}\to  F^a_{\mu\nu} - f^{abc}\alpha^b F^c_{\mu\nu}.
\end{equation}
The trace of the square of the field strength operator, $F^a_{\mu\nu} F^a_{\mu\nu}$, is however invariant, ensuring that Lagrangian density 
\begin{equation}
\Lag_{YM} = - \frac{1}{4} F^a_{\mu\nu} F^a_{\mu\nu},
\end{equation}
is also invariant. This Lagrangian density describes \textit{pure Yang-Mills theory} or \textit{gluodynamics}, as so far we do not have any fermionic matter in the theory.

The energy momentum tensor for pure Yang-Mills theory is given by
\begin{equation}\label{emtYM}
 T_{\mu\nu}(x)  =\theta_{\mu\nu}(x)+\frac{1}{4}\delta_{\mu\nu}\theta(x)\, ,
\end{equation}
with
\begin{eqnarray}
\theta_{\mu\nu}(x)&=&\frac{1}{4}\delta_{\mu\nu}F_{\rho\sigma}^aF_{\rho\sigma}^a-
F_{\mu\alpha}^aF_{\nu\alpha}^a,\\
\theta(x)&=&\frac{\beta(g_{YM})}{2g}F_{\rho\sigma}^aF_{\rho\sigma}^a,
\end{eqnarray}
where we have separated the traceless part $\theta_{\mu\nu}$  and the anomalous trace $\theta$ proportional to the $\beta$-function of the theory. The $\beta$-function  of the pure Yang-Mills theory is in turn given by
\begin{equation}
\beta(g_{YM})=-\frac{11N_c}{48\pi^2}g_{YM}^3 + O(g_{YM}^4).
\end{equation} 
We can see that the sign of the beta-function is negative, and as a consequence of this the strength of the gauge interaction decreases with energy. This means that perturbation theory works in the Yang-Mills theory only at high enough energies --- the completely opposite of the case of scalar field theory or QED. This property of Yang-Mills theory is  known as \textit{asymptotic freedom}.

\subsection{Quantum Chromodynamics}
Quantum chromodynamics (QCD) is the theory that describes the interactions between quarks and gluons. Formally, it is an $SU(N_c=3)$ Yang-Mills theory coupled to $N_f=6$ flavors of fundamental quarks, and thus has the Lagrangian density
\begin{equation}
\Lag_{QCD}=\bar{\psi}_i(i  \slashed{D}_{ij}-m\delta_{ij})\psi_j - \frac{1}{4} F^a_{\mu\nu} F^a_{\mu\nu}, \label{QCDLag}
\end{equation}
where $\slashed{D}_{ij}=\gamma^\mu (D_\mu)_{ij}$ and 
\begin{equation}
(D_\mu)_{ij} =\delta_{ij} \partial_\mu + i g_{QCD} (T^a)_{ij}A^a_\mu.
\end{equation}
The beta function for this theory, given for the sake of generality for unspecified $N_c$ and $N_f$, reads
\begin{equation}
\beta(g_{QCD})=-\left(\frac{11N_c}{3}-\frac{2}{3}N_f\right)\frac{g_{QCD}^3}{16\pi^2}+ O(g_{QCD}^4).\label{QCDbeta}
\end{equation}

From (\ref{QCDbeta}), we see that the beta function is still negative for the physical case of six quark flavors and three colors, indicating that QCD is, just as pure $SU(N_c)$ Yang-Mills theory, asymptotically free.  For low energies, the strong coupling constant on is the other hand typically of order $g\approx 1$ (even diverging at some critical scale $\mu$), and thus the theory must be treated nonperturbatively. A good example of this is zero-temperature nuclear physics, where the properties of e.g.~protons and neutrons cannot be accessed using simple perturbation theory. It is in fact an extraordinary difficult task to calculate even the proton mass using only the structure of the underlying theory and the quark masses as input. 

Due to asymptotic freedom, we can, however, use perturbative methods to understand processes well above the QCD energy scale $\Lambda_{QCD}\approx 200\ \rm{MeV}$. This is an intriguing consequence of asymptotic freedom, implying that it is e.g.~easier to describe at least some features of ultrarelativistic heavy ion collisions than the structure of a proton.  This is, of course, a very unsatisfactory situation, and there is a great need to develop nonperturbative methods to access the physics of the strong interaction at low energies. In this context, the holographic principle, introduced in the next chapter, is a prime example of a relatively recent nonperturbative method, which has been used to gain insights into the strongly coupled regime of QCD.  

\subsection{Large-$N_c$ Limit \label{thooftlimit}}
It was realized by 't Hooft \cite{'tHooft:1973jz} that $SU(N_c)$ gauge theory simplifies in the  $N_c\to \infty$ limit --- an observation that will turn out to play a major role in holography. In our previous discussion, we absorbed the gauge coupling $g_{YM}$ into the definition of the field strength $F^a_{\mu\nu}$. In order to understand, why Yang-Mills theory simplifies in the large-$N_c$ limit, we will now rewrite the action with a different normalization of the gauge field, leading to the Lagrangian
\begin{equation}
\Lag_{YM} = -\frac{1}{g_{YM}^2} \frac{1}{4} Tr F^2,
\end{equation}
and the partition function
\begin{equation}
Z=\int \FD A_{\mu}\exp\left(-\frac{1}{g_{YM}^2} \frac{1}{4} Tr F^2\right).
\end{equation}

Next, we introduce the 't Hooft  coupling
\begin{equation}
\lambda \equiv g_{YM}^2 N_c\,.
\end{equation}
which allows us to make the important observation that the logarithm of the partition function can be expanded in powers of $1/N_c$ as
\begin{equation}
\log Z=\sum_{h=0}^\infty N_c^{2-2h}f_h(\lambda)= N_c^2 f_0(\lambda)+f_1(\lambda) + \frac{1}{N_c^2}f_2(\lambda) + \dots ,
\end{equation}
where the $f_h(\lambda)$ are functions of $\lambda$ only.  What is remarkable about this expansion is that, at a fixed $\lambda$, Feynman diagrams are organized by their topologies. In particular,  $f_0(\lambda)$ contains only the simplest diagrams that can be drawn on a planar surface without crossing any lines (so-called \textit{planar diagrams}). We can see that in the limit $N_c\to\infty$, the partition function will be dominated by these diagrams and we can neglect the more complicated non-planar diagrams. This is called the \textit{large-$N_c$ limit} or \textit{planar limit} of $SU(N_c)$ Yang-Mills theory.
\subsection{Supersymmetry and $\mathcal{N}=4$ SYM theory\label{symch}}
Poincare symmetry is defined by the generators $J_{ab}$ of the $SO(1,3)$ Lorentz group as well as the generators of the translation group, $P_a$. In addition, we have seen that in QFTs there exist internal symmetries, represented by the generators $T_c$, such as the local $SU(N_c)$ gauge symmetry of Yang-Mills, satisfying the Lie algebra of (\ref{liealg}). It has been shown that it is not possible to combine spacetime and internal symmetries into a larger group, such that $[T_c,P_a]\neq 0$, $[T_c,J_{ab}]\neq 0$. This result is known as the \textit{Coleman-Mandula theorem} \cite{Coleman:1967ad}. 

Despite the above, it turned out that if we generalize the notion of a Lie algebra to a \textit{graded Lie algebra}, with some generators $Q^a_\mu$ satisfying anticommutation laws, the theorem can be evaded. It turns out that these generators $Q^a_\mu$, called supercharges, have a natural representation as spinors, and thus produce another spinor field upon acting on a bosonic field. This means that these exotic generators $Q^a_\mu$ provide a symmetry between bosons and fermions, called \textit{supersymmetry}. For a more detailed explanation, see e.g.~\cite{Martin:1997ns}.

For our limited purposes, it is sufficient to know that supersymmetric transformations  turn fermionic fields into bosonic, and vice versa. 
A supersymmetric theory with $\mathcal{N}$ supercharges has a global $U(\mathcal{N})$ symmetry, called the $R$-symmetry. It turns out that in four dimensions (without gravity) we can have at most four supercharges, i.e.~$\mathcal{N}=4$. A particularly interesting theory is generated when we consider the supersymmetric extension of $SU(N_c)$ Yang-Mills theory \cite{Brink:1976bc} with this maximal number of four supercharges, called the $\mathcal{N}=4$ supersymmetric Yang-Mills (SYM) theory \cite{Kovacs:1999fx}. 

The bosonic part of the Lagrangian of the ${\mathcal N}=4$ SYM theory is given by
\begin{equation}
\Lag=\frac{1}{g_{YM}^2} \mathrm{Tr} \left(\frac{1}{4} F^{\mu\nu}F_{\mu\nu} + \frac{1}{2}D_\mu\phi^{i}D^\mu\phi^{i} + \left[\phi^{i},\phi^{j}\right]^2  \right). \label{ymsymaction}
\end{equation}
It turns out that in this theory, the $\mathcal{N}=4$  supersymmetry prevents gauge couplings to obtain  corrections at the quantum level, and  in fact  makes the theory conformal. Thus the symmetry group of  $\mathcal{N}=4$ SYM theory  is the product of the conformal symmetry group and the $U(4)$ group corresponding to the $R$-symmetry, whose bosonic part is $SO(6)$, 
\begin{equation}
SO(4,2)\times SO(6).
\end{equation}
We will see that this theory will play a highly special role in holography.

Let us finally mention that we have so far discussed how one can add supersymmetry into a gauge theory. However, we can in fact also gauge supersymmetry itself, with the resulting theory being so-called \textit{supergravity}, described by the action (\ref{sugra}). For details of the construction of supergravity theories, see e.g.~\cite{Bagger:1990qh}.

\section{Finite Temperature Field Theory}
Consider next the \textit{Wick rotation}
\begin{equation}
(t,\vec{x})\to (-i\tau,\vec{x}), \label{Wick}
\end{equation}
under which the 3+1-dimensional Minkowski spacetime metric turns into a four-dimensional Euclidean one. Under this transformation, we find that the generating functional (\ref{genfunc}) becomes
\begin{equation}
Z^E[J]= \int \FD \phi \exp \left\lbrace - \int d^4 x [\Lag^E + J(x)\phi(x)]\right\rbrace, \label{partfunc}
\end{equation}
where $\Lag^E$ is the Euclidean Lagrangian density $\Lag^E(\tau)=\Lag(i t)$. With this action, the functional integration converges much better, since the oscillatory exponent $e^{iS}$ is replaced with a damped one, $e^{-S^E}$. While this offers us some computational advantages, there is also another advantage of the Euclidean approach; namely, it allows us to make a direct connection to thermal field theory.

Indeed, eq.~(\ref{partfunc}) is the quantum field theory generalization of the well-known \textit{partition function} of statistical physics in the path integral formalism. We can namely rewrite the above relation as
\begin{equation}
Z^E[0]=\int \FD \phi e^{-S^E[\phi]}=\text{Tr}\; e^{-\beta H}
, \label{partfunc3}
\end{equation}
where we have exploited the form of the quantum mechanical time evolution operator and limited the temporal integration into a finite interval,
\begin{equation}
S^E[\phi]=\int_0^\beta d \tau\int d^3x \Lag^E.
\end{equation}
Here $H$ is the Hamiltonian, and the parameter $\beta$ is related to the temperature of the system, $\beta=1/T$.  

One assumption inherent in the above identification of the Euclidean functional integral as a partition function is that bosonic and fermionic fields satisfy the \textit{Kubo-Martin-Schwinger (KMS) relation}, i.e.~obey (anti)periodic boundary conditions in time,
\begin{equation}
\phi(0,\vec{x})=\pm \phi(\beta,\vec{x}),
\end{equation}
where the + sign corresponds to bosonic and - sign to fermionic fields. This means that we can express the time dependence of the fields in terms of Fourier sums, writing 
\begin{equation}
\phi(\tau,\vec{x})=\sum_n\phi(\omega_n,\vec{x})e^{i\omega_n t},
\end{equation}
where the discrete $\omega_n$ read 
\begin{eqnarray}
\omega_n &=& \frac{2\pi n}{\beta}, \ \ \ \ \ \  \ \ \ \ \  \ \ \  \textrm{bosonic fields}\\ 
\omega_n &=& \frac{2\pi (n+1)}{\beta}, \ \   \ \ \ \  \textrm{fermionic fields}
\end{eqnarray}
with $n$ running over all integers. These frequencies are commonly referred to as \textit{Matsubara frequencies}.

From the partition function (\ref{partfunc}), we can now calculate in principle all equilibrium thermodynamics properties of the system. For example, the pressure $p$ and entropy $S$\footnote{We use standard notation where  $S$ is used for both entropy and action. However, it should be clear from the context which is which.} are given simply by the relations
\begin{eqnarray}
p=\frac{\partial T \ln Z}{\partial V},\\
S=\frac{\partial T \ln Z}{\partial T}. \label{entrop}
\end{eqnarray}

\subsection{Euclidean Correlators\label{introec}}

It is important to stress that the Euclidean approach is best suited for the description of systems in thermal equilibrium. When performing the Wick-rotation (\ref{Wick}),  we trade the real-valued time coordinate $t$ for the Euclidean time $\tau$, which is not directly related to the dynamics of the system. A consequence of this is that we have formally only one type of correlation function available, namely the \textit{Euclidean correlator}
\begin{equation}
G^{E}(k)=\int d^4x e^{-i k.x} \av{T^E \Ope(x)\Ope(0)}, 
\end{equation}
where $T^E$ denotes Euclidean time ordering.

Despite the above, also the Minkowski-space correlators $G^R$, $G^A$ and $G^F$, defined above in the $T=0$ context, have their counterparts in thermal field theory \cite{Son:2002sd}. It in fact turns out that Euclidean and Minkowskian correlators are closely related, and e.g.the momentum space retarded correlator $G^R(k)$ can be analytically continued to complex values of $\omega$, where for $\omega=2\pi i T n$ it reduces to the Euclidean propagator 
\begin{equation}
G^{R}(2\pi i T n, \vec{k})=-G^{E}(2\pi  T n, \vec{k}).
\end{equation} 
Similarly, we can write the Euclidean imaginary time correlator in the form
\begin{equation}
G^E(\tau,0)= \int_0^\infty\frac{d\omega}{\pi}\rho(\omega)\frac{\cosh\left[\left(
\frac{\beta}{2} -\tau\right) \pi \omega\right]}{\sinh\left( \frac{\beta}{2}\omega\right)}, \label{euclidean}
\end{equation}
where 
\begin{equation}
\rho(\omega)=\im G^R(\omega,0), 
\end{equation}
is the so-called \textit{spectral density}. 

The relation (\ref{euclidean}) has the remarkable property that we can in principle find Minkowskian correlators from the (typically much simpler) Euclidean ones. In order to do so, we however need to invert the integral relation, meaning that the Euclidean correlator must be known for all frequencies, which turns out to be very often a very difficult task. For example, in lattice field theory, where one uses computer simulations to measure the Euclidean correlators, one typically only obtains the results for a finite number of points and with considerable error bars, making the above analytic continuation in practice impossible.

\section{Hydrodynamic Limit \label{chlimit}}
As the final topic of our field theory chapter, let us briefly comment on the fact that close to thermal equilibrium, many physical systems allow a description using fluid dynamics --- a theory that e.g. in the case of guark gluon plasma physics can be viewed as an effective theory for the long wavelength field modes. It is thus suitable for the description of processes involving large distances and time scales \cite{Son:2007vk},  understood here as large in comparison with the typical microscopic scales, e.g.~the temperature of the system. This means that a hydrodynamic description can be expected to be valid when $(\omega, k)\ll T$.

The equations of hydrodynamics are not derived from an action principle, but rather as conservation laws for the energy momentum tensor $T_{\mu\nu}$
\begin{equation}
\partial_\mu T^{\mu\nu}=0. \label{conservation}
\end{equation}  
Assuming local thermal equilibrium, which implies that the system can be described in terms of a spatially varying `temperature field' $T(\vec{x})$ and a local fluid velocity $u^{\mu}(x)$, we find that there are only four independent variables, since $u^{\mu}u_{\mu}=-1$. This number agrees with the number of equations in (\ref{conservation}), and implies that the energy momentum tensor in eq.~(\ref{conservation}) will be a function of only these four quantities and their derivatives.

At the lowest order in a derivative expansion, we find that the most general form of $T^{\mu\nu}$ is that of an ideal fluid, and that the conservation equation (\ref{conservation}) is equivalent to the relativistic Euler equation. As it is well-known, ideal hydrodynamics cannot describe dissipation processes, since entropy is conserved. This means that if we desire to take dissipation into account, we have to go to next order, where we write the energy momentum tensor in the form
\begin{equation}
T^{\mu\nu}=(\rho + p)u^\mu u^\nu + p g^{\mu\nu}+\sigma^{\mu\nu}, \label{emthydro1}
\end{equation}
where $\sigma^{\mu\nu}$ is the dissipative part of $T^{\mu\nu}$, proportional to the first derivatives of $T(x)$ and $u^\mu$. The most general rotationally invariant $\sigma^{\mu\nu}$ turns out to be given by 
\begin{eqnarray}
\sigma^{\mu\nu}&=&P^{\mu\alpha} P^{\nu\beta} 
\left[
\eta\left(\partial_\alpha u_\beta+\partial_\beta u_\alpha -\frac{2}{3}g_{\alpha\beta} \partial_\lambda u^\lambda \right)+\zeta g_{\alpha\beta} \partial_\lambda u^\lambda 
\right] \nonumber,\label{emthydro2}
\end{eqnarray}
where $\eta$ and $\zeta$ are the \textbf{shear} and \textbf{bulk} viscosities, respectively. Using this form of the energy momentum tensor we find that the conservation equations (\ref{conservation}) give us a relativistic generalization of the Navier-Stokes equations.

\subsection{Kubo Formulae for Viscosities}
In a field theory, the energy momentum tensor (\ref{emtYM}) is expressed as a function of the field strengths $F^a_{\mu\nu}$, while in hydrodynamics it is given by (\ref{emthydro1})-(\ref{emthydro2}). Since hydrodynamics is just an effective theory at large time and space scales, these two must be related. This relation is found using \textit{linear response theory}, where we couple some sources $J_a(\vec{x})$ to bosonic operators $\Ope_a$, such that action reads
\begin{equation}
S=S_0 + \int_x J_a(x) \Ope_a(x).
\end{equation} 
If we assume that the expectation values of the operators $\av{\Ope_a(x)}$ vanish when evaluated with the action $S_0$, and $J$ can be considered small, then the expectation values are given by
\begin{equation}
\av{O_a(x)}=-\int_y G_{ab}^R(x,y) J_b(y),
\end{equation}
where $G_{ab}^R(x,y)$ is the retarded Green's function of the operator $\Ope_a$. 

If we now apply linear response theory to the hydrodynamics setting, we find (after some work) that the shear and bulk viscosities are given by the relations
\begin{eqnarray}
\eta &=& \lim_{\omega\to 0}\frac{\rho_s(\omega,T)}{\omega}\, ,\label{eta}\\
\zeta &=& \lim_{\omega\to 0}\frac{\rho_b(\omega,T)}{\omega}\, ,\label{zeta}
\end{eqnarray}
where \begin{equation}
\rho_{s,b} (\omega,T)=\im G_{s,b}^R(\omega,\vec{k}=0)\, ,
\end{equation}
and we have defined the \textit{shear and bulk channel} retarded correlators
\begin{eqnarray}
G_s^R(\omega,\vec{k}=0) &=&-i\int\! {\rm d}^4x\,  e^{i \omega t} \theta (t) \langle[
T_{12}(t,\vec{x}),T_{12}(0,0) ]\rangle \label{Gsdef}\, ,\\
G_b^R(\omega,\vec{k}=0) &=& -i\int\! {\rm d}^4x\,  e^{i \omega t} \theta (t) \langle[\frac{1}{3}
T_{ii}(t,\vec{x}),\frac{1}{3}T_{jj}(0,0) ]\rangle \label{Gbdef} \, .
\end{eqnarray}
These relations are called the \textit{Kubo formulae} for the viscosities \cite{Kubo:1957mj}.

\setcounter{equation}{0}
\chapter{Introduction to Holography\label{sec3}}
Holography \cite{'tHooft:1993gx,Susskind:1994vu} is a puzzling idea stating that a theory of quantum gravity in some manifold $\cal{M}$  can be fully described by a non-gravitational field theory living on the boundary $\partial\cal{M}$. At first sight, it is surprising that such relation between a theory with gravity and one without it could exist.  In particular, there are two very curious aspects of this principle.

First, gravity is very different from other field theories. In the modern theory of gravity, general relativity, the motion of a particle in a gravitational field is understood as inertial motion in curved spacetime. This is known as the \textit{equivalence principle} and, as a consequence,  we do not have any appropriate concept of force as in other field theories. Thus it is quite surprising that gravity could be equivalent to some non-gravitational  field theory.

The second surprising aspect of this principle is that  gravity and its equivalent field theory live on spacetimes of different dimensionality. We will see that this can be understood as a consequence of the very peculiar structure of gravity, where the partition function is given only by a boundary term. As a consequence, the entropy  of black holes scales not with volume as it does in any other local field theory, but with the surface of the black hole. Indeed, a study of black hole entropy  was one of the primary motivations behind the holographic principle.

In this chapter, we will present a specific realization of the holographic principle, the AdS/CFT correspondence, which is the exact equivalence between a certain string theory in Anti-de Sitter space and the conformal $\mathcal{N}=4$ supersymmetric Yang-Mills (SYM) theory in four-dimensional Minkowski space \cite{Maldacena:1997re,Aharony:1999ti}. While we can obtain some insights into the holographic nature of gravity already from general relativity, string theory gives us an exact equivalence of partition functions in both gravity and the corresponding dual field theory, and also provides us with an exact correspondence between observables in these two theories.

In addition to being the most important realization of the holographic principle, the AdS/CFT correspondence  has another interesting property. It turns out that it is a \textit{strong-weak} duality, i.e. it equates a strongly coupled theory on one side with a weakly coupled one on the other side. This is actually the most interesting property of the AdS/CFT correspondence, since it allows us to look into the strongly coupled regimes of both gravitational and gauge theories. In this thesis, we concentrate on using this correspondence to understand strongly coupled field theories in terms of weakly interacting gravity.

At the end of this chapter, we use the AdS/CFT correspondence to calculate some properties of hot SYM theory in the hydrodynamic limit. In particular, we are interested in the viscosities of the theory, and we derive the well-known result that theories with gravitational duals have a universal shear viscosity to entropy ratio.

There are a great number of reviews and introductory texts to this topic. Let us mention here a few of them that we have found helpful \cite{Petersen:1999zh,CasalderreySolana:2011us,Nastase:2007kj,McGreevy:2009xe,Polchinski:2010hw}.
\section{The Holographic Principle \label{hp}}
In general relativity, the gravitational interaction is understood in terms of the curvature of a spacetime manifold. The geometry of a manifold is in turn determined entirely by the metric tensor $g_{\mu\nu}$ and the  \textit{Levi-Civita connection}, $\Gamma^\rho_{\ \mu\nu}$. 

The action that describes the interaction of matter and gravity is given by
\begin{equation}
S=S_{EH}+S_M, \label{gravaction}
\end{equation}
where $S_{M}$ is the action for  matter and $S_{EH}$ is  
the  \textit{Einstein-Hilbert action} describing gravitation. In $D+1$ dimensions it reads
\begin{equation}
S_{EH}=\frac{1}{16\pi G_{D+1}}\int d^{D+1} x \sqrt{-g} R, \label{EHaction}
\end{equation}
where
\begin{equation}
R=R^{\mu}_{\ \mu}, \quad R_{\mu\nu}=R^{\rho}_{\ \mu \rho \nu},
\end{equation}
are the \textit{scalar curvature} and \textit{Ricci tensor}, respectively. Both of these are  contractions of the \textit{Riemann tensor}
\begin{equation}
R^{\rho}_{\ \sigma\mu\nu}=\partial_\mu \Gamma^\rho_{\ \nu\sigma}- \partial_\nu \Gamma^\rho_{\ \mu\sigma} + \Gamma^{\rho}_{\ \mu\lambda}\Gamma^{\lambda}_{\ \nu\sigma}- \Gamma^{\rho}_{\ \nu\lambda}\Gamma^{\lambda}_{\ \mu\sigma}.   \label{Riemann}
\end{equation}
The \textit{Einstein field equations} are then obtained by varying the action (\ref{gravaction}) with respect to the metric $g_{\mu\nu}$ as
\begin{equation}
R_{\mu\nu}-\frac{1}{2}g_{\mu\nu}R=8 \pi G_{D+1}T_{\mu\nu}, \label{eeq}
\end{equation}
where $T_{\mu\nu}$ is the  metric energy momentum tensor as defined in the previous chapter.

We can think of the Riemann tensor $R^{\rho}_{\ \sigma\mu\nu}$ as playing a  similar role as the field strength $F^a_{\mu\nu}$ in Yang-Mills theory, but there is one crucial difference. In Yang-Mills theory, the action contains only first derivatives of gauge fields, and consequently the equations of motion are second order differential equations.  This is very different in  the gravitational case, where the action (\ref{EHaction}) contains  second derivatives of the metric but the Einstein equations (\ref{eeq}) are also of second order.

The reason for this is that all of the second derivatives in the Einstein-Hilbert action are surface terms, and in the case of a closed manifold, i.e.~a manifold which is both compact and without a boundary, they do not contribute to the equations of motion.  For a manifold $\cal{M}$ with boundary $\partial\cal{M}$, it was shown by  York \cite{York:1972sj}, and later by Gibbons and Hawking \cite{Gibbons:1976ue}, that the appropriate action is 
\begin{equation}
S_G= S_{EH} + S_{GHY}=\frac{1}{16\pi G_{D+1}} \int_{\cal{M}}d^{D+1} x \sqrt{-g}   R + \frac{1}{8\pi G_{D+1}} \int_{\partial \cal{M}} d^{D} x\sqrt{-\gamma} K, \label{gravac}
\end{equation}
where  $S_{GHY}$ is the \textit{Gibbons-Hawking-York boundary term}, $\gamma_{\mu\nu}$ is the induced metric on the boundary and $K$ is the trace of the exterior curvature of the boundary. The addition of this boundary term to the the action ensures that the variational principle is well-defined.

In the  vacuum, the Einstein-Hilbert action has the  peculiar property that it vanishes as a consequence of the equations of motion. This means that the  on-shell gravitational action is given entirely by the Gibbons-Hawking-York boundary term
\begin{equation}
\left. S_G\right|_{\text{on shell}}= S_{BHY}. \label{act1}
\end{equation}
The action (\ref{gravac}) is important in a construction of  the partition function for gravity. The Euclidean partition function can be written \cite{Gibbons:1976ue}  as 
 \begin{equation}
Z^E=\int \mathcal{D}  [g_{\mu\nu}] \exp  \{ - S_G^E \},
\end{equation} 
where $ \mathcal{D} [g_{\mu\nu}]$ represents functional integration over all possible metric configurations. In the following section, we show an interesting example where we evaluate  this partition function for a black hole, and we will see how this leads to the \textit{Bekenstein-Hawking entropy}, which scales with the surface of the black hole. 

One way to understand entropy is by the amount of information which is needed in order to fully specify the system, and thus we naturally expect that it will scale with the volume of the space. The scaling of the black hole entropy with the surface lead 't Hooft \cite{'tHooft:1993gx} to conjecture that gravity could be in its nature holographic, i.e.~it could be entirely characterized by its behavior at the boundary of the space. This property has become known as the \textbf{holographic principle} \cite{'tHooft:1993gx,Susskind:1994vu}.
\begin{quote}
\textit{A theory of quantum gravity in a manifold $\cal{M}$ is equivalent to a non-gravitational field theory living on the corresponding boundary $\partial\cal{M}$.}
\end{quote}
This is indeed a very important insight into gravity, and it changes completely our view of the gravitational interaction. Let us note that the holographic principle was originally formulated for theories of quantum gravity, but its consequences even reach the level of classical general relativity.

So far, the way we have stated holographic principle is rather vague, as it only says that gravity is equivalent to \textit{some} field theory. In order to make the statement more useful, we have to make this relation somewhat more concrete. This was achieved in 1997 by Maldacena \cite{Maldacena:1997re} who found an explicit realization of the principle in string theory, i.e.~the AdS/CFT correspondence, which we will introduce below in section \ref{adscft}. 

However, before going  into  string theory let us show two applications of what we have just discussed. First, we calculate  the Bekenstein-Hawking entropy of a black hole using the partition function of gravity,  and then we show the importance of the boundary term in the definition of the gravitational energy momentum tensor.


\subsection{Black Hole Entropy \label{chapentr}}
The best-known solution of Einstein equations is the \textit{Schwarzchild solution}, which is a static and  spherically symmetric solution to the vacuum Einstein equations
\begin{equation}
R_{\mu\nu}=0. \label{Einsteinvac}
\end{equation}
In four dimensions, the  metric of this solution is
\begin{equation}
ds^2=- \left(1-\frac{2  M}{r}\right)dt^2+ \left(1-\frac{2M}{r}\right)^{-1}dr^2+r^2d\Omega^2_{2}, \label{schwarz}
\end{equation}
where $d \Omega_2^2$ represents the metric of a  two-sphere. This solution describes a \textit{black hole}, which is characterized by the existence of an \textit{event horizon}, the apparent singularity at $r_h=2M$. Excellent references for black hole physics are \cite{Misner:1974qy} and for more mathematical aspects \cite{Townsend:1997ku}.

Here we would like to discuss only one aspect of black holes, that is their entropy and temperature. We start from the observation that $r=2M$ is only an apparent singularity that can be removed by a coordinate transform. In \textit{Kruskal coordinates}, the metric (\ref{schwarz}) becomes regular at the horizon
\begin{equation}
ds^2=\frac{32M^3}{r}e^{-\frac{r}{2M}}\left(-dV^2+dU^2 \right)+r^2 d\Omega^2_2,  
\end{equation}
where the coordinates $U$ and $V$ are defined as
\begin{eqnarray}
U&=&\left(\frac{r}{2M}-1\right)^\frac{1}{2}e^\frac{r}{4M}\sinh\left( t/4M\right),\label{krusk1}\\
V&=&\left(\frac{r}{2M}-1\right)^\frac{1}{2}e^\frac{r}{4M}\cosh\left( t/4M\right).\label{krusk2}
\end{eqnarray}
Let us now perform a  Wick transformation, $t\to -i \tau$. From (\ref{krusk1})-(\ref{krusk2}), it follows then that in Euclidean time the Schwarzchild solution is periodic in $\tau$ with a period $\beta=8\pi M$. This means that using Euclidean formalism we can assign to the black hole a temperature
\begin{equation}
T_H=\frac{1}{\beta}=\frac{1}{8\pi M}.
\end{equation} 
This is  the famous \textit{Hawking temperature} \cite{Hawking:1974rv,Hawking:1976ra} of a black hole. From the existence of  a temperature, we can calculate directly the entropy of the black hole, as entropy is given by $dS=1/T dM$, where $M$ is the mass (energy) of the black hole.

However, let us choose a different approach, where we calculate the entropy using the Euclidean partition function of gravity. First, we need to evaluate the gravitational action, which is given only by the boundary term, since $R=0$ by the equations of motion. In order to evaluate the boundary term, we need to specify the boundary $\partial \cal{M}$, which we choose to be the hypersurface $r=r_0$ for some constant $r_0$. Following \cite{Gibbons:1976ue}, we find that the  Euclidean action is
\begin{equation}
S^E_G=S^E_{BHY}=4\pi M^2 + O(M^2 r_{0}^{-1}).
\end{equation}
In the asymptotic limit $r_{0}\to \infty$, only the first term survives.  We can  then evaluate the partition function, and using (\ref{entrop}) we find the entropy of the black hole 
\begin{equation}
S=4\pi  M^2 =\frac{A}{4 }, \label{bhent}
\end{equation}
where $A$ is the area of the surface of the black hole. Alternatively, we can restore the gravitational constant 
\begin{equation}
S=\frac{A}{4 G_4}, \label{bhent1}
\end{equation}
 This is the Bekenstein-Hawking entropy  of a black hole.  For the original derivation,  see \cite{Bekenstein:1972tm}.

This entropy formula has an interesting consequence, as it says that the maximal amount of information a system can store is proportional to the external surface of the system. In order to see this, consider  some region of space with volume $V$ and external surface $A$. Now, start adding some matter into this region. Then, at some point we achieve some maximum energy density, $E_{max}$ and to this configuration  corresponds the maximal entropy, $S_{max}$.  Since entropy is from a microscopic viewpoint interpreted as the amount of information needed to fully specify the system, in a local field theory we  expect  that it would be proportional to  the volume $V$.

However, in gravity the situation is different. It is important to realize that the maximal energy density  $E_{max}$, which  can be achieved in a given region, is the energy density of a black hole. Indeed, adding more matter will only increase  the size of the black hole. From (\ref{bhent}), it follows that the entropy of a black hole is proportional to the surface $A$. Thus the maximal entropy of the space region is given by the size of the largest black hole it can contain, and thus will be also proportional to the surface area. In the spherical case, this is  known as the \textit{spherical entropy bound} \cite{Susskind:1994vu}:
\begin{equation}
S_{max}< \frac{A}{4 }, \label{spherical}
\end{equation}
For generalizations of this entropy bound, see  \cite{Bousso:1999xy,Bousso:2002ju,Page:1993up}. 

The holographic principle is an elegant way to understand the above, since it says that gravity is  equivalent to some non-gravitational field theory living on the boundary. The boundary field theory is typically considered to be a local field theory, where entropy scales with the dimensionality of the boundary, i.e.~with the surface area. Thus, the holographic principle explains why this holds for gravity in the bulk spacetime.

\subsection{Holographic Renormalization \label{holoren}}
An interesting problem in the theory of gravity is the definition of the energy of a gravitational field. Since the Einstein-Hilbert action vanishes on shell, the  canonical energy momentum tensor (\ref{canten}) associated with this action  is  zero and the same holds for the metric energy momentum tensor (\ref{metem}). In  general relativity,  there were some attempts to solve this problem by so-called energy momentum pseudotensors, which are supposed to  represent the local energy of the gravitational field, but these objects do not transform as tensors \cite{Misner:1974qy} and thus the energy of the gravitational field is not observer independent.
 
It is in fact the boundary term in the action that allows us to define a physical  energy momentum tensor for the gravitational field. This is the well-known Brown-York tensor \cite{Brown:1992br}
\begin{equation}
T^{BY}_{\mu\nu}=-\frac{2}{\sqrt{-\gamma}}\frac{\delta S_G}{\delta \gamma_{\mu\nu}}.\label{BYtensor}
\end{equation}
We can see  that the Brown-York tensor is formally similar to the metric energy momentum tensor (\ref{metem}), but the variation is performed with respect to the boundary metric $\gamma_{\mu\nu}$ and the resulting tensor is defined only on the boundary. We call such a definition \textit{quasilocal}.

A problem with the Brown-York tensor is that it typically diverges when we take the boundary, where it is defined, to infinity.  In \cite{Brown:1992br}, a solution for this was proposed, namely that we should subtract the energy momentum tensor defined on the boundary with the same metric $\gamma_{\mu\nu}$, but in some reference spacetime, such as flat space. However, it turns out that it is not possible to do this for a general boundary geometry and reference spacetime.

For asymptotically Anti-de Sitter spacetimes (see Appendix \ref{append} for a review of AdS space), an interesting solution to this problem emerged that does not involve any subtractions from  reference spacetimes \cite{Balasubramanian:1999re}. In this work, it was realized that, in asymptotically AdS spacetimes,  we can interpret the divergences of the Brown-York tensor with the divergences of a field theory defined on the boundary. Moreover, they found that the trace of the gravitational energy momentum tensor  equals to the trace anomaly of a  conformal field theory living on the boundary.

We will see that this is exactly the situation encountered in  the AdS/CFT correspondence, where gravity in the AdS space is dual to a conformal field theory on the boundary. The work of \cite{Balasubramanian:1999re} was performed after Maldacena discovered the AdS/CFT correspondence, but it does not rely on it. Thus, we can consider it to be an independent motivation leading to the duality between gravity in AdS space and conformal field theories living on its boundary.

As an alternative to the subtraction scheme of Brown and York \cite{Brown:1992br}, it was proposed in \cite{Balasubramanian:1999re} that divergences can be removed by adding counterterms into the action. This is a procedure very similar to the renormalization procedure in  quantum field theory described in the previous chapter, and thus the term \textit{holographic renormalization} was coined for it.

Holographic renormalization is based on the simple observation that we can freely add to the gravitational action (\ref{gravac}) a term $S_{ct}$ that does not affect the equations of motion
\begin{equation}
S=S_{EH+\Lambda}+S_{BHY}+S_{ct},
\end{equation}
where $S_{EH+\Lambda}$ is now the Einstein-Hilbert action  including a negative cosmological constant. This term vanishes due to the equations of motion and thus only last two terms give non-zero contribution to the Brown-York tensor. A requirement to cancel divergences determines uniquely form of $S_{ct}$. 

Let us now consider only the case of an $AdS_5$ space, where $S_{ct}$ is given by
\begin{equation}
S_{ct}=-\int_{\partial\mathcal{M}}\frac{3}{\Lag}\sqrt{-\gamma}\left(1-\frac{\Lag^2}{12}R \right).
\end{equation}
We can then calculate the Brown-York energy momentum tensor 
\begin{equation}
T^{\mu\nu}=\frac{1}{8\pi G_5}\left[K^{\mu\nu}-K\gamma^{\mu\nu}-\frac{3}{\Lag}\gamma^{\mu\nu}-\frac{\Lag}{2}G^{\mu\nu}\right],
\end{equation}
where $G^{\mu\nu}$ is the Einstein tensor, $G^{\mu\nu}=R^{\mu\nu}-(1/2)\gamma^{\mu\nu}R$, of the boundary metric. We find that the trace of this expression is given by
\begin{equation}
T^\mu{}_\mu=-\frac{\Lag^3}{8\pi G_5}\left[-\frac{1}{8}R^{\mu\nu}R_{\mu\nu}+\frac{1}{24}R^2 \right], \label{trac}
\end{equation}
which can be recognized (up to a constant) to be exactly the trace anomaly of conformal field theory. If we use the relation between $\Lag^3/(8\pi G_5)$ and the corresponding coupling constant in $\mathcal{N}=4$ SYM from the AdS/CFT correspondence, we see that eq.~(\ref{trac}) agrees exactly with the prediction of conformal field  theory. A truly remarkable result. For futher discussion of the  renormalization of the gravitational action, see e.g. \cite{Skenderis:2002wp,Myers:1999psa}.



\section{The AdS/CFT Correspondence \label{adscft}}
In the previous section, we have motivated the holographic principle using  properties of  classical general relativity and a study of  black hole thermodynamics. However, the holographic principle as we have described it, is somewhat  vague, since it does not specify exactly, which field theory should be equivalent to which gravitational theory and under what circumstances this equality holds.

The first explicit realization of the holographic principle was found by Maldacena in 1997 \cite{Maldacena:1997re}, when he found an exact duality in string theory, known today as the \textit{AdS/CFT correspondence}. 
Maldacena conjectured that type IIB string theory on $AdS_5\times S^5$ is equivalent to four-dimensional conformal $\mathcal{N}=4$ Super Yang-Mills (SYM) theory. Later, Witten \cite{Witten:1998qj} and Gubser \textit{et al.} \cite{Gubser:1998bc}, established an exact correspondence between  the partition functions and fundamental observables  of these theories. The AdS/CFT correspondence has later been generalized to a situation where both supersymmetry and conformal invariance are broken. These models are generally known as \textit{gauge/gravity dualities}. 

We begin our discussion of the AdS/CFT correspondence with a short review of string theory, and then move on to introduce the duality and its generalizations.
\subsection{String Theory: A Quick Review \label{stringtheory}}
Unlike quantum field theory, where elementary particles are considered to be point-like objects, string theory considers them one-dimensional objects, i.e.~strings. In string theory, we have two characteristic parameters, the string tension $T$  and a dimensionless  coupling constant $g_s$ that controls the  strength of the interaction. We can write the string tension in the form
\begin{equation}
T\equiv \frac{1}{2\pi \alpha'}, \quad \quad \alpha'\equiv l_s^2,
\end{equation} 
where  $l_s$ is the  string length. The action of a pointlike particle in a relativistic theory is proportional to the length of the particle's worldline. In string theory, it is correspondingly proportional to the area of the worldsheet, i.e.~the two-dimensional generalization of the wordline, 
\begin{equation}
S=-T\int d^2\sigma \sqrt{-g},\label{NambuGoto}
\end{equation}
where $\sigma^1,\sigma^2$ are coordinates defined on the worldsheet and $g$ is the determinant of the worldsheet metric. 

Strings are extended objects and thus we have to discuss their endpoints. There are two possibilities for this, as the string can be either closed  or open. In the case of open strings, we have to specify boundary conditions for the endpoints, and there are again two possibilities: We  impose either the \textit{Dirichlet} or \textit{von Neumann} boundary conditions. We will also see that a  special role  is played in string theory  by objects called $D$-branes, which are extended objects at which open strings can end with Dirichlet boundary conditions. 

In the case of closed strings, we do not have to specify boundary conditions, since  their endpoints are identified and they can freely move in the bulk spacetime. Quantizing the string action (\ref{NambuGoto}), we find the spectrum of the excitations of the string. In the case of a closed string, we find also  a massless particle of spin two, i.e.~a graviton.  This is the reason why  the theory of closed strings is considered to be a theory of gravity. Moreover, since string theory allows us to calculate quantum corrections and solves certain problems with divergences that plague other approaches to quantize gravity, we consider string theory to be a theory of quantum gravity.

During the quantization of the above action, we find the surprising property that the  spacetime Lorentz group is anomalous, unless the string lives in 26 spacetime dimensions. This number of spacetime dimensions is called the \textit{critical dimension}, and such a theory is called \textit{critical string theory}.  However, string theory with the action (\ref{NambuGoto}) describes only bosons, and in order to provide a full description of Nature, we need to introduce fermions into the theory. This is done using  supersymmetry, which surprisingly decreases the critical dimension of the theory to only 10.

In order to understand the basics of the AdS/CFT conjecture, it is sufficient for us to consider only type IIB string theory, which in its   low-energy limit reduces to supergravity (SUGRA) given by the action
\begin{equation}
S_{SUGRA}=\frac{1}{16\pi G_{10}} \int d^{10}x \sqrt{-g}\left(
R-\frac{1}{2}\partial^\mu\phi\partial_\mu\phi -\frac{1}{2}\frac{1}{5!}F_5^2 + \dots 
\right),\label{sugra}
\end{equation}
where $R$ is the Ricci scalar,  $\phi$ is the dilaton field and $F_5$ is the field strength for the four form $C_4$, and dots represent fermionic terms and other so-called Ramond-Ramond forms that are all irrelevant for our consideration.  The gravitational constant, $G_{10}$, can be  expressed in terms of the  string length and coupling constant as
\begin{equation}
16\pi G_{10}=(2\pi)^7g_s^2l_s^8.\label{g10}
\end{equation}
An important point that we would like to stress  is that the string coupling is not  a free parameter here, but is given by the expectation value of the dilaton field $\phi$ as $g_s=e^\phi$. When we talk about the string coupling constant, we in fact mean its asymptotic value at infinity, i.e. $g_s=e^{\phi_\infty}$. 
 
\subsection{The Maldacena Conjecture \label{maldacena}}
The Maldacena conjecture follows from the observation of the  dual role that $D3$-brane solutions play in string theory. In the low energy limit of  closed string theory, they are  solutions of the  supergravity action. In  open string theory, they play the role of objects on which open strings can end up, and they give the rise to a non-Abelian gauge-theory. Next, we will  discuss these two perspectives separately.

\subsubsection*{Closed String Perspective: $D3$-branes as Spacetime Geometry}
 In 1995, Polchinski \cite{Polchinski:1995mt} discovered that $Dp$-branes can be identified with the \textit{black $p$-branes}  solutions of SUGRA. The black $p$-branes are higher dimensional generalizations of  black hole  solutions known from general relativity. 
We can find the $D3$-brane solution  considering the metric ansatz 
\begin{equation}
ds^2=-B^2(r)dt^2+E^2(r)d\vec{x}^2 + R^2(r) dr^2 + G^2(r)r^2 d\Omega_5^2,
\end{equation}
where  $d\Omega_5^2$ is an element of the unit 5-sphere, and
\begin{equation}
d\vec{x}^2=\sum_{i=1}^3 \left(dx^i\right)^2.
\end{equation}
Writing  down the field equations for the action (\ref{sugra}) and imposing self-duality for the field strength, $F_5=\star F_5$, we  can find a solution (see \cite{Petersen:1999zh} for a detailed derivation). 

Instead of writing the solution for a single $D3$-brane, we can write it directly for $N$ coincident branes,
\begin{equation}
ds^2=H^{-\frac{1}{2}}\left(-{f}dt^2+ d\vec{x}^2\right)+ 
H^{\frac{1}{2}} \left({f}^{-1}dr^2+ r^2 d\Omega_5^2 \right), 
\end{equation}
\begin{equation}
{f}=1-\left(\frac{r_h}{r}\right)^4, \quad
H=1+\left(\frac{\mathcal{L}}{r}\right)^4, \quad 
e^\phi=1, \quad 
\int_{S^5} F_5=N,  \quad\label{solution5}
\end{equation}
where the last equation says that due to the existence of  $D3$-branes, the field strength $F_5$ must be quantized. Observe that the dilaton field is in this example constant and thus the string coupling $g_s=e^{\phi}$ is also a constant. See also section~\ref{nonconfmodel} for a generalization where the dilaton field is allowed to  vary over  space. 

The  length parameter $\mathcal{L}$ appearing above  is given by 
\begin{equation}
\mathcal{L}^4=4\pi g_sN l_s^4.
\end{equation}
In the limit $\mathcal{L} \gg r>r_h$, we can neglect the $1$ in the function $H$ and write it as
\begin{equation}
H^{\frac{1}{2}}\approx \left(\frac{\mathcal{L}}{r}\right)^2.
\end{equation}
Then  the  $r^2$-term in front of $d\Omega_5^2 $ cancels out and we find that the  metric for $N$ coincident $D3$-branes reads
\begin{equation}
ds^2=\frac{r^2}{\mathcal{L}^2} \left(-f dt^2+d\vec{x}^2 \right)+ \frac{\mathcal{L}^2}{r^2}\frac{dr^2}{f} +  \mathcal{L}^2 d\Omega_5^2, \label{branesol1}
\end{equation}
\begin{equation}
f= 1-\left(\frac{r_h}{r}\right)^4 . \label{branesol2}
\end{equation}
We can see that the first part of (\ref{branesol1}) is the metric of a  black hole in $AdS_5$ space in the planar limit (see Appendix A), and the  second part is the metric of a 5-dimensional sphere with radius $\mathcal{L}$.

In the limit $r_h\to 0$,  we find that the  metric (\ref{branesol1}) simplifies to
\begin{equation}
ds^2=\frac{r^2}{\mathcal{L}^2}\left(-dt^2 + d\vec{x}^2 \right) +\frac{\mathcal{L}^2}{r^2}dr^2+   \mathcal{L}^2 d\Omega_5^2. \label{branesol3}
\end{equation}
Using the coordinate transformation $r\to \Lag^2/z$, we rewrite this in the form
\begin{equation}
ds^2=\frac{\mathcal{L}^2}{z^2}\left(-dt^2 + d\vec{x}^2  + dz^2\right)+   \mathcal{L}^2 d\Omega_5^2, \label{branesol4}
\end{equation}
where the first part can be recognized as  the metric of empty $AdS_5$ space. Thus, in this special limit, the metric of $N$ coincident $D3$-branes reduces to the topological product of pure $AdS_5$ space and an $S^5$ sphere,
\begin{equation}
AdS_5\times S^5. \nonumber
\end{equation}

Remember that classical SUGRA is a valid approximation of type IIB string theory only if we can neglect both quantum and string effects. String effects are negligible, if the string length $l_s$ can be considered small when compared with the other length scale in the problem, the AdS radius $\mathcal{L}$. Quantum effects can on the other hand be considered negligible, if a dimensionless combination of the gravitational constant $G_{10}$ and  the AdS radius $\mathcal{L}$ is small enough. These two conditions read
\begin{equation}
l_s\ll\mathcal{L}, \quad \quad \frac{G_{10}}{\mathcal{L}^8}\ll 1. \label{validity}
\end{equation}
\subsubsection*{Open String Perspective:  $D3$-branes  and Gauge Fields}
In  open string theory, the $D3$-branes  are viewed as  four-dimensional hypersurfaces, on which open strings can end up with Dirichlet boundary conditions \cite{Polchinski:1995mt}. The action in this case is given by 
\begin{equation}
S=S_{brane} + S_{bulk} + S_{int},
\end{equation} 
where $S_{brane}$ is the action of a four-dimensional theory  on the brane, $S_{bulk}$ is the action describing the interaction in the bulk space, and $S_{int}$ is the interaction term between these two. It turns out that the bulk theory is SUGRA coupled to the massive string modes.

A crucial observation is that in the  low-energy limit $\alpha'\to 0$, the massive string modes drop out and the bulk theory reduces to free supergravity,  while the interaction term,  proportional to the $(\alpha')^2$, can be neglected. Thus, we obtain two decoupled (non-interacting) theories, free SUGRA in the bulk with the action $S_{bulk}=S_{SUGRA}$ and a field theory on the brane, $S_{brane}$.

It turns out that in a low energy limit, if we consider $N$ parallel coinciding $D3$-branes, the theory on the brane is given by the Lagrangian \cite{Witten:1995im} 
\begin{equation}
\Lag=\frac{1}{4 \pi g_s} \mathrm{Tr} \left(\frac{1}{4} F^{\mu\nu}F_{\mu\nu} + \frac{1}{2}D_\mu\phi^{i}D^\mu\phi^{i} + \left[\phi^{i},\phi^{j}\right]^2  \right), \label{symaction}
\end{equation}
which can be recognized as the Lagrangian of   $\mathcal{N}=4$ SYM theory with the symmetry group $SU(N_c)$, with the coupling constant
\begin{equation}
g_s=4\pi g_{YM}^2. \label{ce}
\end{equation} 
 The number of colors $N_c$ can be identified with the number of the $D3$-branes, i.e. $N=N_c$.

\subsubsection*{The Conjecture  and the 't Hooft Limit}
Above we saw that the physics of $D3$-branes, particularly when we consider $N$ coincident branes, can be described by type IIB string theory in $AdS_5\times S^5$ in the closed string perspective or by  $\mathcal{N}=4$ SYM in the open string theory perspective. The AdS/CFT correspondence  \cite{Maldacena:1997re} is the conjectured equivalence of these two theories
\begin{equation}
\left\lbrace \mathcal{N}=4\; SU(N)\; \text{SYM theory} \right\rbrace=
\left\lbrace \text{IIB string theory in}\; AdS_5\times S^5  \right\rbrace \label{correspondence}
\end{equation}
with the parameters of the theories related by 
\begin{equation}
g_s= \frac{g_{YM}^2}{4\pi}, \quad \quad  \left(\frac{\mathcal{L}}{l_s}\right)^4=g_{YM}^2 N_c=\lambda, \label{constrela}
\end{equation}
where $\lambda$ is the 't Hooft coupling. 

Let us next consider the  symmetries of the theories on both sides of the conjecture. In section~\ref{symch}, we  found that the  symmetry group of  $\mathcal{N}=4$ SYM theory is $SO(4,2)\times SO(6)$, where $SO(4,2)$ represents the conformal symmetry and $SU(4)\approx SO(6)$  the $R$-symmetry due to the presence of $\mathcal{N}=4$  supercharges. The symmetry group of the gravitational theory is on the other hand a product of the symmetry groups of Anti-de Sitter space $AdS_5$ and a five-sphere $S^5$. From Appendix A, we know that $AdS_5$ has a symmetry group $SO(2,4)$, while the rotation symmetry group of $S^5$ is clearly $SO(6)$. We can then see that the symmetry groups of the theories on both sides of the conjecture (\ref{correspondence}) match exactly.  Moreover, an analogous statement can  also be made about the fermionic degrees of freedom, which we have so far neglected \cite{CasalderreySolana:2011us}. 

The equality (\ref{correspondence}) has been conjectured to be true for any value of  the couplings $g_s$ and $g_{YM}$ \cite{Maldacena:1997re}. However, we do not understand either of these theories, when the coupling parameters become arbitrarily large. In (\ref{validity}), we have specified, under what circumstances classical SUGRA is a good approximation of type IIB string theory, while  using (\ref{constrela}) we can find out the parameter values this corresponds to  in $\mathcal{N}=4$ SYM theory.  In  closed string theory, we can express the parameters $G_{10}$ and $l_s$ in terms of the 't Hooft coupling and the number of colors as
\begin{equation}
\frac{G_{10}}{\mathcal{L}^8} \propto \frac{1}{N_c^2}, \quad \quad
\frac{l_s^4}{\mathcal{L}^4}\propto \frac{1}{\lambda}.
\end{equation}
Thus, the condition (\ref{validity}) corresponds to the limit 
\begin{equation}
N_c\to\infty , \quad \quad \lambda\gg 1,
\end{equation}
which is the 't Hooft limit of strongly coupled  $\mathcal{N}=4$ SYM (see section \ref{thooftlimit}). This is a very important result, since it says that we can understand (at least one) strongly coupled gauge theory (in the large-$N_c$ limit) using classical SUGRA.  This is a truly remarkable result, and also our  main motivation to study the AdS/CFT correspondence.

The opposite argumentation is also valid, i.e.~to weakly coupled  gauge theory corresponds a strongly coupled string theory. This property makes the AdS/CFT correspondence a \textit{strong/weak duality}. 

\subsection{The AdS/CFT Correspondence \label{witten}}
In section \ref{hp}, we introduced the holographic principle, which states that a gravitational theory is equivalent to \textit{some} non-gravitational theory living on the boundary. The Maldacena conjecture that  we have just described specifies exactly what gravity theory (type IIB string theory) is equivalent to what field theory ($\mathcal{N}=4$ SYM theory) and gives us an exact relation between the coupling constants of these theories. However, in order for this to be useful, we would like to  have a  correspondence between the fundamental observables of these  theories. This was achieved by Witten \cite{Witten:1998qj} and Gubser \textit{et al.} \cite{Gubser:1998bc}, who formulated the AdS/CFT correspondence as an exact equivalence  of the partition functions and  specified the  mapping between observables in these  theories.

In  Euclidean space, we can write the result of the above procedure as an equivalence of partition functions 
\begin{equation}
Z^E_{\mathcal{N}=4}[\phi_0]=Z^E_{IIB, \ AdS\times S^5}[\phi], \label{duality1}
\end{equation}
where $\phi_0$ is the boundary value of the bulk field $\phi$, i.e. $\phi_0=\phi|_{boundary}$. The left hand side is the partition function of  $\mathcal{N}=4$ SYM theory, where the boundary values of the fields $\phi_0$ act as sources for the gauge invariant  operators $\Ope(x)$. Now, we proceed in two steps. First, we explain what exactly  these partition functions are, and then  elaborate on the correspondence between the fields $\phi$ and the gauge operators $\Ope(x)$.

On the left-hand side of (\ref{duality1}), we have the partition function of  $\mathcal{N}=4$ SYM theory that can be schematically written as  
\begin{equation}
Z^E_{\mathcal{N}=4 }[\phi_0]=\int \FD [\dots ] \exp \left[ - S^E_{\mathcal{N}=4} + \int d^4 x \Ope (x) \phi_0(x) \right],
\end{equation}
where $ \FD [\dots ]$ denotes a functional integration over all  SYM fields and $S^E_{\mathcal{N}=4}$ is the Euclidean action of  $\mathcal{N}=4$ SYM  theory.   Schematically, we write this as
\begin{equation}
Z^E_{\mathcal{N}=4 }[\phi_0]= \left\langle \int d^4 x \Ope (x) \phi_0(x)  \right\rangle,
\end{equation}
where $\left\langle   \right\rangle$ means that we have integrated over all SYM fields.

The right-hand side of (\ref{duality1}) is the partition function of type IIB string theory. As we have seen in the previous section, in the limit where we can neglect both quantum and string effects, type IIB string theory simplifies to classical SUGRA. Thus, in the limit
\begin{equation}
\lambda\to \infty, \quad \quad N_c\to \infty, \label{lim55}
\end{equation}
we can write eq.~(\ref{duality1}) as
\begin{equation}
\left\langle \int d^4 x \Ope (x) \phi_0(x)  \right\rangle=
e^{-S^E_{SUGRA}(\phi)}, \label{duality}
\end{equation} 
where $S^E_{SUGRA}$ is the Euclidean action of SUGRA (\ref{sugra}), and the left-hand side represents the partition function of  $\mathcal{N}=4$ SYM.

The partition function formalism allows us to calculate   $n$-point correlation functions as functional derivatives using the formula (\ref{npoint}). If we now use the correspondence between the partition functions (\ref{duality}), we can rewrite the Euclidean form of eq. (\ref{npoint})  as
\begin{equation}
\av{\Ope(x_1),\dots,\Ope(x_n)}_E= \left. \frac{\delta^n S^E_{SUGRA}}{\delta \phi(x_1),\dots, \delta\phi(x_n)} \right|_{\phi\to \phi_0 }. \label{oper}
\end{equation}
The correlators obtained using this formula potentially   contain  divergences, which can be avoided in two ways. The first one is to choose boundary conditions such that we avoid any divergences, see section~\ref{exec} for a worked-out example. Another method is  holographic renormalization, described in section~\ref{holoren}, where we renormalize the action by adding appropriate local counter-terms.

All in all, we have obtained  a truly remarkable result, as using the formulas above we may calculate renormalized $n$-point correlation functions in strongly coupled gauge theory using classical SUGRA. A natural question to ask now is,  which field $\phi$ corresponds to which gauge operator $\Ope$. This is known as the\textit{ field/operator correspondence}. 

\subsubsection{The Field/Operator Correspondence}
Let us first consider the simplest case, where the bulk field is some scalar field $\phi$ that corresponds to some gauge invariant operator $\Ope$.
We are primarily interested in the case, where the  boundary is four-dimensional, but let us for a while consider $D+1$ dimensional Euclidean AdS space with the metric
\begin{equation}
ds^2_{AdS_{D+1}}=\frac{\Lag^2}{z^2}\left(\delta_{\mu\nu}x^\mu x^\nu +dz^2\right), \quad \quad \mu=0,1, \dots, D-1, \label{metric7}
\end{equation}
The action of a massive scalar field $\phi$ then reads
\begin{equation}
S^E=K\int dz dx^D\sqrt{-g} \left[(\partial \phi)^2+m^2\phi^2\right] , \label{action7}
\end{equation}
where $K$ is some normalization constant, $m$ the mass of the scalar field, and in $(\partial \phi)^2$, a  summation  in Euclidean metric is understood. In the case of $AdS_{D+1}$, we can solve this system using the method of Green's functions. For a step-by-step derivation, in the cases of both massless and massive scalar fields, see \cite{Petersen:1999zh}; for a derivation in momentum space, see section~\ref{exec} below.

Going through the above exercise, we find that the two-point correlation function has the form
\begin{equation}
\av{\Ope(x_1)\Ope(x_2)}_E \sim \frac{1}{(x_1-x_2)^{2\Delta}}, \label{scalcorr}
\end{equation}
which can be identified as the correlator of the conformal field theory operator (\ref{confcorr}) with scaling dimension 
\begin{equation}
\Delta=\frac{D}{2}+\sqrt{m^2\Lag^2+\frac{D^2}{4}}. \label{scaldim}
\end{equation}
Let us next consider an operator with the scaling dimension $\Delta=4$. From the above, we find that on the gravity side, this corresponds to a massless scalar field that can be recognized    as the dilaton field in the SUGRA action (\ref{sugra}). In the case of conformal  $\mathcal{N}=4$ SYM, we  find that the  operator with the scaling dimension $\Delta=4$ is  $\Ope=\text{Tr} F^2$ \cite{Witten:1998qj}. Thus, we have found the first example of the correspondence between a bulk field and a gauge invariant operator. Moreover, this matching of the field and the dual operator can be further confirmed by calculating the anomaly of the current corresponding to  $R$-symmetry in  $\mathcal{N}=4$ SYM theory \cite{Witten:1998qj}. This is an important check, as the anomaly is a nonperturbative effect.

\subsubsection*{Higher Spin Fields}
The above discussion can be generalized from the scalar field $\phi$ to  fields of higher spin. In particular, we would like to understand how the field/operator coupling 
\begin{equation}
\Ope (x) \phi_0(x),
\end{equation}
is changed when  $\phi_0(x)$ is some higher spin field. An especially interesting situation occurs when we consider a spin two particle, a graviton. Then it can be found that the graviton couples to the energy momentum tensor on the boundary \cite{Gubser:1997se,Gubser:1997yh}
\begin{equation}
T^{\mu\nu}(x)h_{\mu\nu}(x,z=0), \label{gravem}
\end{equation}
where $h_{\mu\nu}$ denotes a perturbation to the AdS metric. Using the  formula (\ref{npoint}), we then find the expectation value of the energy momentum tensor in the boundary theory to satisfy
\begin{equation}
\av{T^{\mu\nu}(x)}=\lim_{z\to 0} \frac{\delta S^E_{SUGRA}}{\delta g_{\mu\nu}(x,z)}.
\end{equation}
We note that this is not actually a tensor, but only tensor density. Instead, we can construct the full tensor from 
\begin{equation}
\av{T^{\mu\nu}(x)}=\lim_{z\to 0} \frac{2}{\sqrt{-g}}\frac{\delta S^E_{SUGRA}}{\delta g_{\mu\nu}(x,z)}.
\end{equation}
This expression can be recognized as the Brown-York energy momentum tensor of a gravitational field that we  discussed in section \ref{holoren}.

The above discussion can be easily generalized to $n$-point correlation functions of the energy momentum tensor. In particular, we are interested in the correlators $\av{T_{12}T_{12}}$ and $\av{T_{\mu\mu}T_{\nu\nu}}$, which, as we have seen, play a very important roles in hydrodynamics. However, in hydrodynamics we are interested in  real-time  correlators, while what we have discussed so far are Euclidean correlators. Let us nevertheless demonstrate the calculation of Euclidean correlators in the simple case of a scalar field and leave the calculation of the full correlators to section \ref{hlimit}.
\subsection{Euclidean Correlators: Example \label{exec}}
We will now demonstrate the extraction of Euclidean correlators in the simplest case of a massive scalar field $\phi$ with the action (\ref{action7}) on the $AdS_{D+1}$ background (\ref{metric7}). Our main motivation here is to show how to remove divergences by the correct choice of  boundary conditions.

The  metric (\ref{metric7}) is a function of only  the $z$-coordinate  and is translationally invariant in the other directions. Thus, we can introduce a Fourier decomposition  in these directions of the form 
\begin{equation}
\phi(z,x^\mu)=\int \frac{d^D k}{(2\pi)^D}e^{ik.x}\phi(z, k^\mu),
\end{equation}
where $\mu=0,1,\dots, D-1$ and $k^\mu=(\omega,\vec{k})$, with $\omega$ and $\vec{k}$ being the energy and the spatial momentum, respectively. 

The equation of motion, found from the action (\ref{action7}), reads 
\begin{equation}
z^{D+1}\partial_z\left(z^{1-D }\partial_z \phi \right)-(k^2z^2+m^2\Lag^2)\phi^2=0, \quad
\quad k^2=\omega^2+\vec{k}^2.  \label{eqm1}
\end{equation}
This equation can be solved by the method of separation of variables, writing
\begin{equation}
\phi(z,k)=f_k(z)\phi_0(k),
\end{equation}
where $\phi_0(k)$ is the boundary value of the scalar field and $f_k(z)$ is a solution to the mode equation
\begin{equation}
z^{D+1}\partial_z\left(z^{1-D }\partial_z f_k \right)-(k^2z^2+m^2\Lag^2)f_k^2=0,  \label{eqm3}
\end{equation}
with unit value at the boundary, $f_k(0)=1$. Near the boundary  $z\to 0$, there are two solutions of  (\ref{eqm3}), one of which behaves as $\propto z^{D-\Delta}$ and the  other as $\propto z^{\Delta}$.  The condition of unit boundary value implies that we choose the first one, and thus have
\begin{equation}
\lim_{z\to 0} z^{\Delta-D} f_{k}(z)= 1. \label{bc1}
\end{equation}
In the limit $z\to \infty$, there are again two solutions, $\propto z^{\pm \omega}$. We require the solution of the equation of motion to be regular everywhere inside the AdS space, and thus pick the solution with the minus sign. This is our second boundary condition, and together with (\ref{bc1}), they completely  determine the solution of the equation of motion.
 
The on-shell action (i.e.~one evaluated with the classical solution)  reduces to the surface term
\begin{equation}
S^{E}=\left. \int \frac{d^D k}{(2\pi)^D} \phi_0(k) \mathcal{F}(k,z)\phi_0(k) \right|_{z=0}, \label{action8}
\end{equation}
where 
\begin{equation}
\mathcal{F}(k,z)=K \sqrt{-g} g^{zz}  f_{-k}(z)\partial_z  f_{k}(z).
\end{equation}
Using the formula (\ref{oper}), we can now find the $n$-point correlation functions, in particular the two-point correlator  
\begin{equation}
G^{E}(k)=\av{\Ope(k)\Ope(0)}_E=-\left. \mathcal{F}(k,z)\right|_{z=0}. \label{eucorr}
\end{equation}
There is one aspect of the calculation above that we would like to stress. If we use the formula (\ref{oper}), the resulting $n$-point correlation functions may contain divergences. However, our choice of boundary conditions ensures that the resulting correlators (\ref{eucorr}) are finite. There exists also a more formal approach,   holographic renormalization,  that we have discussed in  section~\ref{holoren}. There,  we add local counter terms into the action to cancel out the divergences. For details,  see appendix B of  \cite{CasalderreySolana:2011us}, where the authors demonstrate this method on the example that we have just discussed.

\subsection{Finite Temperature}
So far, we have considered only a very special case, where the underlying spacetime was a topological  product of empty $AdS_5$ space and an $S^5$ sphere. We have seen that this very symmetric situation is constrained by both conformal symmetry and supersymmetry. This makes it a very good scenario to test  the general correspondence, but alone it is of little phenomenological interest. It turns out, however, that it is possible to relax some of the underlying assumptions and break both supersymmetry and conformal invariance.

It has been shown that supersymmetry is always broken at finite temperature \cite{Buchholz:1997mf}, and this result can be extended also to the AdS/CFT correspondence. To this end, recall that we have found a general solution of the black 3-brane metric given by the equations (\ref{branesol1}) and (\ref{branesol2}). There, we however neglected the black hole and obtained that the $D3$-brane metric became a product of empty $AdS_5$ space and an $S^5$ sphere.

If we instead keep the black hole in the $AdS_5$ space, we can write the Euclidean form of  the $AdS_5$ part of the $D3$-brane metric in the form
\begin{equation}
ds^2_{AdS_5}=b^2(z)\left(f(z)d\tau^2 + d\vec{x}^2 +\frac{dz^2}{f(z)} \right), \label{adsbh}
\end{equation}
where
\begin{equation}
b(z)=\frac{\Lag}{z}, \quad\quad f(z)=1-\frac{z^4}{z_h^4}.
\end{equation}
As we have shown in section \ref{chapentr},  we can assign a temperature and an  entropy to a black hole. In the case of a  black hole in $AdS_5$ space (in the planar limit), we find that the temperature and entropy density are given by (see Appendix A  for details)
\begin{equation}
T=\frac{1}{\pi z_h}, \quad \quad s=\frac{b^3(z_h)}{4  G_{5} }. \label{te}
\end{equation}
It is natural to identify the temperature of the $AdS_5$ black hole with the temperature of  $\mathcal{N}=4$ SYM theory, since the temperature is a coupling independent quantity. We can also try to match the entropy of the $AdS_5$ black hole and the $\mathcal{N}=4$ SYM theory \cite{Gubser:1996de}, but we have to keep in mind that the entropy generally runs with the coupling strength. Let us now look into this issue in some more detail.

The $5$-dimensional gravitational constant is given by 
\begin{equation}
G_{5}=\frac{\pi}{2}\frac{\Lag^3}{N_c^2},
\end{equation}
and thus we can write the black hole entropy  density (\ref{te}) as 
\begin{equation}
s=\frac{1}{2} \pi^2 T^3  N_c^2\, , \label{entstrong} 
\end{equation}
Now, we can compare this with the Stefan-Boltzmann entropy density of an ideal $\mathcal{N}=4$ SYM theory plasma in the large-$N_c$ limit,
\begin{equation}
s_{\mathcal{N}=4}=\frac{2}{3} \pi^2 T^3  N_c^2\, . \label{entweak}
\end{equation}
We can see that up to a numerical factor $3/4$, these expressions agree. As we have noted, this disagreement is not surprising, since the formula (\ref{entstrong}) is the entropy density in a strongly coupled theory, while (\ref{entweak}) is the  entropy  at  weak coupling. If we calculate the leading  $1/\lambda$-correction to the  entropy of  $\mathcal{N}=4$ SYM theory  \cite{Gubser:1998nz}, we find that this ratio becomes 
\begin{equation}
\frac{s}{s_{\mathcal{N}=4}}=\frac{3}{4} + \frac{1.69}{\lambda^{\frac{3}{2}}}+ \dots ,
\end{equation} 
and thus we can conjecture that including all $1/\lambda$-corrections would lead to an exact equality between the entropies.  It is worth noting that there exists lattice QCD data \cite{Aoki:2005vt,Bazavov:2009zn} that partially justifies this, as it suggests that the entropy density ratio of the gravitational theory and a strongly coupled gluon gas is close to 1. 

\subsubsection*{Euclidean Finite-Temperature Correlators}
To obtain a Euclidean finite temperature correlator, we use the prescription from section~\ref{exec}, but replace the metric of empty AdS space with the metric of a  black hole in AdS space. In $D=4$ it is given by (\ref{adsbh}), leading to the equation of motion (\ref{eqm1}) then reading 
\begin{equation}
z^{5}\partial_z\left(z^{-3} f(z)\partial_z \phi \right)-\left(\frac{\omega^2 z^2}{f(z)}+\vec{k}^2 z^2+m^2\Lag^2 \right)\phi=0.  \label{eqm4}
\end{equation}
As a consequence of the presence of the black hole, there will be a horizon at $z=z_h$, and the solutions of eq.~(\ref{eqm4}) are then defined on the interval $z\in [0,z_h]$, in contrast to $z\in [0,\infty)$ in the zero temperature case. 

Close to the horizon, we find that there exist two solutions of the equation of motion that behave as $\propto(z-z_h)^{\pm\omega}$. Just as in the zero-temperature case, we choose the one with the minus sign, in order for the solution of (\ref{eqm4}) to behave regularly at all values of $z$. The boundary condition at $z=0$ remains same as in the zero-temperature case. The action and the  correlation functions are finally given by the formulae (\ref{action8}) and (\ref{eucorr}). Notice that in Euclidean space, there is no contribution to the correlator (\ref{eucorr}) from the horizon.
%
%
%
%
%
%
%
%
%
%
%
%
%
%
%
\section{The Hydrodynamic Limit \label{hlimit}}
In  section~\ref{chlimit}, we  showed how one can relate the bulk and shear viscosities (\ref{eta})-(\ref{zeta}) to the retarted correlators $T_{\mu\mu}$ and $T_{12}$. Now, we would like to calculate these quantities in a strongly coupled field theory using the AdS/CFT correspondence. To accomplish this, we however need to modify our  prescription for the evaluation of the correlators.

\subsection{Minkowski Space Correlators \label{rtc}}
To obtain Minkowski space (or real-time) correlators, we replace the Euclidean AdS black hole (\ref{adsbh}) with a Minkowski one 
\begin{equation}
ds^2_{AdS_5}=b^2(z)\left(-f(z)dt^2 + d\vec{x}^2 +\frac{dz^2}{f(z)} \right). \label{adsbh1}
\end{equation}
The equation of motion for a scalar field is then similar to the Euclidean case (\ref{eqm4}), only the sign in front of $\omega^2$ is  flipped
\begin{equation}
z^{5}\partial_z\left(z^{-3 } f(z)\partial_z \phi \right)+\left(\frac{\omega^2 z^2}{f(z)}-\vec{k}^2 z^2-m^2\Lag^2 \right)\phi=0.  \label{eqm5}
\end{equation} 
The boundary condition at the boundary $z=0$ remains the same as in the Euclidean case, but close to the horizon there are now two solutions that behave like $\propto (z-z_h)^{i\omega}$. These are both stable solutions that oscillate rapidly, but with constant amplitude. So now regularity is no longer a sufficient condition to fully determine the solution in the bulk. 

To understand how to proceed, we refer to section~\ref{introec}, where we discussed that in Euclidean space there is only one independent Euclidean correlator, while in Minkowski space there are more of them. According to \cite{Son:2002sd}, this multiplicity of correlators in Minkowski space corresponds exactly to the different choices of boundary conditions at the horizon. We are primarily interested in retarded correlators due to their role in linear response theory and hydrodynamics; according to \cite{Son:2002sd}, these quantities are available by choosing an \textit{infalling} boundary condition at the horizon, i.e.~the one that behaves as $\propto (z-z_h)^{-i\omega}$. Physically, this can be understood as the wave propagating in the bulk getting absorbed at the black hole horizon. 

The on-shell scalar action in Minkowski space reads 
\begin{equation}
S=\left. \int \frac{d^4 k}{(2\pi)^4} \phi_0(k) \mathcal{F}(k,z)\phi_0(k) \right|_{z=0}^{z=z_h}, \label{maction1}
\end{equation}
If we were to define Minkowski correlators similarly as in the Euclidean case, i.e.~by taking functional derivatives of the action (\ref{maction1}), we would obtain
\begin{equation}
G(k)=-\left. \mathcal{F}(k,z)\right|_{z=0}^{z=z_h}-\left. \mathcal{F}(k,z)\right|^{z=0}_{z=z_h}. \label{mk1}
\end{equation}
The problem is, however, that this is a completely real quantity, while the correlator is expected to be complex. The mode function satisfies $f^\star_k(z)=f^\star_{-k}(z)$, and it can be shown that the imaginary part of $\mathcal{F}(k,z)$ is independent of $z$ and thus the imaginary part of $G(k)$ trivially zero. In \cite{Son:2002sd}, a solution to this problem was proposed, conjecturing that the retarded correlator is  given by
\begin{equation}
G^R(k)=-2\left. \mathcal{F}(k,z)\right|_{z=0}.\label{corrdef}
\end{equation}
This conjecture was later rigorously proven in \cite{Herzog:2002pc} using the \textit{Schwinger-Keldysh} formalism. From now on, we will be interested only in retarded correlators, and thus we will always use this formula.
\subsection{Shear Viscosity\label{confshear}}
As an important application of the above machinery, we would now like to calculate the bulk and shear viscosities in $\mathcal{N}=4$ SYM theory.  Due to the conformal symmetry of the theory, we however know that $T^\mu_\mu=0$, from where it automatically follows that the bulk viscosity is identically zero. This means that only the shear viscosity $\eta$ will be non-trivial.

To find the value of $\eta$, we start with the metric (\ref{adsbh1}) and introduce a perturbation to its $12$ component,
\begin{equation}
g_{12}= h_{12},
\end{equation}
which  according to eq.~(\ref{gravem}) couples to the $T_{12}$ component of the energy momentum tensor. We now insert this perturbed metric to the Einstein equations, expand the result to linear order in $h_{12}$, and find the equation of motion
\begin{equation}
\ddot{h}_{12}+\left(\frac{d}{dz}\log b^3f\right)
\dot{h}_{12}+\frac{\omega^2}{f^2}h_{12}=0. \label{fluctshearcft}
\end{equation}
Evaluating the gravitational action with this solution, and applying the prescription of eq.~(\ref{corrdef}), we find that the imaginary part of the retarded correlator $G^R_s$ is in the low-energy limit given by
\begin{equation}
G^R_s(\omega\to 0)=\frac{1}{16\pi G_5} b^3 (z_h) i \omega.
\end{equation}
Using the Kubo formula (\ref{eta}), this gives us the shear viscosity in the form
\begin{equation}
\eta=\frac{1}{16\pi G_5} b^3 (z_h),
\end{equation}
from which we obtain using eq.~(\ref{te})
\begin{equation}
\frac{\eta}{s}=\frac{1}{4\pi}. \label{entrratio}
\end{equation}

We skipped some details of the calculation here, as in chapter~\ref{ch6} we will provide a step-by-step derivation of the result in a somewhat more general setting, where the functions $b(z)$ and $f(z)$ in the metric  (\ref{adsbh1}) only asymptotically reproduce the AdS black hole form. 
\subsection{Universality and the Viscosity Bound Conjecture}
The above $1/(4\pi)$ result for the ratio of the shear viscosity and the entropy was first derived in \cite{Kovtun:2004de}, where the shear viscosity  was related to the absorption cross section of gravitons. This method is rather independent of the details of the underlying theory, and led to the conjecture that the ratio (\ref{entrratio}) might be universal for all theories with gravity duals. For the case of models with two-derivative actions, this was later proved via a direct AdS/CFT calculation of the corresponding correlation function in \cite{Buchel:2004qq}. Finally, in \cite{Buchel:2004di,Iqbal:2008by} the same result was obtained using the so-called \textit{membrane paradigm} \cite{Thorne:1986iy}, where only universal properties of black hole horizons are used. 

The constant value of the shear viscosity to entropy ratio has been explicitly verified also  for theories dual to $Dp$-brane \cite{Kovtun:2003wp} and M-brane \cite{Herzog:2002fn} constructions, as well as other setups. See also fig.~\ref{figshear1} (right) for our result that nicely illustrates the constancy of the ratio (\ref{entrratio}) within IHQCD.

One can naturally ask, how this ratio changes, if one proceeds beyond the two-derivative approximation, e.g.~by including finite coupling corrections in the calculations. In the $\mathcal{N}=4$ SYM theory, the ratio $\eta/s$ has been calculated to the next order in a $1/\lambda$ expansion, with the result
\begin{equation}
\frac{\eta}{s}=\frac{1}{4\pi}\left(1+\frac{135\zeta(3)}{8\lambda^{\frac{3}{2}}} \right).
\end{equation}
An important observation is the plus sign in front of the correction term, which indicates that as the field theory coupling becomes weaker, the $\eta/s$ ratio increases. This is consistent with calculations within perturbative QCD, which predict large values for the $\eta/s$ ratio in the weakly coupled regime, as we have discussed in chapter~\ref{ch1}.

The above observations have even lead to the conjecture that the limit (\ref{entrratio}) might actually be a universal lower bound for the ratio of the shear viscosity and entropy \cite{Kovtun:2004de}, i.e. 
\begin{equation}
\frac{\eta}{s}\geq\frac{1}{4\pi} \label{entrratio1}
\end{equation} 
in all quantum field theories, implying that a liquid cannot be arbitrarily close to being an ideal one. In \cite{Kovtun:2004de}, it was argued that any known liquid satisfies this bound, and as shown in \cite{Son:2007vk}, this includes even superfluid helium that is known to flow without dissipation. There are however known violations of this limit originating from both higher derivative gravity actions \cite{Brigante:2007nu} and anisotropies \cite{Rebhan:2011vd}.
\section{Non-Conformal Holography \label{nonconfmodel}}
So far, we have discussed, how the AdS/CFT correspondence can be used to study the conformal $\mathcal{N}=4$ SYM theory. However, our ultimate goal is to use holography to study the physics of strongly interacting matter in the real world, i.e.~QCD.  To this end, a natural question to ask is, whether there is any similarity between QCD and $\mathcal{N}=4$ SYM theory. We know that QCD is a non-conformal theory and, moreover, it exhibits confinement and chiral symmetry breaking. However, as was argued e.g.~in \cite{Liu:2006he}, at temperatures above about twice the critical temperature of the deconfinement phase transition $T_c$, QCD is at the same time deconfined, relatively strongly coupled and almos conformal. Thus, (the strongly coupled limit of) $\mathcal{N}=4$ SYM theory may in fact not be such a bad approximation to  QCD at temperatures $T\gtrsim 2 T_c$. However, one may also ask, whether it is possible to have an even better holographic model, i.e.~go beyond the $\mathcal{N}=4$ SYM approximation?
  
Since the original work of Maldacena, there have been several proposals to modify the original conjecture in order to understand either gravity beyond the SUGRA approximation or the field theory beyond strongly coupled  $\mathcal{N}=4$ SYM theory. We call all these generalizations \textit{gauge/gravity dualities}. This is a very rich subject, and it is clearly beyond the scope of this thesis to go through all of the recent developments. Instead, we will briefly review just a few important approaches towards a holographic description of QCD, and refer the interested reader to \cite{Aharony:1999ti,CasalderreySolana:2011us,Polchinski:2010hw,Mateos:2007ay,Kiritsis:2009hu} for more details.

\subsection{Top-Down Models}
In the so-called \textit{top-down} approach, we modify some essential features in the gravity side of the AdS/CFT correspondence in order for the dual field theory to approach QCD. The simplest such model was proposed by Witten in \cite{Witten:1998zw}, where he started with     finite temperature $\mathcal{N}=4$ SYM theory living in the manifold $R^3\times S^1$, with the $S^1$ corresponding to the Euclidean time with period $\beta=1/T$. Then, according to \cite{Witten:1998zw}, we can perform a Kaluza-Klein reduction along the circle and all fermionic modes  acquire a tree-level mass of order $1/\beta$. The scalars in $\mathcal{N}=4$ SYM theory are periodic, but acquire masses at the quantum level thorough their couplings, while the gauge bosons remain massless. As a consequence, the theory will have a mass gap and exhibit confinement. 

In three dimensions, this picture can be obtained from the Euclidean metric of a black hole (\ref{adsbh}), where we analytically continue one of the spatial directions to Lorentzian time $x_3\to i t$
\begin{equation}
ds^2=\frac{\Lag^2}{z^2}\left(-dt^2 + dx_1^2+dx_2^2 + f(z)d\tau^2 \right) +\frac{\Lag^2}{z^2f}dz^2.
\end{equation}
Now the field theory lives in three dimensions $t$, $x_1$ and $x_2$, while $\tau$ is a compact spatial direction. This theory is at zero temperature, but there is an additional $z$-direction that ends at  $z=z_h$, analogously to the case of the black hole metric (\ref{adsbh}). Thus this theory will develop the mass gap of order $M\sim1/z_h$. We can solve the equation of motion for a classical field in this metric and find normalizable modes with masses of order $M$ that can be identified with the glueballs of the theory. 

The discussion above can be easily generalized to obtain a confining four-dimensional YM theory with a mass gap. In order to do so, we simply have to replace $D3$-branes with $D4$-branes and redo the steps described above. A problem in this model is that there are Kaluza-Klein (KK) modes from the compactification in $\tau$-direction, exhibiting masses comparable to those  of the glueballs. This property, dubbed \textit{KK contamination}, is something what we do not expect in a realistic model of QCD. 

A continuation of the above model is the so-called \textit{Sakai-Sugimoto model} of \cite{Sakai:2004cn,Sakai:2005yt}, where we introduce quarks into the theory by inserting pairs of $D8-\bar{D}8$-branes to the gravity setup However, as discussed in \cite{Aharony:2006da}, there are again problems with the KK contamination, and theory deviates from QCD at scales above the mass scale of the KK modes.

As discussed in \cite{Kiritsis:2009hu}, we expect QCD to be described by a five-dimensional holographic model, which automatically implies that we need to use non-critical string theory. However, the problem of non-critical string theory is that the curvature of the manifold is often of the order of string size scale, and thus we need to go beyond the low-energy approximation. This is in contrast with critical string theory in the AdS/CFT correspondence, where the SUGRA approximation can be applied. Due to this reason, it is hard to make progress towards a proper top-down gravity dual of QCD; see however e.g.~\cite{Klebanov:2004ya,Csaki:2006ji,Kuperstein:2004yf,Bigazzi:2005md} for some attempts in this direction.
 
\subsection{Bottom-Up Models}
A more phenomenological approach to finding a gravity dual to QCD is the   \textit{bottom-up} approach, sometimes dubbed $AdS/QCD$. The simplest example of these is the \textit{hard-wall} model of
\cite{Polchinski:2001tt,Erlich:2005qh}, where on the gravity side one has a constant dilaton field in the AdS background. In this model, the AdS $z$-coordinate is cut off in both the UV and IR regimes, leading to confinement through specific boundary conditions in IR. While this model has been successful in predicting a semi-phenomenological meson spectrum, it exhibits a rather serious problem, as its glueball spectrum is quadratic instead of linear.

Another simple holographic model is the  \textit{soft-wall} model of \cite{Karch:2006pv}, where one introduces a non-constant dilaton, usually with a power law profile in the $z$ coordinate. This usually fixes the issues with the glueball spectrum, but an outstanding problem  is that the equations of motion for both the dilaton field and the metric are not satisfied. Thus, the self-consistency of this approach is rather questionable. For some further attempts in this direction, see e.g.~ \cite{Andreev:2006ct,Kajantie:2006hv}.

\chapter{Improved Holographic QCD\label{sec4}}
We concluded our last chapter with a brief discussion of various non-conformal holographic models of QCD. In this thesis, we are particularly interested in one such model of QCD, called  \textit{Improved Holographic QCD}  (IHQCD), proposed by Elias Kiritsis  and his collaborators \cite{Gursoy:2007cb,Gursoy:2007er}. We will see that it combines elements of both top-down and bottom-up approaches towards holographic QCD. 

In this chapter, we introduce  this model and show how to calculate  thermodynamic observables within it, as well as briefly discuss the numerical methods that we have used in this exercise. The main result of this thesis, a  calculation of the correlators of the energy momentum tensor in the IHQCD model, is left to the next chapter.

\section{Introduction}
In QFTs, the most important operators are typically those of the lowest dimension possible.  In the case of QCD \cite{Kiritsis:2009hu},  these are the operators of  dimension $\Delta=4$, namely the scalar operator $\Tr [F^2]$, the pseudo-scalar operator $\Tr [F\wedge F]$, and the energy momentum tensor $T_{\mu\nu}$. We expect its dual theory to live in five dimensions, and thus it needs to be described by non-critical string theory. If we keep only the lowest-dimensional terms in the theory, then according to  \cite{Gursoy:2007cb,Gursoy:2007er}, we can write the effective action of non-critical string theory in the string frame in the form
\begin{equation}
S=M_p^3\int d^5x\sqrt{-g_S}\left[ e^{-2\phi}\left(R+4(\partial_\mu\phi)^2 +\frac{\delta c}{l_s^2}\right) -\frac{1}{2.5!}F_5^2-\frac{1}{2}F_1^2-\frac{N_f}{l_s^2}e^{-\phi}
\right], \label{stringaction}
\end{equation}
where $M_p$ is the Planck mass  and, analogously to the original AdS/CFT correspondence, the metric $g_{\mu\nu}$ is dual to the energy momentum tensor $T_{\mu\nu}$ and the dilaton field $\phi$  to the scalar operator $\Tr [F^2]$. The function $F_5$ finally gives a quantization condition for the number of branes, and generates the gauge group $SU(N_c)$ on the field theory side. 

In addition to the terms explained above, we have in the action the $RR$-form  $F_1=\partial_\mu a$, where $a$ is the axion field that is dual to the pseudo-scalar operator $\Tr [F\wedge F]$, while the last term corresponds to $D4-\bar{D}4$ brane pairs that represent quarks in the system. We have also defined
\begin{equation}
\delta c= 10-D=5,
\end{equation}
as the central charge of non-critical string theory. From here, we can pass to the Einstein frame using a conformal transformation
\begin{equation}
(g_{S})_{\mu\nu}=e^{\frac{4}{3}\phi}g_{\mu\nu}.
\end{equation}
We define 
\begin{equation}
\lambda=N_c e^\phi, \label{ldef}
\end{equation}
and solving for the five-form that controls the number of $D3$-branes, rewrite the action as
\begin{equation}
S=M_p^3N_c^2\int d^5x\sqrt{-g} \left[R -\frac{4}{3}\frac{ (\partial\lambda)^2}{\lambda^2}  -\frac{\lambda^2}{2N_c^2}\left(\partial a\right)^2+V(\lambda) \right],
\label{ihqcdaction1}
\end{equation}
where  $V(\lambda)$ is given by
\begin{equation}
V(\lambda)=\frac{\lambda^{\frac{4}{3}}}{l_s^2} \left[\delta c - \frac{N_f}{N_c} \lambda -\frac{1}{2}\lambda^2 \right]. \label{pot3}
\end{equation}

We  observe that the axion field is suppressed by $1/N_c^2$, and thus in the large-$N_c$  limit it has no effect on the geometry and can be neglected in our discussion\footnote{Note that the axion does not affect the geometry, but is still present in our model. Recently,  fluctuations of the axion field were considered in \cite{Gursoy:2012bt} in order to determine  the Chern-Simons diffusion rate of the model.}. In  the 't Hooft limit, we keep the number of flavors $N_f$ constant, while $N_c \to \infty$. We  observe that in this limit, the ratio $N_f /N_c \to 0$ and thus the term representing quarks in the model is suppressed. In this thesis we are interested in this limit, where QCD is effectively quarkless\footnote{However, there exists in addition the so-called \textit{Veneziano limit} \cite{Veneziano:1979ec}, where one considers a  situation in which also $N_f\to \infty$, with $N_f/N_c\sim 1$. A holographic model of QCD in the Veneziano limit was recently studied in \cite{Jarvinen:2011qe}.}.

We can relate the Planck mass  to the $5D$ gravitational constant via 
\begin{equation}
G_5=\frac{1}{16\pi M_p^3N_c^2},
\end{equation}
and we observe that the gravitational constant is small in the large-$N_c$ limit. Using this relation, we rewrite the action as
\begin{equation}
S=\frac{1}{16\pi G_5}\int d^5x\sqrt{-g} \left[R -\frac{4}{3}\frac{ (\partial\lambda)^2}{\lambda^2} +V(\lambda) \right] .
\label{ihqcdaction}
\end{equation}

Consider next the metric ansatz 
\begin{equation}
ds^2=b^2(z)\left(-dt^2 + d\vec{x}^2 + dz^2 \right),\label{met13}
\end{equation}
where $b(z)$ is some function of the $z$-coordinate. A non-constant dilaton potential results in a running coupling $\lambda$, with a  beta function
\begin{equation}
\beta=\frac{d \lambda}{d \ln E},
\end{equation}
where $E$ is some energy scale. We  identify the energy scale of the dual field theory, in an analogy with the AdS/CFT correspondence, with the conformal factor $b(z)$ via
\begin{equation}
E(z)= E_0 b(z),
\end{equation}
where $E_0$ is some constant. Thus, the beta function of IHQCD is given by 
\begin{equation}
\beta=\frac{d \lambda}{d \ln b} \label{beta13}.
\end{equation}
We will now discuss the equations of motion for both the metric and the dilaton field. Then, in section \ref{potchoice}, we show how  to determine the dilaton potential, using the IR and UV solutions of the equations of motion. 
\subsection{The Equations of Motion \label{bggeo}}
Using the  IHQCD action (\ref{ihqcdaction}), we can find the equations of motion for both the metric and  the  dilaton field.   Since we are interested in QCD at a finite temperature, we use a  black hole metric  ansatz 
\begin{equation}
ds^2=b^2(z)\left(-f(z)dt^2 + d\vec{x}^2 + \frac{dz^2}{f(z)} \right),\label{adsbh2}
\end{equation}
which is analogous to  (\ref{adsbh1}), but now with  $b(z)$ and $f(z)$ being  some generic functions of the $z$-coordinate.
\subsubsection{Einstein equations}
Varying  the action (\ref{ihqcdaction}) with respect to the metric $g_{\mu\nu}$ gives us the Einstein equations 
\begin{equation}
G_{\mu\nu}= \frac{1}{2}T_{\mu\nu}, \label{eeeeq}
\end{equation}
where the left-hand side is the Einstein tensor, obtained from a variation of $R$ with respect to the metric $g_{\mu\nu}$. The non-zero components of the Einstein tensor for the metric ansatz (\ref{adsbh2}) are
\begin{eqnarray}
G_{tt} &=& -\frac{3 f\left(\dot{b} \dot{f}+2 f \ddot{b}\right)}{2b }, \\
G_{\vec{x}\vec{x}} &=& \frac{6 \dot{b} \dot{f}+6 f \ddot{b}+b \ddot{f}}{2 b}, \\
G_{zz} &=& \frac{3 \dot{b} \left(4 f \dot{b}+b \dot{f}\right)}{2 b^2 f},
\end{eqnarray}
where the dots denote  derivatives with respect to the $z$-coordinate.
On the right hand side of (\ref{eeeeq}) is the energy momentum tensor  with the non-zero components 
\begin{eqnarray}
T_{tt} &=&  \frac{4}{3}f^2 {\dot{\phi}}^2 - b^2fV(\lambda), \\
T_{\vec{x}\vec{x}} &=&  - \frac{4}{3}f {\dot{\phi}}^2 + b^2V(\lambda),\\
T_{zz} &=& \frac{4}{3} {\dot{\phi}}^2 +  \frac{b^2}{f}V(\lambda).
\end{eqnarray}
Adding all these together, we may write the  Einstein equations (\ref{eeeeq}) in the form
\begin{eqnarray}
3 \dot{b} \dot{f}+b \ddot{f}  &=&0, \label{eeq1}\\
6\frac{{\dot{b}}^2}{b^2} - 3\frac{\ddot{b}}{ b }   &=& \frac{4}{3} {\dot{\phi}}^2, \label{eeq2}\\
6\frac{\dot{b}^2}{b^2} + 3\frac{\ddot{b}}{ b } +\frac{3\dot{b}\dot{f}}{fb}&=&\frac{b^2}{f}V(\lambda).\label{eeq3}
\end{eqnarray}
For an arbitrary $z$ and a generic potential $V(\lambda)$, these equations are too complicated to solve analytically, and only numerical solutions are possible. However, in both the UV and IR limits it is possible to perform analytic expansions, as we will show in the following section.

\subsubsection{The Equation of Motion for the Dilaton  }
The equation of motion for the dilaton field is obtained by varying  the action (\ref{ihqcdaction}) with respect to the  field $\phi$, producing
\begin{equation}
\square\phi +{{\partial V}\over{\partial\phi}}=0. \label{dileq2}
\end{equation}
For the metric (\ref{adsbh2}), the five-dimensional Laplace operator is given by
\begin{equation}
\square\phi=\frac{1}{\sqrt{-g}}\partial_\mu \left(
\sqrt{-g} g^{\mu\nu}\partial_\nu \phi
\right)
=\frac{f}{b^2}\ddot{\phi}+\left(3\frac{\dot{b}f}{b^3}+\frac{\dot{f}}{b^2}\right)\dot{\phi},
\end{equation}
and thus we find
\begin{equation}
\ddot{\phi}+\left(3\frac{\dot{b}}{b}+\frac{\dot{f}}{f}\right)\dot{\phi} + \frac{b^2}{f}\frac{\partial V(\phi)}{\partial\phi}=0.
\end{equation}

\section{The Dilaton Potential \label{potchoice}}
A crucial role in IHQCD is played by the potential $V(\lambda)$ that makes the coupling $\lambda$ run and the beta function of IHQCD (\ref{beta13}) to become nontrivial. As discussed in \cite{Gursoy:2007cb}, there are higher order $\alpha'$ corrections to the string action (\ref{stringaction}) that will cause the dilaton potential to contain higher order terms in $\lambda$. In  \cite{Gursoy:2007cb}, there is a suggestion on how to derive them exactly. However, it is questionable whether such a top-down derivation of the potential would reproduce the behaviour of QCD. 

In IHQCD, we instead introduce the dilaton potential by hand, and choose its form to mimic certain properties of QCD. This is the bottom-up part of the IHCD model, which indeed is an interesting mixture of both approaches. Next, we will discuss solving the Einstein equations (\ref{eeq1})-(\ref{eeq3}) in both their UV and IR limits, and demonstrate how we can construct the dilaton potential from its expected physical behavior in these limits.

\subsection{The UV Solution}
Consider a vacuum version of  the Einstein equations (\ref{eeq1})-(\ref{eeq3}), obtained by setting $f=1$,
 \begin{eqnarray}
6\frac{{\dot{b}}^2}{b^2} - 3\frac{\ddot{b}}{ b }   &=& {4 \over 3} {\dot{\phi}}^2, \label{eeqv1}\\
6\frac{\dot{b}^2}{b^2} + 3\frac{\ddot{b}}{ b}&=&b^2V(\phi),\label{eeqv2}
\end{eqnarray}
and  define the  so-called \textit{superpotential}
\begin{equation}
W=-\frac{\dot{b}}{b^2}.\label{spdef}
\end{equation}
With the help of this, we write
\begin{equation}
\dot{\lambda}=-\beta(\lambda)b W, \label{betdef}
\end{equation}
using which equations (\ref{eeqv1})-(\ref{eeqv2}) can further  be rewritten  as first order differential equations
 \begin{eqnarray}
 b\dot{W}&=& \frac{4}{9}\dot{\phi}^2,\label{eeqvv1}\\
12\frac{\dot{b}^2}{b^2}-3b\dot{W} &=& b^2V(\phi). \label{eeqvv2}
\end{eqnarray}
Using $\phi=\ln \lambda$, we now find
\begin{equation}
b\dot{W}=\frac{4}{9}\frac{\dot{\lambda^2}}{\lambda^2},
\end{equation}
from where we obtain, using (\ref{betdef}),  the relation
\begin{equation}
\frac{dW}{W}=-\frac{4}{9}\frac{\beta(\lambda)}{\lambda^2}d\lambda.
\end{equation}
We can integrate this relation to give
\begin{equation}
W(\lambda)=W(0)\exp \left[
-\frac{4}{9}\int d\lambda \frac{\beta(\lambda)}{\lambda^2},
\right], \label{vsol1}
\end{equation}
and from (\ref{eeqv2}), find the relation between the dilaton potential and  the beta function
\begin{equation}
V(\lambda)=12W^2(\lambda)\left[1-\left(\frac{\beta(\lambda)}{3\lambda}\right)^2\right].\label{vsol2}
\end{equation}
The last two equations imply that the dilaton potential is fully determined by the beta function. 

The UV solution of the Einstein equation corresponds to  the limit $z\to 0$. In QCD, the  beta function can be found using  perturbative methods, as described in the second chapter. We find that up to two loops, the beta function of quarkless QCD is given by the expansion
\begin{equation}
\beta(\lambda)=-b_0 \lambda_c^2-b_1\lambda_c^3 + \dots, \quad\quad \lambda\to 0, \label{betadef61}
\end{equation}
where  $\lambda_c$ is the 't Hooft coupling on the field theory side, and the coefficients $b_0$ and $b_1$ are given by 
\begin{equation}
b_0=\frac{22}{3(4\pi)^2},\quad \quad b_1=\frac{51}{121}b_0^2.
\end{equation}
In IHQCD, we match the holographic  beta function (\ref{beta13}) to the beta function of  quarkless QCD, with the identification $\lambda=\lambda_c/\left(8\pi^2\right)$.

Using the equations (\ref{vsol1}) and (\ref{vsol2}), we find that the above expansion of the beta function has a mapping to the small $\lambda$ limit of the potential
\begin{equation}
V(\lambda)=12 W^2(0)\left[1+\frac{8}{9}b_0\lambda + \left(\frac{23}{81}b_0^2+\frac{4}{9}b_1\right)\lambda^2+\dots \right]. \label{uvpot}
\end{equation}
Here, we fix the value of the constant $W(0)$ to $1/\Lag$ by the requirement to reproduce the pure $AdS_5$ metric in the limit $\lambda\to 0$.

The above relation determines the  potential $V(\lambda)$ in the UV region, and using it we can find the UV solution of the Einstein equations. We can expand equations (\ref{spdef})-(\ref{betdef}) in powers of $\lambda$, and keeping only the first non-trivial terms, find
\begin{equation}
\lambda(z)=-\frac{1}{b_0 \log \Lambda z},\label{lambdaz}
\end{equation}
where $\Lambda$ is the integration constant of (\ref{betdef}). Similarly, from (\ref{spdef}) it follows that the  solution to the lowest order reads
\begin{equation}
b(z)=\frac{\Lag}{z}\left[1+\frac{4}{9}\frac{1}{\log \Lambda z}\right]. \label{b15}
\end{equation}
This agrees with our experience from  four-dimensional large-$N_c$ gauge theory, where we expect a  running coupling of the form  $\lambda\sim 1/\log E$, where $E$ is the energy scale. The integration constant $\Lambda$ has a dimension of mass and  is generated dynamically in the model. We identify it  with the QCD scale $\Lambda_{QCD}$.
\subsubsection{Scheme Dependence}
In practical calculations within QFTs, there is always some residual scheme dependence left, related to the choice of the renormalization scheme and scale, and thus the parametrization of the coupling constant. However, any physical observable must clearly be scheme independent. In the holographic setting, different reparametrizations of the coupling constant is related to the freedom to perform radial diffeomorphisms, i.e.~redefine the holographic coordinate $z$. This kind of dependence can be reduced by picking a specific frame for the metric \cite{Gursoy:2007cb}, as we did when choosing to work with eq.~(\ref{adsbh2}).

In holography, the scheme dependence related to the definition of the coupling constant can also be understood in terms of redefinitions of the bulk fields. This was well demonstrated in \cite{Gursoy:2007cb}, where a relation between the $\lambda$ field in the bulk and the 't Hooft coupling on the field theory side was studied. In particular, it was shown that considering various deformations of the IHQCD action amount to changing this relation.

Similarly, the relation between $E$ and $\lambda(z)$, given by the $\beta$-function, is changed under redefinitions of the bulk fields. However, as was shown in \cite{Gursoy:2007cb}, we can identify the holographic $\beta$-function with the one on the field theory side, assuming we only take the first two terms in the expansion (\ref{betadef61}). This is the reason, why we do not include higher-loop terms in $\beta$-function. 

For a more detailed discussion of scheme dependence issues within IHQCD, see \cite{Gursoy:2007cb,Gursoy:2007er,Gursoy:2010fj}.
\subsection{The IR Solution}
While the UV limit of QCD can be easily studied using perturbative methods, there is only  little analytical understanding of the IR regime of QCD. However, the most important feature of QCD in the IR regime is confinement. In our holographic picture, the IR regime  of the dual theory is identified  with large values of the $z$-coordinate, which  corresponds to the limit $\lambda\to \infty$. 

The confinement criterion is introduced in IHQCD using holographic Wilson loops \cite{Maldacena:1998im}. Applying this method to the IHQCD background, we obtain that the  criterion is satisfied if the function $b(z)$ is given by \cite{Gursoy:2007cb}
\begin{equation}
b(z)\to e^{-\left(\frac{z}{R} \right)^\alpha + \dots}, \label{bir}
\end{equation}
where $R$ is some IR scale and $\alpha\geq 1$. This implies that the dilaton potential and the beta function should behave in the IR  as 
\begin{eqnarray}
V(\lambda)&=&\left(\log \lambda \right)^{\frac{\alpha-1}{\alpha}}\lambda^{\frac{4}{3}}+\dots, \label{irpot}\\
\beta(\lambda)&=&-\frac{3}{2}\lambda\left[1+\frac{3}{4}\frac{\alpha-1}{\alpha}\frac{1}{\log \lambda}+ \dots\right].
\end{eqnarray}
The value of the parameter $\alpha$  can be fixed by calculating the low-energy particle spectrum and matching the results with lattice QCD.
\subsubsection{The Gluebal Spectrum}
In quarkless QCD, the low-energy particle spectrum of the theory is composed of the bound states of  gluons, called the \textit{glueballs}. In a confining theory, we expect this spectrum to be gapped and  discrete.

The discrete glueball spectrum of IHQCD is obtained by considering  fluctuations of fields around the background spacetime and demanding some reasonable boundary conditions. There are two different types of fluctuations, of which the spin-zero fluctuation is obtained as a combination of the metric and the dilaton field that is invariant under radial diffeomorphisms. The spin-two fluctuations are on the other hand related to metric fluctuations. The equation of motion for the scalar fluctuation is given by
\begin{equation}
\ddot{\phi}+3\frac{\dot{b}}{b}\dot{\phi}-\left(\frac{\ddot{X}}{X}+3\frac{\dot{b}}{b}\frac{\dot{X}}{X}\right)+m^2\phi=0, \label{eqscal}
\end{equation}
where
\begin{equation}
X=\frac{\beta(\lambda)}{3\lambda},
\end{equation}
while the equation of motion for the spin-two fluctuation is given by (\ref{eqscal}) with $X=0$. 

To analyse the glueball spectrum, it is preferable to transform the fluctuation equation (\ref{eqscal}) to the form of the Schr\"odinger equation. To this end, we introduce the function
\begin{equation}
\psi=b^{\frac{3}{2}}\phi,
\end{equation}
for which the fluctuation equation (\ref{eqscal}) takes the  form 
\begin{equation}
-\ddot{\psi}+\tilde{V}\psi=m^2\psi,
\end{equation}
where 
\begin{equation}
\tilde{V}=\frac{3}{2}\frac{\ddot{b}}{b}+\frac{3}{4}\frac{\dot{b}^2}{b^2}+\frac{\ddot{X}}{X}+3\frac{\dot{b}}{b}\frac{\dot{X}}{X}.\label{tildev}
\end{equation}
In the UV, this potential behaves as
\begin{equation}
\tilde{V}=\frac{15}{4z^2}+\dots, \quad z\to 0,
\end{equation}
leading to the UV solution 
\begin{equation}
\psi=C_1(m)z^{-\frac{3}{2}} +C_2(m)z^{\frac{5}{2}}, \quad z\to 0.
\end{equation}
To avoid a singularity in the limit $z\to 0$, we require $C_1(m)=0$. 

Moving then on to the IR limit, we study the background (\ref{bir}), using which we find that the  potential (\ref{tildev}) takes the form
\begin{equation}
\tilde{V}\sim \frac{9}{4 R^2}\left(\frac{z}{R} \right)^{2(\alpha-1)}.\label{potir}
\end{equation}
We thus observe that the potential  has a local maximum in the UV and then decreases. In the IR, there is also a local maximum, giving  the potential the  shape of a potential well, and  leading  to a discrete spectrum. 

Next, let us study the discrete spectrum of the bound states resulting from this potential well. After some work, we find that the quantization condition is given by the integral \cite{Gursoy:2007er}
\begin{equation}
n \pi=\int_{z_1}^{z_2}dz \sqrt{m_n^2-\tilde{V}},
\end{equation}
where $z_1$ and $z_2$ are the  turning points of the potential.  To obtain high energy excitations, we consider $m_n\gg \tilde{V}$ and let the potential  be given by the asymptotic limit (\ref{potir}). We then obtain the result
\begin{equation}
m\sim n^{\frac{\alpha-1}{\alpha}},
\end{equation}
Experiments and lattice simulations suggest that in QCD we have $m_n^2\sim n$. Thus, in order to reproduce this in IHQCD, we need to set $\alpha=2$.
\subsection{Construction of the Potential}
So far, we have found that  the dilaton potential is given in the UV limit  by (\ref{uvpot}), and in  the IR limit by (\ref{irpot}), with $\alpha=2$. The potential that we use in IHQCD is  one that smoothly   interpolates between these two  limits. There are naturally multiple choices for the potential, since the asymptotic behaviour does not uniquely determine the form of the function everywhere. 
 
The potential used in the original IHQCD papers \cite{Gursoy:2007cb,Gursoy:2007er} had the form
\begin{equation}
V(\lambda)=\frac{12}{\Lag^2}\left\lbrace 1+V_0\lambda + V_1\lambda^\frac{4}{3}\left[\log\left(1+ V_2\lambda^\frac{4}{3} +V_3\lambda^2\right) \right] \right\rbrace . \label{dilpot2}
\end{equation} 
where the coefficient $V_0$, $V_1$ and $V_2$ were determined by matching with the QCD beta function, leading to
\begin{equation}
V_0=\frac{8}{9}b_0, \quad \quad V_1\sqrt{V_2}=\left(\frac{23}{81}b_0^2+\frac{4}{9}b_1\right),
\end{equation}
while $V_3$ is related to the IR behavior. This means that the potential (\ref{dilpot2}) has two free parameters, $V_1$ and $V_3$. The parameter $V_1$ controls, how fast the thermodynamic quantities $p/T^4$, $\epsilon/T^4$, $s/T^3$  approach their asymptotically free values, while $V_3$ controls the latent heat density. According to \cite{Gursoy:2008za}, the best match with  lattice data is achieved, if we choose 
\begin{equation}
V_1=14, \quad \quad V_3=170,
\end{equation}
Another choice is the potential used in \cite{Jarvinen:2011qe}, given by
\begin{equation}
V(\lambda)=\frac{12}{\mathcal{L}^2}\left[
1+V_0\lambda + V_1\sqrt{V_2}\lambda^2\frac{ \sqrt{1+\ln(1+\lambda)}}{(1+\lambda)^{2/3}}
\right]\, . \label{V}
\end{equation}
This is the potential that we have used in all of our calculations. It does not contain any free parameters  that would have to be determined by matching with lattice thermodynamics as in case of (\ref{dilpot2}). All observables of interest derived using this potential, such as bulk thermodynamic quantities and energy momentum tensor correlators, coincide with those calculated using eq.~(\ref{dilpot2}).

\section{Thermodynamics}
In the large-$N_c$ limit, the partition function of the model described above can  be approximated by  the saddle-point approximation that is given by the classical solution of  the  Einstein-dilaton field equations. If there exist more  saddle points,  the partition function is given by a sum over them
\begin{equation}
Z(\beta)\approx e^{-S_1^E(\beta)}+e^{-S_2^E(\beta)}+\dots,
\end{equation}
where $S_i^E$ are the Euclidean actions evaluated with  each classical solution at a temperature $T=1/\beta$. There are two possible types of solutions preserving $SO(3)$ invariance
\begin{list}{}{}
\item[1] \textbf{Thermal gas solution,}
\begin{equation}
ds^2=b_0^2(z)\left(d\tau^2 + d\vec{x}^2+dz^2\right), 
\end{equation}
with $z\in (0,\infty)$. This is the Euclidean version of  the vacuum solution and it exists for all temperatures. This solution is related to the confining phase of the dual field theory.
\item[2]  \textbf{Black hole solutions,}
\begin{equation}
ds^2=b^2(z)\left(f(z)d\tau^2+d\vec{x}^2+\frac{dz^2}{f(z)}\right),
\end{equation}
with $z\in (0,z_h)$, where $z_h$ is the black hole horizon. This solution is identified with the deconfined phase of the gauge theory \cite{Gursoy:2008za}. 
\end{list}
Studying the transition between these solutions, we can study the transition between the confined and deconfined phases of the gauge theory, see \cite{Gursoy:2008za} for more details. 

The partition function is dominated by the solution that minimizes the free energy, which we identify with the Euclidean action. This quantity is by definition
\begin{equation}
\mathcal{F}=E-TS,
\end{equation}
where $E$, $T$ and $S$ are identified with the energy, temperature and entropy of the black hole. The temperature and entropy read in turn
\begin{equation}
T=-\frac{\dot{f}(z_h)}{4\pi}, \quad \quad S=s.V_3=\frac{1}{4G_5}b^3(z_h) V_3, \label{entrop66}
\end{equation}
where $V_3$ is the three-dimensional volume and $s$ the entropy density.

The on-shell action is in general divergent close to the boundary. We can avoid these divergences either by the method of holographic renormalization, as we discussed in  section \ref{holoren}, or we can  calculate the difference between the free energy of the thermal gas  and black hole phases , whereby the divergent parts cancel each other, as was shown in \cite{Gursoy:2010fj}.   We choose this second method and  find that the  thermal gas and black hole solutions with the same temperature differ at $O(z^4)$ \cite{Gursoy:2010fj}:
\begin{eqnarray}
b(z&=&b_0(z)\left[1+ \mathcal{G}\frac{z^4}{\Lag^3}\right],\\
\lambda(z)&=&\lambda_0(z) \left[1+\frac{45}{8}\mathcal{G}\frac{z^4\log \Lambda z}{\Lag^3} \right],\\
f(z)&=&1-\frac{C}{4}\frac{z^4}{\Lag^3},\\
\end{eqnarray}
where $C$ and $\mathcal{G}$ are constants that are related to the enthalpy $TS$ and the gluon condensate $\av{\Tr F^2}$ via
\begin{equation}
C=16\pi G_5 \frac{TS}{V_3}, \quad \quad \mathcal{G}=\frac{11G_5}{360\pi}\left( \av{\Tr F^2}_T -\av{\Tr F^2}_0\right),
\end{equation}
where  the last term is a difference between $\av{\Tr F^2}$ at finite and zero temperature. The solution that dominates the partition function is the one that minimizes the free energy.
The free energy difference between the black hole and  the thermal gas solutions reads 
\begin{equation}
\frac{\Delta\mathcal{F}}{V_3}= 16\pi G_5\left(
15\mathcal{G}-\frac{C}{4}\right).
\end{equation}
Studying the free energy of the system, we then find that for the black hole phase there is some minimum temperature, $T_{min}$, while the thermal gas phase exists for all temperatures. At temperatures $T\geq T_{min}$  both phases exist, but up to $T=T_c$ the thermal gas solution dominates. At $T=T_c$, there is a first order phase transition to the black hole phase, and the system remains in the deconfined phase for all $T>T_c$. Since we are interested only in the deconfined phase, from now on we consider only the black hole solution.

Using the free energy, we can calculate other thermodynamic observables, such as the pressure, specific heat and speed of sound
\begin{equation}
p=-\mathcal{F}, \quad C_v=-T\frac{\partial^2 \mathcal{F} }{\partial T^2}, \quad
 c_s^2= \frac{S}{C_v}.
\end{equation}
We have a formula (\ref{entrop66}) for the entropy in terms of the gravitational constant. To relate it to the  field theory result,  we need to find a mapping between parameters on the gravity and field theory sides. We evaluate the  formula for the entropy in the UV, where we know that it should match the entropy of a free gluon gas in the large-$N_c$ limit. We find that they agree if we make the identification of the parameters as
\begin{equation}
\frac{\Lag^3}{4 G_5}=\frac{4N_c^2}{45\pi}.\label{concon}
\end{equation}

In figure~\ref{figthermo}, we finally plot the pressure, the  energy density and, what is particularly interesting, their difference 
\begin{equation}
\frac{\epsilon-3p}{T^4}.
\end{equation}
This quantity is known as the \textit{nonconformality factor} or \textit{trace anomaly} and is proportional to the trace of the energy momentum tensor and thus measures the deviation of the system from conformal field theory (where $(\epsilon-3p)=0$ by definition). We observe that our holographic  result for the non-conformality factor is in a qualitative agreement with lattice QCD simulations \cite{Panero:2009tv}.

For more details on the thermodynamics of pure Yang-Mills theory, see \cite{Gursoy:2009jd}, and for the thermodynamics of quasiconformal theories in a similar framework, \cite{Alanen:2010tg}. Finally, for  discussion of some further issues within the thermodynamics of IHQCD, see e.g.~\cite{Veschgini:2010ws,Megias:2010tj}.

\begin{figure}[t]
\centering
\includegraphics[width=0.75\textwidth]{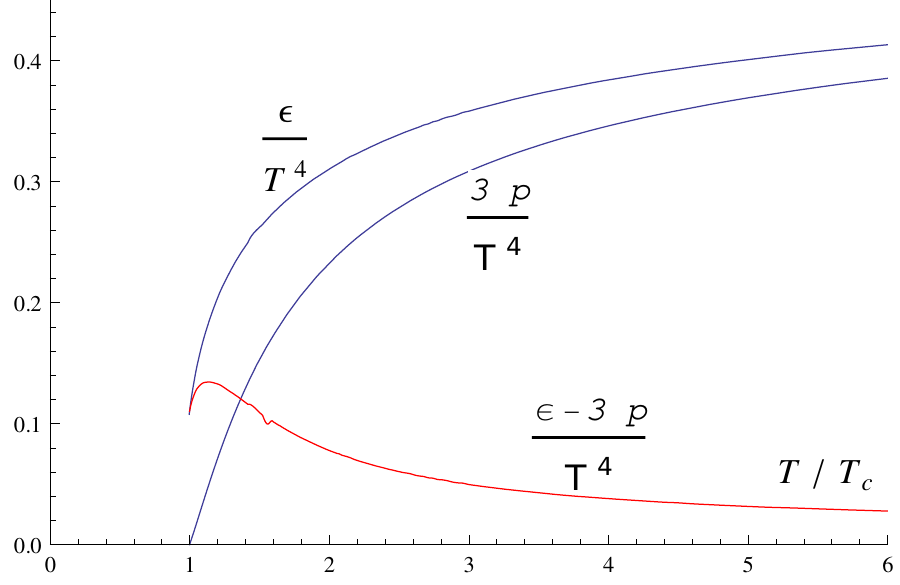}$\;\;\;$ 
\caption{\small The pressure $p/T^4$, energy density $\epsilon/T^4$ and their difference $(\epsilon-3p)/T^4$. Note that the form of the curve of the nonconformality factor is in agreement with the prediction of large-$N_c$ lattice simulations found in \cite{Panero:2009tv}.}
\label{figthermo}
\end{figure}  

\section{Numerical Integration\label{numee}}
As we have already noted, the system of differential equations (\ref{eeq1})-(\ref{eeq3}), with the potential of eq.~(\ref{V}), is too complicated to solve analytically with the exception of  some asymptotic limits. In the general case, we thus need to use some method of numerical integration, which in practice is most straightforwardly implemented in a software capable of also symbolic manipulations, such as Mathematica. In this process, we follow the method of numerical integration described in  \cite{Alanen:2010tg,Alanen:2011hh}, which we will present below.

We start by rewriting the equations (\ref{eeq1})-(\ref{eeq3}) as a system of first order differential equations, using the superpotential (\ref{spdef}),
\begin{eqnarray}
\dot{W}&=& 4bW^2-\frac{1}{f}\left(W\dot{f}+\frac{1}{3}b V\right), \label{Weq1}\\
\dot{b}&=&-b^2 W,\\
\dot{\lambda} &=& \frac{3}{2}\lambda\sqrt{b \dot{W}},\\
\ddot{f}&=&3 \dot{f}b W\label{Weq4}.
\end{eqnarray}
To numerically solve this system, we need to specify five integration constants by choosing the boundary conditions appropriately. We choose to implement the boundary conditions at the horizon, since there we can relate the integration constants with the thermodynamic observables described in the previous section. This allows us then to parametrize the solution of the system (\ref{Weq1})-(\ref{Weq4}) in terms of thermodynamic variables.

In practice, the numerical integration of the above system cannot start from $z=z_h$, as $f(z)$ is singular there. This means that we have to start at some initial $z=z_i=z_h-\epsilon$, where $\epsilon$ is a small, but nonzero number. An analytic calculation then gives us the initial values of the functions
\begin{eqnarray}
\lambda_i &=& \lambda_h+\frac{3}{8}\lambda_h^2 b_h^2\frac{V'(\lambda_h)}{\dot{f}_h}\epsilon, \\
b_i&=&b_h+b_h^2W_h\epsilon,\\
W_i&=&W_h-\frac{1}{16\dot{f}_h^2}b_h^3\lambda^2_h\left(V'(\lambda_h)\right)^2\epsilon,  \\
f_i&=& f_h-\dot{f}_h\epsilon,\\
\dot{f}_i&=& \dot{f}_h-3\dot{f}_hb_hW_h\epsilon,
\end{eqnarray}
where the prime denotes differentiation with respect to $\lambda$, and we have used a notation where $\lambda_i =\lambda(z_i)$,  $\lambda_h=\lambda(z_h)$, and similarly for other functions. 

We now need to determine the values of all five functions at the horizon. We have by definition $f_h=0$, while from the regularity of the second term in (\ref{Weq1}) we obtain
\begin{equation}
W_h=-\frac{b_h V(\lambda_h)}{3\dot{f}_h}.
\end{equation}
The variable $b_h$ appearing here will be traded for the constant of integration $\Lambda$ in eq.~(\ref{b15}), while $\dot{f}_h$ will be traded for the temperature $T$ and $\lambda_h$ will be used to parametrize the solutions. Let us next show in detail, how all of this is implemented.

We start from arbitrary values of $b_h$, $\dot{f}_h$ and $\lambda_h$,  and find the numerical solution for the system  (\ref{Weq1})-(\ref{Weq4}) in some range $z_{max}<z<z_i$, where $z_{max}$ is some maximum value at which $b(z_{max})$ will diverge. We then implement the following three scaling procedures that trade the $b_h$ and $\dot{f}_h$ for the physical parameters.
\begin{list}{}{}
\item[1.] Scale $W(z_{max})$ to one and define a scaling factor $S_1=W(z_{max})$. Use it to define a new set of solutions with $\lambda_1=\lambda$, $W_1=W/S_1$, $b_1=S_1b$, $f_1=S_1^2 f$. This scales $f_1(z_{max})=1$.
\item[2.] Shift $z_{max}$ to zero. Define $\lambda_2(z)=\lambda_1(z+z_{max})$, $W_2(z)=W_1(z+z_{max})$, $b_2(z)=b_1(z+z_{max})$, $f_2(z)=f_1(z+z_{max})$ and  $S_2=z_{max}$. 
\item[3.] Scale $z$ in such a way that for any $\lambda_h$, the relation (\ref{lambdaz})  holds for some $\Lambda$. For small $z$, we know that  
\begin{equation}
\lambda_2(z)=-\frac{1}{b_0 \log(\Lambda_2 z)}=-\frac{1}{b_0 \log(\Lambda(\Lambda_2 z/\Lambda))}.
\end{equation}
To have $\lambda_h$ unchanged, we find that the third scaling is $S_3=\Lambda_2/\Lambda$, and obtain $\lambda_3(z)=\lambda_2(z/S_3)$, $W_3(z)=W_2(z/S_3)$, $b_3(z)=b_2(z/S_3)/S_3$ and $f_3(z)=f_2(z/S_3)$, where $\Lambda_2=\exp\left(-1/b_0\lambda_2(z)\right)/z$.
\end{list}
To construct a concrete solution, we choose a numerical value for $\Lambda$ and a small UV value for $\lambda$ (we used $\Lambda=1/200$ and $\lambda_{UV}=1/50$). We then consider a set of values of $\lambda_h$ in the form of a table, and integrate  the system (\ref{Weq1})-(\ref{Weq4}) numerically for each value of $\lambda_h$. Due to eq.~(\ref{entrop66}), we can relate the temperature with the horizon value $\lambda_h$. Thus to each $\lambda_h$ corresponds some temperature $T$, and  our set of solutions of the system (\ref{Weq1})-(\ref{Weq4}) can be identified as solutions corresponding to different temperatures.

\chapter{Energy Momentum Tensor Correlators in IHQCD\label{ch6}}
We now proceed to present the most important new results of the thesis, a calculation of energy momentum tensor correlators in IHQCD. These results were published in our recent papers \cite{Kajantie:2011nx,Kajantie:2013gab} and also presented in conferences in Munich \cite{Vuorinen:2013fd} and Barcelona \cite{Krssak:2013jla}. Here, we  present a more thorough calculation that includes some details that were omitted in the original papers. We divide our discussion into two separate sections. In the first one, we consider the energy momentum tensor correlator in the shear channel, i.e. the correlator of the operator $T_{12}$. In the second section, we consider the correlator in the bulk channel, i.e.~the one that corresponds to the $T_{ii}$.

In  chapter~3 of this thesis, we discussed the hydrodynamic limit of the AdS/CFT correspondence and the  shear viscosity in the conformal $\mathcal{N}=4$ SYM theory (the bulk viscosity was trivially zero there, due to the conformal invariance). We found that the ratio of the shear viscosity to entropy is universal in all holographic models with a two-derivative  gravity action.

According to the Kubo formula (\ref{eta6}), the shear viscosity is given by the low-frequency limit of the imaginary part of the retarded correlator $\av{T_{12}T_{12}}$. This means that to distinguish between various holographic models in the shear channel, we need to consider the full correlator, i.e.~to go beyond the low-frequency limit.  We will see that for temperatures close enough to the critical temperature $T_c$, we find a sizeable difference between  the shear  correlator  in IHQCD and the conformal $\mathcal{N}=4$ SYM theory.

Another reason to consider the full correlator is that we would like to make a comparison of  our holographic results with other field theory methods, such as perturbative and lattice QCD. As we will see, perturbative QCD fails in the low-frequency  and  low-temperature limits. However, in the  region of validity of perturbative calculations, the matching of our holographic calculations with those of perturbative QCD is an important test of IHQCD. 

The full correlators are also required, if we want to confront our holographic results with those of lattice QCD. This is because in lattice QCD we have only Euclidean correlators available, which can be determined from our Minkowskian correlators as  integrals over all frequencies  using the formula (\ref{euclidean}). 

\section{Shear Channel}
First, let us  consider the retarded correlator of the energy momentum tensor   in the shear channel, i.e.
\begin{equation}
G_s^R(\omega)=-i\int\! {\rm d}^4x\,  e^{i \omega t} \theta (t) \langle[
T_{12}(t,\vec{x}),T_{12}(0,0) ]\rangle,\label{Gsdefhol}
\end{equation}
and in particular the corresponding spectral density that is given by the imaginary part of the above function,
\begin{equation}
\rho_{s} (\omega)=\mathop{\mbox{Im}} G_{s}^R(\omega)\, .
\end{equation} 
The shear viscosity is obtained from here using the  Kubo formula (\ref{eta}),  
\begin{equation}
\eta=\lim_{\omega\to 0} \frac{\rho_s(\omega)}{\omega}.\label{eta6}
\end{equation}
Next, we present a detailed calculation of the correlator using the method of  holographic renormalization,  inspired by \cite{Buchel:2004di}. For an analogous calculation using a different coordinate system, but leading to the same result, see \cite{Gubser:2008sz}. 
At the end of this section,  we confront our IHQCD results with those of  conformal $\mathcal{N}=4$ SYM theory and perturbative QCD. 
\subsection{Shear Channel Correlator\label{scc}}
We start with the background metric
\begin{equation}
ds^2=b^2(z) \left(-f(z) dt^2 + d\vec{x}^2 +\frac{dz^2}{f(z)}\right), \label{met61}
\end{equation}
and consider an action that consists of a bulk term, given by  (\ref{ihqcdaction}), and a counterterm part consisting of the Gibbons-Hawking-York boundary term and other counterterms required to cancel divergences,
\begin{equation}
S=S_{bulk} + S_{ct}. \label{actfull}
\end{equation}
Explicitly, these two terms take the forms
\begin{equation}
S_{bulk}=\frac{1}{16\pi G_5}\int d^5x\sqrt{-g} \left[R -\frac{4}{3}\frac{ (\partial\lambda)^2}{\lambda^2} +V(\lambda) \right], \label{actbulk} 
\end{equation}
and 
\begin{equation}
S_{ct}=\frac{1}{16\pi G_5}\int d^4 x \sqrt{-\gamma}\left[2K+\frac{6}{\Lag} + \frac{\Lag}{2}R(\gamma)\right]. \label{actct}
\end{equation}
The boundary term does not affect the  equation of  motion, and thus the functions $b(z)$, $f(z)$, $\lambda(z)$ are solutions of the Einstein equations derived from the bulk term (\ref{Weq1})-(\ref{Weq4}). They are solved numerically by the method that we  described at the end of previous chapter.

\subsubsection*{Fluctuation Equation}
Let us now introduce a small perturbation around the 12-component of the background metric (\ref{met61}),  i.e.
\begin{equation}
g_{12}= h_{12}.
\end{equation}
According to (\ref{gravem}), the perturbation $h_{12}$ couples to the $T_{12}$ component of the energy momentum tensor. 

For simplicity, let us consider only the case of a vanishing spatial momentum  and a harmonic time-dependence of the perturbation, i.e. write 
\begin{equation}
h_{12}(x,z)=h_{12}(z) e^{i\omega t}.\label{harmdep}
\end{equation}
The equation of the motion for $h_{12}$, or  the fluctuation equation, is found  by expanding the bulk term up  to the second-order in $h_{12}$.  Remember that the  boundary term does not affect the equations of motion, but instead plays an important role in the evaluation of the action and in finding the correlators. 

We find that the bulk action reads
\begin{equation}
S_{bulk}=\frac{1}{16\pi G_5} \int d^5 x \left[
L(h_{12},\dot{h}_{12},\ddot{h}_{12}) - 2\partial_z\left(fb^2\dot{b}\right) + \mathcal{O}(h^4_{12}) \right], \label{lagshear}
\end{equation}
in which $L$ is the Lagrangian 
\begin{equation}
L(h_{12},\dot{h}_{12},\ddot{h}_{12})=A\ddot{h}_{12}h_{12}+B\dot{h}_{12} \dot{h}_{12}+C\dot{h}_{12}h_{12}+Dh_{12}h_{12},
\end{equation}
where the coefficients are given by
\begin{equation}
A=2fb^3, \quad B=\frac{3}{4}A, \quad C=A\frac{d}{dz}\log \left(b^4f\right), \quad D=\frac{d}{dz}\log \left(b^2\dot{b}f\right) +\frac{b^3\omega^2}{2f},
\end{equation}
and $\partial_z\left(fb^2\dot{b}\right)$ is a surface term that we chose to separate from the $L$ The equation of motion (EOM) is then found as a variation of $L$ with respect to $h_{12}$,
\begin{equation}
\ddot{h}_{12} +\frac{d}{dz}\left(b^3 f \right)\dot{h}_{12}+
\frac{\omega^2}{f^2}h_{12}=0. \label{fluctshear1}
\end{equation}

\subsubsection*{Evaluation of the Action}
Using the equation of motion, we can rewrite the bulk action in the form
\begin{eqnarray}
S_{bulk}&=&\frac{1}{16\pi G_5} \int d^5 x \left\lbrace 
  \left[\frac{1}{2} (\text{EOM}) h_{12}+\partial_z\left(B \dot{h}_{12}h_{12}+\frac{1}{2}(C-\dot{A}) h_{12}^2\right)\right] \right. \nonumber\\  &-& \left. 2\partial_z\left(fb^2\dot{b}\right) + O(h^4_{12}) \right\rbrace.
\end{eqnarray}
For the on-shell solution, the only non-vanishing contribution comes from the surface terms. 

The second contribution to the action (\ref{actfull}) comes from the counterterm action (\ref{actct}). The boundary metric $\gamma_{\mu\nu}$ is induced by the surface $z=\text{constant}$,  with the exterior curvature and normal vector  given by 
\begin{equation}
2K=n^z\gamma^{\mu\nu}\partial_z \gamma_{\mu\nu}, \quad \quad n^z=\sqrt{g^{zz}},
\end{equation}
from where  we find that up to second order in $h_{12}$ 
\begin{eqnarray}
\sqrt{-\gamma}&=&b^4\sqrt{f}\left(1-\frac{1}{2}h_{12}h_{12}\right),\\
\sqrt{-\gamma}2K&=&b^3f \left(8\frac{\dot{b}}{b}+\frac{\dot{f}}{f}\right)-2b^3f\left[
\dot{h}_{12}h_{12} +\frac{1}{4}\left( 8\frac{\dot{b}}{b}+\frac{\dot{f}}{f}\right) h_{12}h_{12}\right],\\
R(\gamma)&=&\frac{\omega^2}{2b^2 f}h_{12}h_{12}.
\end{eqnarray}
Adding the bulk and counter term actions together, we find that on shell the full action is given entirely  by the surface terms. 

Next, we  go  to  Fourier space using the transformation
\begin{equation}
h_{12}(x,z)=\int \frac{d^4k}{(2\pi)^4} e^{kx}h_{0}(k)h_k(z)
\end{equation}
where $h_{k}^\star=h_{-k}$, and $h_k$ is normalized as
\begin{equation}
h_k(0)=1.\label{boundnorm}
\end{equation}
With the  help of this, we find that the full action (\ref{actfull}) can be written as 
\begin{equation}
S=\frac{1}{16\pi G_5}\int\frac{d^4k}{(2\pi)^4} \left[\frac{1}{2} fb^3\left(\dot{h}_k h_{-k}-\dot{h}_{-k} h_{k} \right) + (\dots)  h_{k} h_{-k} \right]_{z= 0}^{z= z_h}.
\end{equation}
Using  the prescription (\ref{corrdef}), we find that the retarded correlation function in the shear channel becomes
\begin{equation}
G_s^R(\omega)=\frac{1}{16\pi G_5} \left[ fb^3\dot{h}_k h_{-k} + (\dots) h_{k}h_{-k} \right]_{z\to 0}.\label{shearcorr2}
\end{equation}
As we are interested only in the imaginary part of the correlator, we can neglect the  second, manifestly real term, leading to
\begin{equation}
\rho_s(\omega)=\im G_s^R(\omega)=\frac{1}{16\pi G_5} fb^3 \im \dot{h}_k h_{-k}. \label{shearcorr3}
\end{equation}
To simplify the above result, we can further rewrite the imaginary part as  
\begin{equation}
\im \dot{h}_k h_{-k}=\frac{\dot{h}_k h_{-k}-\dot{h}_{-k} h_{k}}{2i}=\frac{W(h_k,h_{-k})}{2i},
\end{equation}
where $W(h_k,h_{-k})$ is the Wronskian of the two independent solutions $h_{k}$ and $h_{-k}$. The advantage of this method is that the Wronskian can be obtained from the fluctuation equation (\ref{fluctshear1}) simply by integrating $\dot{W}/W=-P$, where $P$ is the coefficient of the $\dot{h}_{12}$-term  in (\ref{fluctshear1}). Performing the integration, we find
\begin{equation}
W(h_k,h_{-k})=\frac{W_0(\omega)}{b^3 f}, \label{wronsk1}
\end{equation} 
where $W_0$ is  independent of the  $z$-coordinate. Inserting (\ref{wronsk1}) into  (\ref{shearcorr3}), the spectral density becomes
\begin{equation}
\rho_s(\omega)=\frac{1}{16\pi G_5} \frac{W_0(\omega)}{2i},
\end{equation} 
from where we explicitly see that the quantity is independent of the $z$-coordinate. To determine the $z$-independent constant $W_0$, we however need to evaluate the Wronskian at some value of $z$. We choose to do so at the horizon, where we  enforce the infalling boundary condition. 

Close to the horizon, we can expand  the fluctuation equation (\ref{fluctshear1}) in powers of $(z_h-z)$, starting with $(z_h-z)^p$, where $p$ is the characteristic exponent. In our case we find that $p=\pm i \omega /\dot{f}_h$, where $\dot{f}_h=\dot{f}(z_h)$. The infalling boundary condition corresponds to the choice of the  plus sign, and thus at the horizon
\begin{equation}
h_{k}(z\to z_h)=(z_h-z)^{i \omega/\dot{f}_h}. \label{boundhor1}
\end{equation}
We wish to enforce this boundary condition as close  to the horizon as possible. However, this is often challenging in practical calculations, since we are not able to go  arbitrarily close to the horizon due to numerical reasons. Instead,  we  find the solution to the fluctuation equation (\ref{fluctshear1}) as an analytic expansion around $z_h$, writing
\begin{equation}
h_{k}(z\to z_h)=(z_h-z)^{i \omega/\dot{f}_h}(1+d_1 (z_h-z)+ d_2(z_h-z)^2+\dots) \label{boundhor}.
\end{equation}
where $d_1$ and $d_2$ are exactly calculable coefficients. From our experience, it is sufficient to include terms up to $d_3$. 
 
Now, we proceed to evaluate the Wronskian in the near-horizon limit, and determine that  $W_0(\omega)$ is given by
\begin{equation}
W_0(\omega)=2i\omega b_h^3.
\end{equation}
Using this, we find
\begin{equation}
\rho_s(\omega)=\frac{1}{16\pi G_5} \omega b_h^3,
\end{equation}
where we still need to ensure proper normalization (\ref{boundnorm}), and subsequently find
\begin{equation}
\rho_s(\omega)=\frac{1}{16\pi G_5}  b_h^3 \frac{\omega}{\left| h_k(0)\right|^2}. \label{rhoshear}
\end{equation}
In our calculation we have set $z_h=1$, so that all dimensionful
quantities are expressed in units of $z_h$. In the conformal case, $z_h=1/(\pi T)$, from where it follows that   dimensionful quantities like $\omega$ should be replaced by
\begin{equation}
\omega\to \frac{\omega}{\pi T}.
\end{equation}
In practice,  it turns out to be  helpful to scale everything by this factor; from now on, we will always use these rescaled quantities.

\subsection{Holographic Results}
We will now use the result in eq.~(\ref{rhoshear}) to calculate the spectral density in the shear channel of IHQCD, and using the same method, the same quantity  in the $\mathcal{N}=4$ SYM theory \cite{Teaney:2006nc,Kajantie:2010nx}. Starting from the $\omega\to \infty$ limit, we scale the $z$-coordinate as $z\to z'=\omega z$, using which we find that the large-$\omega$ limit corresponds to the small-$z$ solution of the fluctuation equation (\ref{fluctshear1}), 
\begin{equation}
\ddot h_{12}-{3\over z}\dot h_{12}+h_{12}=0 \,. \label{fluctlarge}
\end{equation}
This equation can be solved analytically and gives us the large-$\omega$ limit of the shear spectral density as
\begin{equation}
\rho_{s}(\omega)\stackrel[\omega\to \infty]{}{\rightarrow}\frac{\Lag^3}{4\pi G_5}\frac{\pi }{32}\omega^4=\frac{N_c^2}{360\pi}\omega^4, \label{rhoas61}
\end{equation}
where we have used the relation (\ref{concon}) to express the spectral density in field theory units. The conformal theory turns out to have the same  large-$\omega$ limit of the spectral density \cite{Kajantie:2010nx}.

\begin{figure}[t]
\centering
\includegraphics[width=0.48\textwidth]{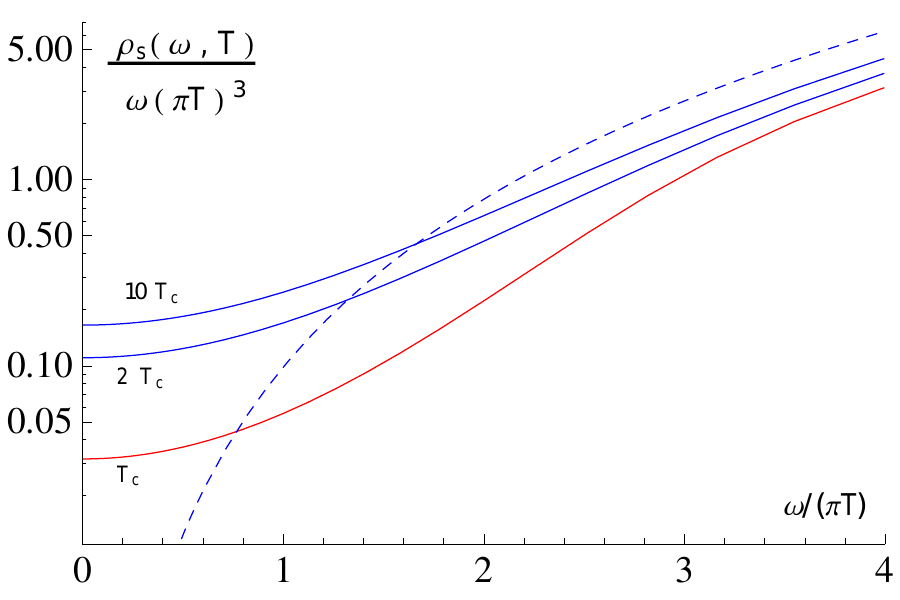}
$\;\;\;$
\includegraphics[width=0.48\textwidth]{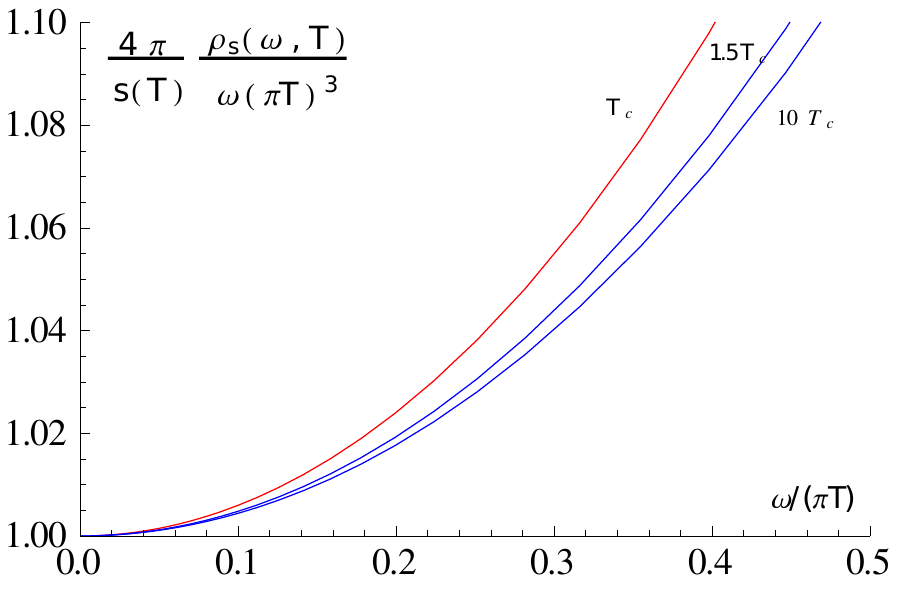}
\caption{\small Left: The ratio of the IHQCD  spectral density  $\rho_s(\omega)$ and frequency $\omega$ in the shear channel, displayed for three different temperatures in units of $\mathcal{L}^3/(4\pi G_5) $. The dashed curve corresponds to the asymptotic limit of $\pi\omega^3/32$.
Right: The ratio of the spectral density and frequency, normalized by the entropy. Using the Kubo formula of eq.~(\ref{eta}), we observe that the shear viscosity over entropy obtains the universal value of $1/(4\pi)$.}
\label{figshear1}
\end{figure}

In fig.~\ref{figshear1} (left) we first plot the ratio of the shear spectral density and frequency, in units of $\mathcal{L}^3/(4\pi G_5) $,  for three different temperatures.  The dashed curve represents the asymptotic limit, towards which the IHQCD curves are logarithmically approaching. In fig.~\ref{figshear1} (right), we then plot the same quantity normalized by the entropy and the factor of $4\pi$. According to the Kubo formula (\ref{eta6}), the shear viscosity can be read off from the value of these curves at $\omega\to0$. We  see that in IHQCD we recover the universal prediction  $\eta/s=1/(4\pi)$.

In fig.~\ref{figshear2}, we plot the spectral density in IHQCD and the conformal $\mathcal{N}=4$ SYM theory for two different temperatures, one of which is the critical one and the other slightly higher, $T=1.5T_c$. We observe that close to the critical temperature, there is a sizeable effect from the nonconformality of IHQCD that disappears with increasing temperature. This is consistent with our expectations that QCD becomes conformal at high enough temperatures. From here we see that in the case of the shear spectral density this approach is extremely fast, signaling that for temperatures $\gtrsim 1.5 T_c$, IHQCD becomes an effectively conformal theory.

\begin{figure}[t]
\centering
\includegraphics[width=0.48\textwidth]{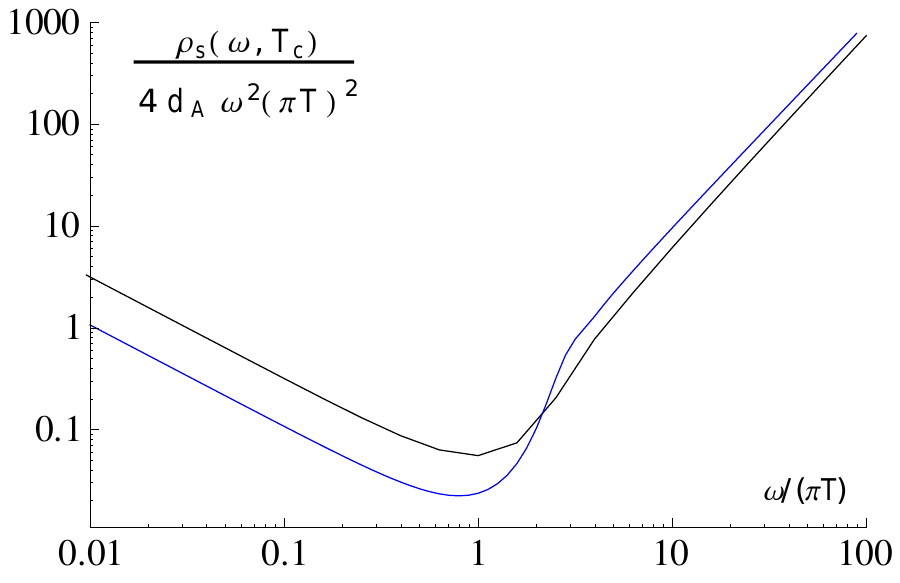}
$\;\;\;$
\includegraphics[width=0.48\textwidth]{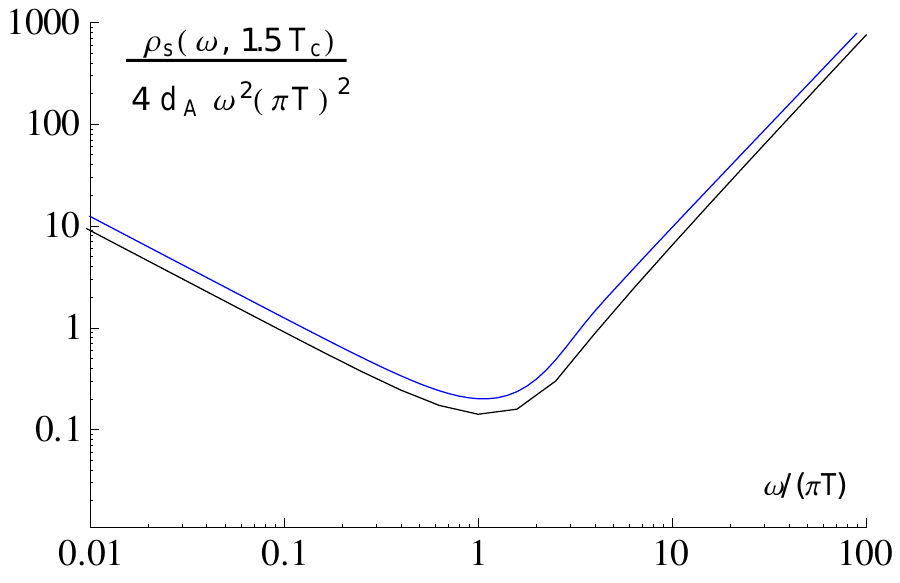}
\caption{\small The ratio of the shear spectral density and $\omega^2$ in IHQCD (black curve) and $\mathcal{N}=4$ SYM theory (blue curve) for two different temperatures. Observe the sizeable difference between the IHQCD and SYM theory curves at the critical temperature that  disappears with increasing temperature.}
\label{figshear2}
\end{figure}
\subsection{Perturbative Limit}
Next, we would like to compare our IHQCD results with those of perturbative QCD that were calculated in \cite{Zhu:2012be,Laine:2011xm}. To this end, let us briefly review  
 what is known about the behavior of the shear spectral functions in weakly coupled SU($N_c$) Yang-Mills theory. This is helpful in particular for the analysis of the UV  behavior of our results, as due to asymptotic freedom all physical correlators are expected to reduce to their perturbative limits as $\omega\to\infty$.

At the moment, most spectral functions  are known up to and  including NLO  in perturbation theory.  In the shear case, we can read off the result from eq.~(4.1) of \cite{Zhu:2012be}, obtaining (note an additional factor of -1/16 due to differing definitions of the shear operator)
\begin{eqnarray}
\rho^{\text{pert}}_s(\omega,T)=d_A\frac{\omega^4}{160\pi}\bigl( 1 + 2 n_{\frac{\omega}{2}} \bigr)\Bigg\{1-\frac{10\lambda}{16\pi^2}\bigg(\frac{2}{9}+\phi_T^\eta({\omega\over T})\bigg)\Bigg\}+
{\mathcal O}(\lambda^2),\label{rhosper0} 
\end{eqnarray}
where $d_A\equiv N_c^2-1$, $n_x\equiv 1/(e^{x/T}-1)$, and $\phi_T^\eta(\omega/T)$ is some dimensionless function that can be evaluated numerically and in the  $\omega\to\infty$ limit behaves like $T^6/\omega^6$. In the large-$N_c$ limit, we can set $d_A\approx N_c^2$.
An important thing to note is that even at high temperatures --- and thus weak coupling --- the perturbative calculations are not valid in the limit of very small $\omega$. This is due to the multitude of soft scales that enter the calculation at small momentum exchange and require complicated resummations \cite{Arnold:2003zc,Zhu:2012be,Laine:2011xm}.

We find the large-frequency limit of the perturbative spectral density (\ref{rhobper0}) to read
\begin{equation}
\rho_{s}^{\text{pert}} \stackrel[\omega\to \infty]{}{\rightarrow} \frac{N_c^2}{160\pi}\,\omega^4. 
\end{equation}
Comparing with the IHQCD asymptotics (\ref{rhoas61}) we find that the ratio of the IHQCD and perturbative asymptotics of the spectral density  is
\begin{equation}
\frac{\rho_{s}^{\text{pert}}(\omega)}{\rho_{s}(\omega)}\stackrel[\omega\to \infty]{}{\rightarrow}\frac{9}{4}. \label{fak1}
\end{equation}

In fig.~\ref{figlarge1}, we plot the shear spectral density in both IHQCD and perturbative QCD. We observe that up to the overall factor (\ref{fak1}), the IHQCD spectral density approaches the perturbative limit for all frequencies $\omega \gtrsim 2$. This is consistent with our expectation that perturbative QCD fails in the limit of very small $\omega$, and gives us confidence that IHQCD captures at least the qualitative features of the physics of Yang-Mills theory even in the weakly coupled UV limit. The discrepancy in the overall factor (\ref{fak1}) is in fact not surprising at all, and is likely related to the fact that the IHQCD action is of a two-derivative form.

\begin{figure}[t]
\centering
\includegraphics[width=0.7\textwidth]{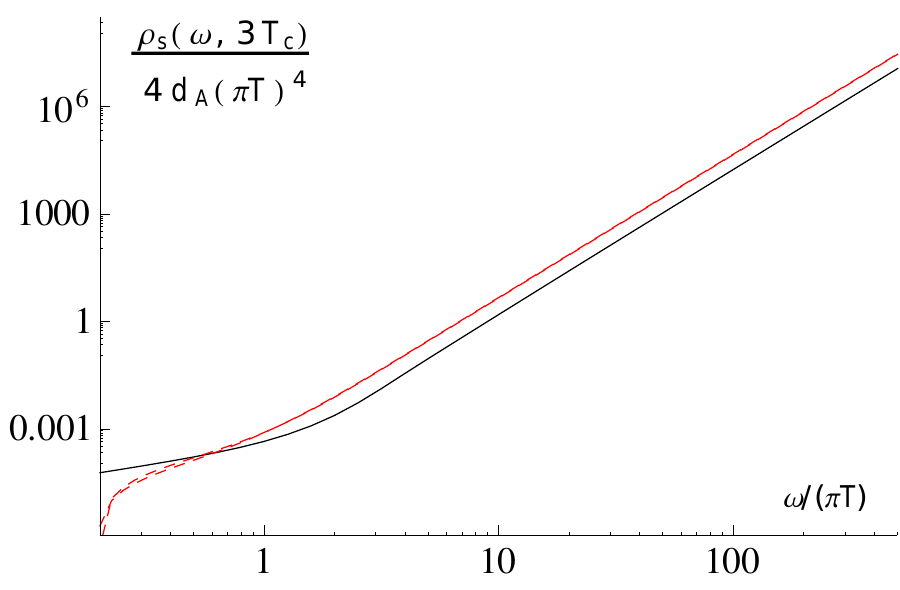}
\caption{\small The YM theory spectral density in the shear channel for $T=3 T_c$, normalized by $4d_A=4(N_c^2-1)$. The black curve corresponds to the IHQCD, and the dashed red curves to the 2-loop perturbative QCD results of \cite{Zhu:2012be}.}
\label{figlarge1}
\end{figure}

\section{Bulk Channel}
In the bulk channel,  the  correlator of interest reads
\begin{equation}
G_b^R(\omega)=-i\int\! {\rm d}^4x\,  e^{i \omega t} \theta (t) \langle[\frac{1}{3}
T_{ii}(t,\vec{x}),\frac{1}{3}T_{jj}(0,0) ]\rangle, \label{Gbdefhol}
\end{equation}
while the corresponding spectral density is given by
\begin{equation}
\rho_b(\omega)=\im G_b^R(\omega),
\end{equation}
from where the bulk viscosity is obtained using the Kubo formula (\ref{zeta})
\begin{equation}
\zeta=\lim_{\omega\to 0}\frac{\rho_b(\omega)}{\omega}.\label{zeta61}
\end{equation}
Note, that in the $\mathcal{N}=4$ SYM theory, the energy momentum tensor correlator in the bulk channel is identically zero, due to  conformal invariance. This means that our IHQCD results do not have analogous counterparts in this theory, and that the spectral density can subsequently only be confronted  with the predictions of perturbative QCD. This is indeed what we will do in section~\ref{bulkpersec}. 

In addition to the perturbative results, in the bulk channel there also exist lattice data for the corresponding Euclidean correlators. In  section~\ref{latticesec}, we will thus calculate the bulk channel imaginary time correlator in IHQCD and compare it with the results of lattice simulations. 

Finally, it is important to note that while in our holographic setup we calculate the correlator $\av{T_{ii}T_{jj}}$, in both perturbative and lattice QCD  the  available results are for  the correlator $\av{T_{\mu\mu}T_{\nu\nu}}$. However, as was shown in \cite{Meyer:2010ii}, the correlators of $T_{00}$ reduce to  contact terms, and thus we may simply  replace    $\av{T_{\mu\mu}T_{\nu\nu}}$ by $\av{T_{ii}T_{jj}}$.
\subsection{Bulk Channel Correlator}
The calculation of the bulk channel correlator is very similar to the one in the shear channel. We thus skip many intermediate steps in its derivation, as they are analogous to the steps discussed in section~\ref{scc}\footnote{ For the full derivation, see \cite{Gubser:2008sz}.}.

First, we again introduce perturbations to the metric (\ref{met61}), but since this time  we are interested in the two-point functions  of  $T_{ii}$, we perturb only  the  diagonal terms of the metric, i.e. write 
\begin{eqnarray}
g_{00}&=&-b^2f(1+h_{00}),\\ \label{bulmet1}
g_{ii}&=&b^2(1+h_{ii}),\\
g_{44}&=&\frac{b^2}{f}(1+h_{44}), \label{bulmet3}
\end{eqnarray}
where $i=1,2,3$ and due to the assumed $SO(3)$ invariance, we have $h_{ii}=h_{11}=h_{22}=h_{33}$. These perturbations are functions of $z$ and $t$ only, since we are interested in the  correlator at vanishing spatial momentum.  We again assume  a harmonic time dependence, i.e.~write in analogy with eq.~(\ref{harmdep})
\begin{equation}
h_{\mu\nu}(x,z)=h_{\mu\nu}(z) e^{i\omega t}.\label{harmdepbulk}
\end{equation}

Analogously to the shear case, we next proceed to find the equations of motion for the perturbations. We find that the equation of motion  for $h_{ii}$ decouples from the remaining equations and is given by
\begin{equation}
\ddot{h}_{ii}+ {d\over dz}\log(b^3fX^2)\dot{h}_{ii}+
\left({\omega^2\over f^2} -{\dot f\,\dot X\over fX}\right)h_{ii}=0\,,\label{fluctbulk}
\end{equation}
where we have defined 
\begin{equation}
X(\lambda) \equiv \frac{\beta (\lambda)}{3 \lambda}, \label{Xdef}
\end{equation}
which has the small-$z$ limit
\begin{equation}
X\stackrel[z\to 0]{}{\rightarrow}{1\over\log \Lambda z}\, . \label{expan33}
\end{equation}
Now, we can evaluate the action (\ref{actfull}) up to second order in $h_{ii}$,
and using the prescription (\ref{corrdef}) for  two-point correlation functions, find that the energy momentum tensor correlator in the bulk channel is given by (cf.~eq.~(\ref{shearcorr2}))
\begin{equation}
G_b^R(\omega)=\frac{1}{16\pi G_5} \left[ 6fX^2b^3\dot{h}_k h_{-k} + (\dots) h_{k}h_{-k} \right]_{z\to 0}.
\end{equation}
The fluctuation equation (\ref{fluctbulk}) is solved using purely infalling boundary conditions at the horizon, implemented via an analytic expansion around $z= z_h$, as was done  in (\ref{boundhor}). Then, using again the method of  Wronskians,  we find that the bulk channel spectral density is given by
\begin{equation}
\rho_b(\omega)=\frac{1}{16\pi G_5}  6 X_h^2b_h^3 \frac{\omega}{\left| h_k(0)\right|^2}, \label{rhobulk}
\end{equation}
where $X_h=X(z_h)$.

\subsection{Holographic Results}

\begin{figure}[t]
\centering
\includegraphics[width=0.48\textwidth]{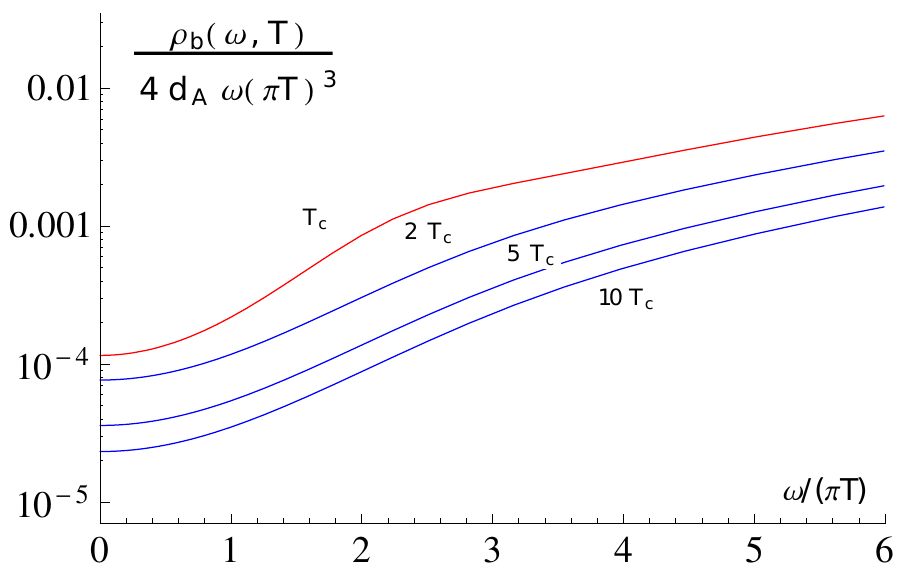}$\;\;\;$ \includegraphics[width=0.48\textwidth]{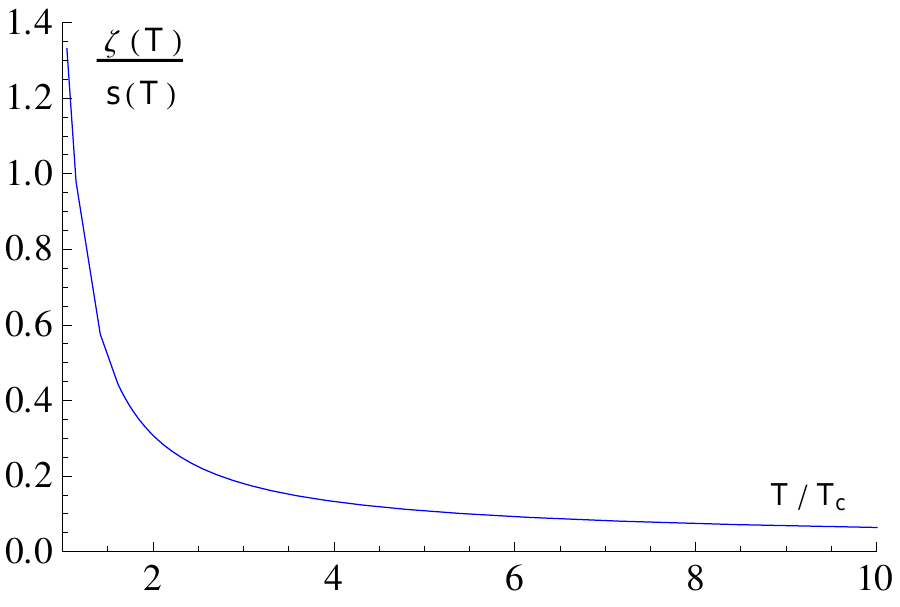}
\caption{\small Left: The ratio of the IHQCD bulk spectral density  $\rho_b(\omega)$ and $\omega$, normalized by $4 d_A$. Right: The ratio of the bulk viscosity (\ref{zeta61}) and entropy as a function of temperature.}
\label{figbulk1}
\end{figure}

We plot the ratio of the spectral density (\ref{rhobulk}) and frequency in  fig.~\ref{figbulk1} (left) for four different temperatures, normalized by the factor $4d_A=4(N_c^2-1)\approx N_c^2$. 
The bulk viscosity is again read off from the values of these curves at $\omega\to0$. Subsequently, we plot the ratio of the bulk viscosity and the entropy in figure~\ref{figbulk1} (right), observing  a dramatic rise in the quantity close to the critical temperature. With increasing temperature, the viscosity decreases and vanishes logarithmically. This is consistent with the observation that QCD behaves like a conformal theory for high temperatures.

In the bulk channel, we are not able to analytically determine the behavior of the spectral density as $\omega\to \infty$. This is due to the presence of the $X$-term in the fluctuation equation (\ref{fluctbulk}).  Analogously to shear case, scaling the radial variable according to $z\to z'=\omega z$, the small-$z$ limit of equation (\ref{fluctbulk}) becomes
\begin{equation}
\ddot h_{ii}+\Bigg\{-{3\over z}\biggl(1+{4\over 9(\log \Lambda z/\omega )^2}\biggr)
+{2\over z|\log\Lambda z/\omega |}\Bigg\}\dot h_{ii}+h_{ii}=0 \, ,
\end{equation}
which without the logarithmic terms would lead to the usual $\omega^4$ behavior of the spectral function. 

Due to the above complications, we are unfortunately not able to determine the large-$\omega$ limit of the bulk spectral density analytically, but must resort to numerics. In fig.~\ref{figbulk2} (left), we first plot the bulk spectral density normalized by the factor $4d_A\omega^4$ as a function of $\pi T_c$. We find that the asymptotic limit is logarithmic in $\omega$, but independent of temperature,
\begin{equation}
\rho_b\stackrel[\omega\to \infty]{}{\rightarrow}\frac{\omega^4}{(\log \omega/T_c)^2}. \label{bulkas}
\end{equation}
Interestingly, we  find a similar result using  perturbative QCD, but there  the logarithmic  behavior is clearly of different origin. Let us thus next present the perturbative QCD results in some detail and discuss this issue further.

\begin{figure}[t]
\centering
\includegraphics[width=0.48\textwidth]{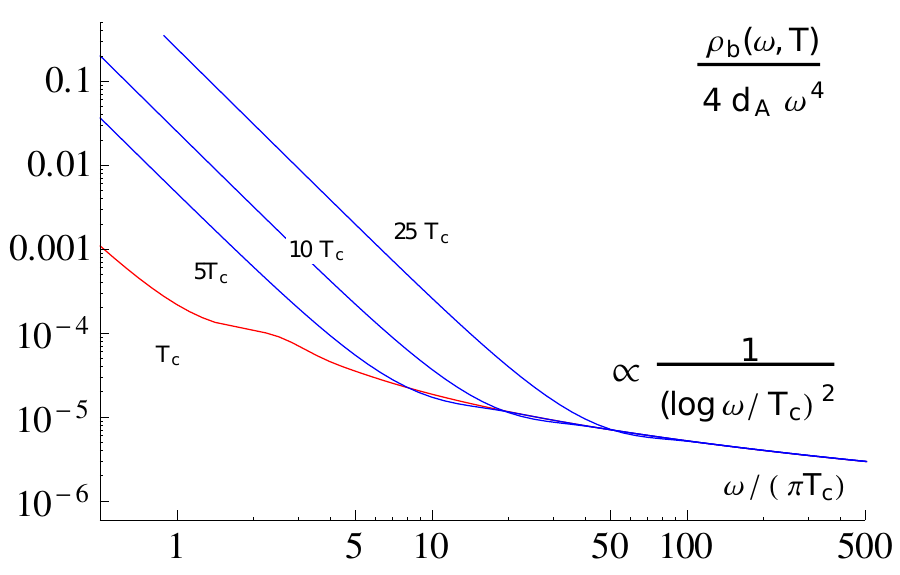}
$\;\;\;$ 
\includegraphics[width=0.48\textwidth]{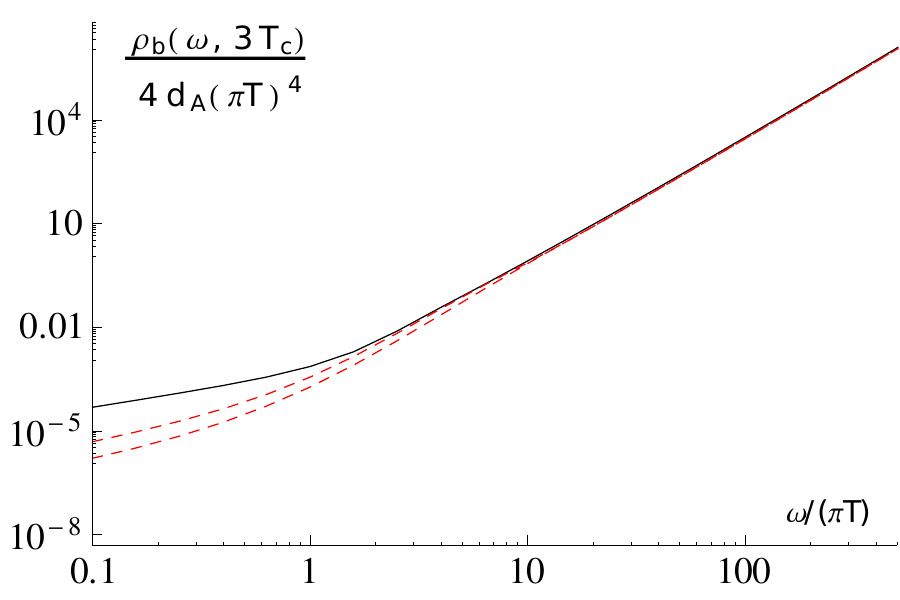}
\caption{\small Left: The ratio $\rho_b/\omega^4$, plotted as a function of $\omega/(\pi T_c)$ for multiple temperatures, normalized by $4 d_A$. For large values of $\omega$, this ratio reduces to the  temperature independent limit $1/(\log\omega/T_c)^2$.
Right: Spectral densities in the bulk  channel for $T=3 T_c$, normalized again by $4d_A$. The solid black curves correspond to the IHQCD and the dashed ones to the 2-loop perturbative QCD results of \cite{Zhu:2012be}.
}
\label{figbulk2}
\end{figure}
\subsection{Perturbative Limit\label{bulkpersec}}
In the bulk channel, the perturbative spectral function consistent with our definitions is obtainable from eq.~(4.1) of \cite{Laine:2011xm}. Multiplying this result by $1/9$ and choosing the constant $c_\theta$ as $g^2 c_\theta = \frac{\beta(\lambda)}{4\lambda}$, where $\beta(\lambda)$ is the beta function of Yang-Mills theory, we obtain
\begin{equation}
\rho_b^{\text{pert}}(\omega,T)=d_A\frac{\omega^4}{576\pi}\frac{\beta(\lambda)^2}{\lambda^2}\bigl( 1 + 2 n_{\frac{\omega}{2}} \bigr)\Bigg\{1+\frac{\lambda}{8\pi^2}\bigg(\frac{44}{3}\ln\frac{\bar{\mu}}{\omega}+\frac{73}{3}+8\phi_T^\theta({\omega\over T})\bigg)\Bigg\}\nonumber +
O(\lambda^4), \label{rhobper0} 
\end{equation}
where $\phi_T^\theta(\omega /T)$ is again some numerical function, see \cite{Laine:2011xm} for details.  
>From here, we can find the large-$\omega$ behavior of the quantity,
\begin{equation}
\rho_b^{\text{pert}}(\omega,T) \stackrel[\omega\to \infty]{}{\rightarrow} \frac{121 d_A^2\omega^4}{324(4\pi)^5}\lambda^2, \label{rhoperas}
\end{equation}
An important difference to the shear channel result is clearly the appearance of the 't Hooft coupling in the leading large-$\omega$ behaviour of eq.~(\ref{rhoperas}). Together with the realization that the renormalization scale, with which the coupling runs, is in the limit $\omega\gg T$ necessarily proportional to $\omega$, this implies that the leading UV behaviour of the bulk spectral function takes the form of a $T$-independent constant times $\omega^4/(\ln\,\omega/\Lambda_\tinymsbar)^2$. 

In fig.~\ref{figbulk2} (right) we plot the bulk spectral density in both IHQCD and perturbative QCD. We find an excellent agreement between our holographic result and the perturbative one in the region where the perturbative calculation can be trusted. What is extraordinary is that these results match even including their overall normalization, as they both reproduce the same logarithmic behavior.

Another surprising issue is clearly the striking difference in the origin of the logarithmic large-$\omega$ limits of the  spectral densities in IHQCD and perturbative QCD. In the latter, this behavior has its origins in the running of the gauge coupling, while in IHQCD  the behavior originates entirely from the fluctuation of the $h_{ii}$ metric component. Indeed, the IHQCD beta function (\ref{beta13}) is $\omega$-independent. A possible resolution to this problem could come from considering fluctuations of both the metric and the dilaton field, as was done for example in \cite{Springer:2010mf,Springer:2010mw}. We, however, leave a closer inspection of this issue to future work.
\subsection{Euclidean Correlators\label{latticesec}}
\begin{figure}[t]
\centering
\includegraphics[width=0.48\textwidth]{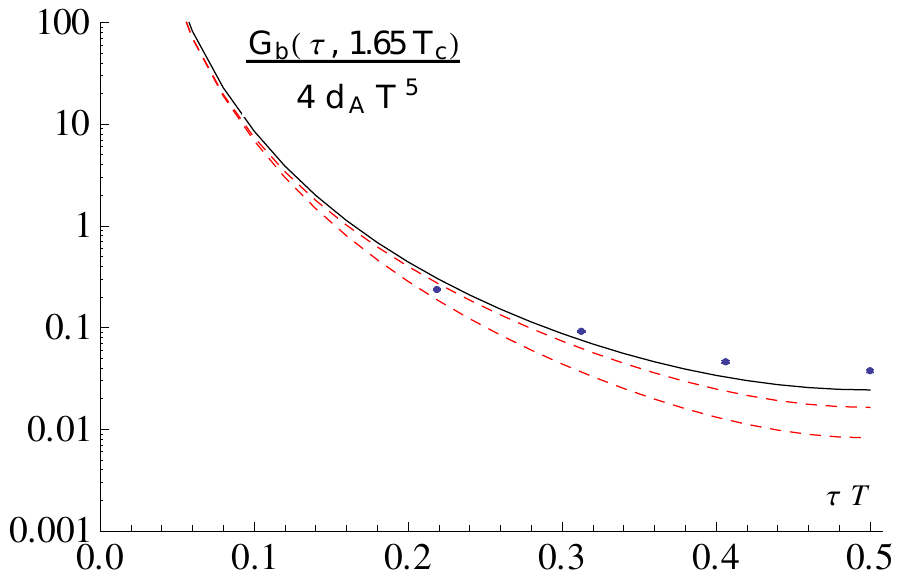}$\;\;\;$ \includegraphics[width=0.48\textwidth]{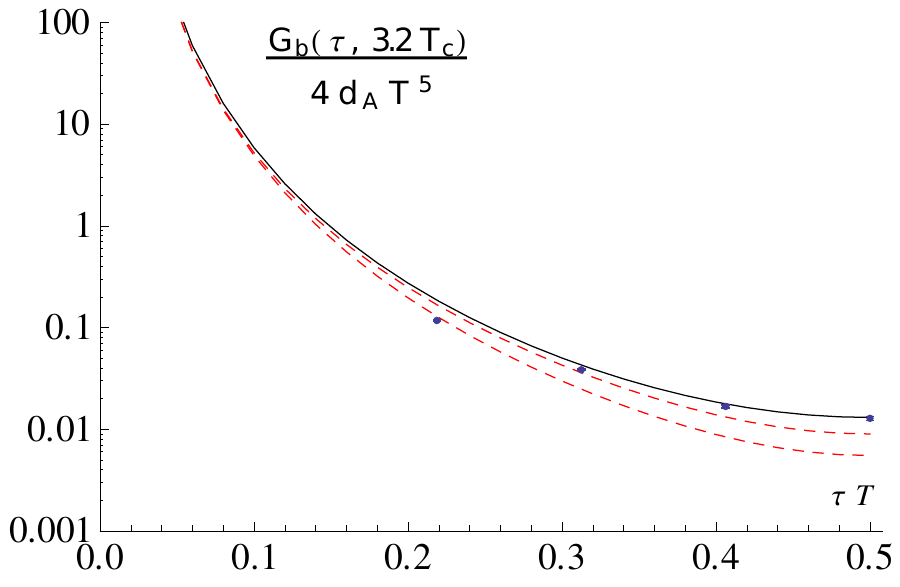}
\caption{\small The imaginary time correlators of the bulk channel, computed for two different temperatures in IHQCD (black curves) and perturbative QCD (red dashed curves) \cite{Laine:2011xm} and compared with the lattice data (blue points) of \cite{Meyer:2010ii}.}
\label{figimaginary}
\end{figure}

Lattice simulations  are the only fundamentally nonperturbative first principles method available to study the strongly coupled regime of QCD. However, due to  technical reasons it is only possible to measure  Euclidean correlators  on the lattice, where they can be obtained from the spectral density (\ref{rhobulk}) via 
\begin{equation}
G_b(\tau,T)=\int_0^\infty\frac{d\omega}{\pi}\rho_b(\omega,T)\frac{\cosh\left[\left(
\frac{\beta}{2} -\tau\right) \pi \omega\right]}{\sinh\left( \frac{\beta}{2}\omega\right)}\, ,\quad \beta\equiv 1/T\, ,
\label{Gtau}
\end{equation}
Using this relation, we can calculate Euclidean correlators in both IHQCD and perturbative QCD, and furthermore compare the results with the lattice data of \cite{Meyer:2010ii}. In fig.~\ref{figimaginary}, we plot the  results of all three methods for two different temperatures,  $1.65T_c$ and $3.2T_c$.  We can see that the holographic results seem to be in better agreement with the lattice data than the perturbative ones. This is true for all temperatures  but the difference is  most pronounced close to $T_c$. This is consistent with our expectations that the perturbative approach fails in the strongly coupled regime.  

In fig.~\ref{figbrhosymmpoint}, we plot the value of the imaginary time correlator at the symmetry point $\tau=1/(2T)$ as a function of temperature, normalized dimensionless by $T^5$. The plot  shows a rapid decrease in the quantity with increasing temperature, which confirms that the system indeed approaches the conformal limit as the temperature is increased.
\begin{figure}[t]
\begin{center}
 \includegraphics[width=0.7\textwidth]{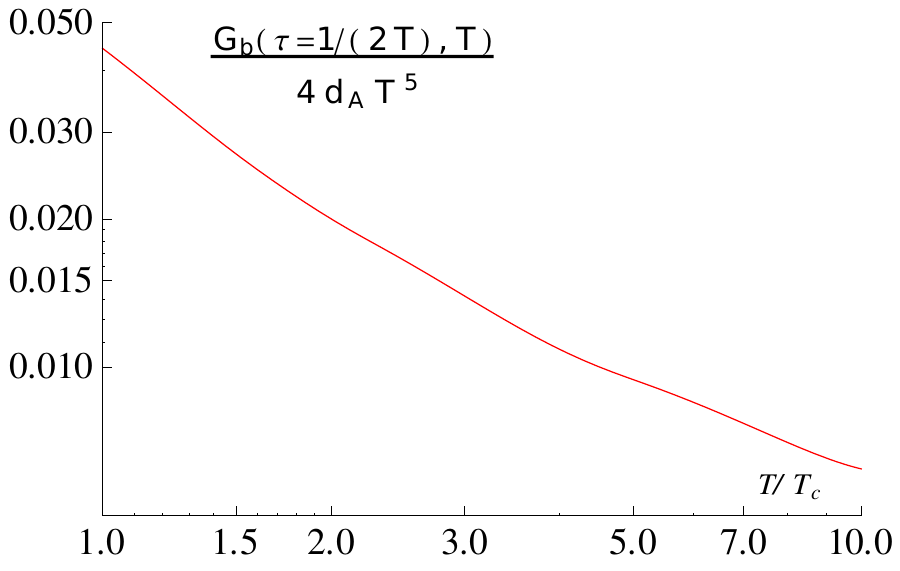}
 \end{center}
 \caption{The imaginary time correlator of eq.~(\ref{Gtau}), normalized by $T^5$ and plotted as a function of temperature at the symmetry point $\tau=1/(2T)$.}
 \label{figbrhosymmpoint}
 \end{figure}
\chapter{Conclusions\label{ch7}}
In this thesis, we have studied the properties of hot, deconfined $SU(N)$ Yang-Mills plasma using a holographic model of the strong interactions. We motivated this by briefly reviewing recent experimental results from the LHC and RHIC colliders, which indicate that the created quark gluon plasma (QGP) should be described rather as a strongly coupled liquid than a gas of weakly interacting quasiparticles. A particularly surprising lesson from these experiments is that the success of hydrodynamical simulations depends crucially on the introduction of a very small, yet nonzero shear viscosity. A first principle, nonperturbative derivation of this parameter using standard quantum field theory methods is, however, very problematic. Due to this reason and the success of holographic methods in the description of the bulk thermodynamics of QCD-like theories, we have chosen to approach transport phenomena in the theory using the AdS/CFT correspondence. 

Specifically, we have decided to use the \textit{Improved Holographic QCD} (IHQCD) model  to study the physics of the QGP. As we have explained, this model is based on a two-derivative approximation of   noncritical string theory and contains a nontrivial potential for the dilaton field that encodes conformal symmetry breaking. We demonstrated that choosing a specific form of the dilaton potential, one is able to mimic most of the crucial properties of QCD. 

The novel scientific results derived in this thesis have to do with an IHQCD determination of energy momentum tensor correlators in the shear $\langle T_{12}T_{12}\rangle$ and bulk $\langle T_{ii}T_{jj}\rangle$ channels at vanishing spatial momentum. The respective viscosities are given by the low-frequency limits of the imaginary parts of these correlators, but the full functions carry much more information about the properties of the theory. In particular, computing the full correlators enables one to perform interesting comparisons between the holographic approach and lattice as well as perturbative QCD.

In the shear channel of the IHQCD model, the ratio of the shear viscosity to entropy  was found to be consistent with the universal prediction $\eta/s=1/(4\pi)$, which was only expected. Beyond this, the full correlator was confronted with its counterpart in conformal $\mathcal{N}=4$ SYM theory, and we observed a sizeable difference between IHQCD and the conformal theory at temperatures close to $T_c$. This difference was seen to disappear rapidly with increasing temperature, which we concluded to be consistent with the expectation that IHQCD (and large-$N$ Yang-Mills theory) becomes effectively conformal in this limit. In addition, we also compared our IHQCD results with state-of-the-art perturbative calculations. We found that up to overall normalization, there is very good agreement between these two approaches in the domain of validity of perturbation theory.

After considering the shear channel, we moved on to use IHQCD to calculate energy momentum correlators in the bulk channel, for which the predictions of the conformal ${\mathcal N}=4$ SYM theory vanish. Again, we observed an approach of the system towards the conformal limit as the temperature was increased. Also, we found that IHQCD reproduces exactly the perturbative logarithmic behavior of the bulk spectral density in the large-$\omega$ limit. It was interesting to observe, how these two approaches converge to the same result, even though the logarithmic terms have very different origins in the two calculations.   
 
Using both the IHQCD and perturbative QCD results, we finally calculated Euclidean correlators, which we compared with state-of-the-art lattice simulations in the bulk channel. We found an extraordinarily good agreement between IHQCD and lattice QCD, and concluded that lattice data  manifestly favors the IHQCD results over the perturbative ones in the region of small to intermediate $\omega$ and  $T$.  

In the near future, we would like to use IHQCD to address the following two problems. First, we would like to determine the correlator of the operator $\Tr F^2$. Note that while in conformal ${\mathcal N}=4$ SYM theory the $T_{12}$ and $\Tr F^2$ operators are identical, while $T_{\mu}^\mu=0$, in a non-conformal theory all these three correlators are independent. In order to solve the problem, one needs to consider  fluctuations of both the metric and the dilaton, similarly to what was done in \cite{Springer:2010mf,Springer:2010mw}. We also expect this calculation to be helpful in understanding the large-$\omega$ behavior of the bulk spectral density.

Another technically very demanding, yet highly interesting problem is to explore the effects of conformal invariance breaking in holographic thermalization. Also here, we plan to tackle the problem within IHQCD, following the approach of \cite{Baier:2012tc,Baier:2012ax,Steineder:2012si,Steineder:2013ana} in the SYM theory.

\appendix
\setcounter{chapter}{0}
\renewcommand{\chaptername}{Appendix}
\renewcommand{\theequation}{\Alph{chapter}.\arabic{section}.\arabic{equation}}
\addcontentsline{toc}{chapter}{\numberline{}Appendix}
\setcounter{equation}{0}
\chapter{Anti-de Sitter Space \label{append}}
The Anti-de Sitter (AdS) space is the maximally symmetric solution to Einstein equations with a negative cosmological constant.
\section{Anti-de Sitter Space}
We begin by considering the Einstein-Hilbert action (\ref{EHaction}) with a cosmological term, that is
\begin{equation}
S_{EH+\Lambda}=\frac{1}{16\pi G_{D+1}}\int dx^{D+1}\sqrt{-g}(R- \Lambda), \quad \quad \Lambda=-\frac{D(D-1)}{\Lag^2},
\end{equation}
where $\Lag$ is the  AdS radius. From here, we find the Einstein equations
\begin{equation}
R_{\mu\nu}-\frac{1}{2}g_{\mu\nu}R=\frac{1}{2}\Lambda g_{\mu\nu},
\end{equation}
and contracting this equation obtain
\begin{equation}
R=\frac{D+1}{1-D}\Lambda,
\end{equation}
allowing us to rewrite the Einstein equations in the form
\begin{equation}
R_{\mu\nu}=\frac{\Lambda}{1-D}g_{\mu\nu}. \label{adseq}
\end{equation}

Spacetimes having the property that the Ricci tensor is proportional to the metric tensor are called \textit{Einstein spaces}. The AdS spacetime is furthermore an example of maximally symmetric Einstein spaces, i.e.~spaces satisfying
\begin{equation}
R_{\mu\nu\rho\sigma}=\frac{R}{(D+1)D}\left(g_{\nu\sigma}g_{\mu\rho}-g_{\nu\rho}g_{\mu\sigma}\right).
\end{equation}
We now assume spherical symmetry, and proceed to solve the Einstein equations to find a metric of the AdS space, resulting in
\begin{equation}
ds^2_{D+1}=-\left(1+\frac{r^2}{\Lag^2}\right)dt^2+ \frac{dr^2}{1+\frac{r^2}{\Lag^2}}+r^2d\Omega^2_{D-1}. \label{ads6}
\end{equation} 
However, there is also a more fundamental way to find the metric of this spacetime that does not rely on field equations, but rather only on symmetries. Similarly to viewing a sphere as a Euclidean space with constant positive curvature, we can namely view Anti-de Sitter space as a Lorentzian space with negative curvature. 

The most natural way to define a $D+1$-dimensional sphere is by embedding it in $D+2$ dimensional space. In the case of the AdS space, the situation is similar. We start from a $D+2$-dimensional Minkowski space with two time directions
\begin{equation}
ds^2=-dt^2 +d\vec{x}^2  +dz^2  - d\tilde{t}^2,
\end{equation}
and define the embedding equation as 
\begin{equation}
-t^2 +\vec{x}^2  +z^2  - \tilde{t}^2=-\Lag^2.
\end{equation}
Solving this equation, we find that the metric of the AdS space reads
\begin{equation}
ds^2_{D+1}=\frac{\Lag^2}{z^2}\left(-dt^2 +d\vec{x}^2  +dz^2\right),
\label{purads}
\end{equation}
which can be proved to be simply a different coordinate system representation of the metric (\ref{ads6}).

Recall now the definition of conformal symmetry from eq.~(\ref{confransdef}). The above form of the metric clearly has advantage that the conformal structure of the AdS space is manifest.

\section{Anti-de Sitter Black Holes}
We can find the black hole solution of the Einstein equations (\ref{adseq}) in perfect analogy to the vacuum Schwarzchild solution \cite{Petersen:1999zh}. We proceed by considering the metric ansatz
\begin{equation}
ds^2_{D+1}=-\tilde{f}(r)dt^2+\frac{dr^2}{\tilde{f}(r)}+r^2d\Omega^2_{D-1}. \label{aap2}
\end{equation} 
Solving the Einstein equations, we then find
\begin{equation}
\tilde{f}(r)=A+\frac{B}{r^{D-2}}+\frac{r^2}{\Lag^2}, \label{app1}
\end{equation}
where $A$ and $B$ are some constants that we can determine through the requirement that eq.~ (\ref{app1}) reproduces the Schwarzchild solution in the limit of a vanishing cosmological constant, i.e.~$\Lag\to\infty$. We find that this is achieved if 
\begin{equation}
A=1, \quad \quad B=w_D M= \frac{16 \pi G_{D+1}}{(D-1)\Omega_{D-1}}M,
\end{equation}
where $M$ can be identified with the mass of the black hole. Thus we can write the metric of the AdS black hole in the form
\begin{equation}
ds^2_{D+1}=-\left(1-\frac{w_D M}{r^{D-2}}+\frac{r^2}{\Lag^2} \right)dt^2+\left(1-\frac{w_D M}{r^{D-2}}+\frac{r^2}{\Lag^2} \right)^{-1}dr^2+r^2d\Omega^2_{D-1}. \label{aap3}
\end{equation}
The function $\tilde{f}(r)$ vanishes at certain values of $r$, which correspond to particular horizons. The largest value where it vanishes is denoted by $r_h$, with the physical space (outside the black hole horizon) corresponding to $r\geq r_h$. 

The above black hole has a temperature \cite{Hawking:1982dh}
\begin{equation}
T=\frac{1}{4\pi}\frac{Dr_h^2+(D-2)\Lag^2}{r_h \Lag^2}.
\end{equation}
The horizon location is on the other hand defined by $\tilde{f}(r_h)=0$, which gives
\begin{equation}
0=1-\frac{w_D M}{r_h^{D-2}}+\frac{r_h^2}{\Lag^2} \approx -\frac{w_D M}{r_h^{D-2}}+\frac{r_h^2}{\Lag^2},
\end{equation}
where the last approximation holds in the limit of a large mass. We can solve the above equation and find the position of the horizon
\begin{equation}
r_h=(Mw_D\Lag^2)^\frac{1}{D}. \label{rplusdef}
\end{equation}
In the large mass limit, we further have
\begin{equation}
ds^2_{D+1}=-\left(\frac{r^2}{\Lag^2}- \frac{w_DM}{r^{D-2}}\right)dt^2+
\left(\frac{r^2}{\Lag^2}- \frac{w_DM}{r^{D-2}}\right)^{-1}dr^2
+r^2d\Omega^2_{D-1}, \label{aap7}
\end{equation}
while in the so-called planar limit, where we consider the radius of the sphere $\Omega_{D-1}$ to be large, we can write
\begin{equation}
r^2d\Omega^2_{D-1}\approx \frac{r^2}{\Lag^2}d\vec{x}^2,
\end{equation}
Using this, above metric takes form
\begin{equation}
ds^2_{D+1}=b^2(r)\left[-f(r)dt^2 +  d\vec{x}^2 \right] + \frac{dr^2}{b(r)^2 f(r)},
\end{equation}
where 
\begin{equation}
b(r)=\frac{r}{\Lag}, \quad \quad f(r)=1-\frac{w_dM\Lag^2}{r^D}=1-\frac{r_h^D}{r^D},
\end{equation}
in the last equality of which we have used the definition (\ref{rplusdef}).

Using now a coordinate transformation $r\to\Lag^2/z$, we can rewrite the above metric as
\begin{equation}
ds^2_{D+1}=b^2(z)\left[-f(z)dt^2 +  d\vec{x}^2 +\frac{dz^2}{ f(z)}\right] ,
\end{equation}
where 
\begin{equation}
b(z)=\frac{\Lag}{z}, \quad \quad f(z)=1-\frac{z^D}{z_h^D}.
\end{equation}
Here, $z_h$ denotes the position of the horizon, and with the physical space corresponding to $z\leq z_h$. We can observe that in the limit $z_h\to \infty$, i.e.~for a vanishing black hole, the metric takes the form of pure AdS space (\ref{purads}). The Hawking temperature of the above solution on the other hand reads
\begin{equation}
T=\frac{1}{4\pi} \frac{D}{z_h}.
\end{equation}  

\lhead[\fancyplain{}{}]
      {\fancyplain{}{{\bfseries\rightmark}} }
\bibliographystyle{utphys}
\bibliography{thesis}
\end{document}